\documentclass{book}
\usepackage{makeidx,epsfig,subfigure}
\usepackage{setspace,graphicx}
\usepackage{cite}
\usepackage{Generic}
\newcommand{\avstress}{\langle\sigma\rangle}
\newcommand{\be}{\begin{equation}}
\newcommand{\ee}{\end{equation}}
\newcommand{\bea}{\begin{eqnarray}}
\newcommand{\eea}{\end{eqnarray}}
\newcommand{\vrel}{v^{\mbox{\small rel}}}
\newcommand{\vm}{v^{\mbox{\small m}}}
\newcommand{\vs}{v^{\mbox{\small s}}}
\newcommand{\Dm}{D^{\mbox{\small m0}}}
\newcommand{\Ds}{D^{\mbox{\small s0}}}
\newcommand{\sigmap}{\sigma^{\mbox{\small pol}}}
\newcommand{\sigmapol}{\sigma^{\mbox{\small pol}}}

\newcommand{\sigmatot}{\sigma}
\newcommand{\etasol}{\eta^{\mbox{\small sol}}}
\newcommand{\etapol}{\eta^{\mbox{\small pol}}}
\newcommand{\gdot}{\dot\gamma}
\newcommand{\gdotbar}{\bar{\gdot}}
\newcommand{\Dt}[1]{D_t{#1}}
\newcommand{\lae}{\stackrel{<}{\scriptstyle{\sim}}}

\newcommand{\vecv}[1]{\mathbf{{#1}}}

\newcommand{\ommax}{\omega_{\rm max}}

\begin{document}

\pagenumbering{roman}


\mainmatter
\pagenumbering{arabic}

\chapter{Rheology of Giant Micelles}


{\bf M. E. Cates}, SUPA School of Physics, University of Edinburgh, JCMB Kings Buildings, Mayfield Road, Edinburgh EH9 3JZ, United Kingdom
\\
and
\\
{\bf S. M. Fielding}, School of Mathematics, University of Manchester, Booth Street East, Manchester M13 9EP, United Kingdom

{\bf Abstract:} Giant micelles are elongated, polymer-like objects
created by the self-assembly of amphiphilic molecules (such as
detergents) in solution. Giant micelles are typically flexible, and
can become highly entangled even at modest concentrations. The
resulting viscoelastic solutions show fascinating flow behaviour
(rheology) which we address theoretically in this article at two
levels. First, we summarise advances in understanding linear
viscoelastic spectra and steady-state nonlinear flows, based on
microscopic constitutive models that combine the physics of polymer
entanglement with the reversible kinetics of self-assembly. Such
models were first introduced two decades ago, and since then have been
shown to explain robustly several distinctive features of the rheology
in the strongly entangled regime, including extreme shear-thinning. We
then turn to more complex rheological phenomena, particularly
involving spatial heterogeneity, spontaneous oscillation, instability,
and chaos. Recent understanding of these complex flows is based
largely on grossly simplified models which capture in outline just a
few pertinent microscopic features, such as coupling between stresses
and other order parameters such as concentration. The role of
`structural memory' (the dependence of structural parameters such as
the micellar length distribution on the flow history) in explaining
these highly nonlinear phenomena is addressed. Structural memory also
plays an intriguing role in the little-understood shear-thickening
regime, which occurs in a concentration regime close to but below the
onset of strong entanglement, and which is marked by a shear-induced
transformation from an inviscid to a gelatinous state.\footnote{This
  is a preprint of an article whose final and definitive form has been
  published in Advances in Physics (c) 2006 copyright Taylor \&
  Francis; Advances in Physics is available online at
  http://journalsonline.tandf.co.uk/. The URL of the article is
  http://journalsonline.tandf.co.uk/openurl.asp?genre=article\&id=doi:10.1080/00018730601082029.}

\section{Introduction}

\begin{figure}[tbp]
\centering
\includegraphics[width=4.0 in]{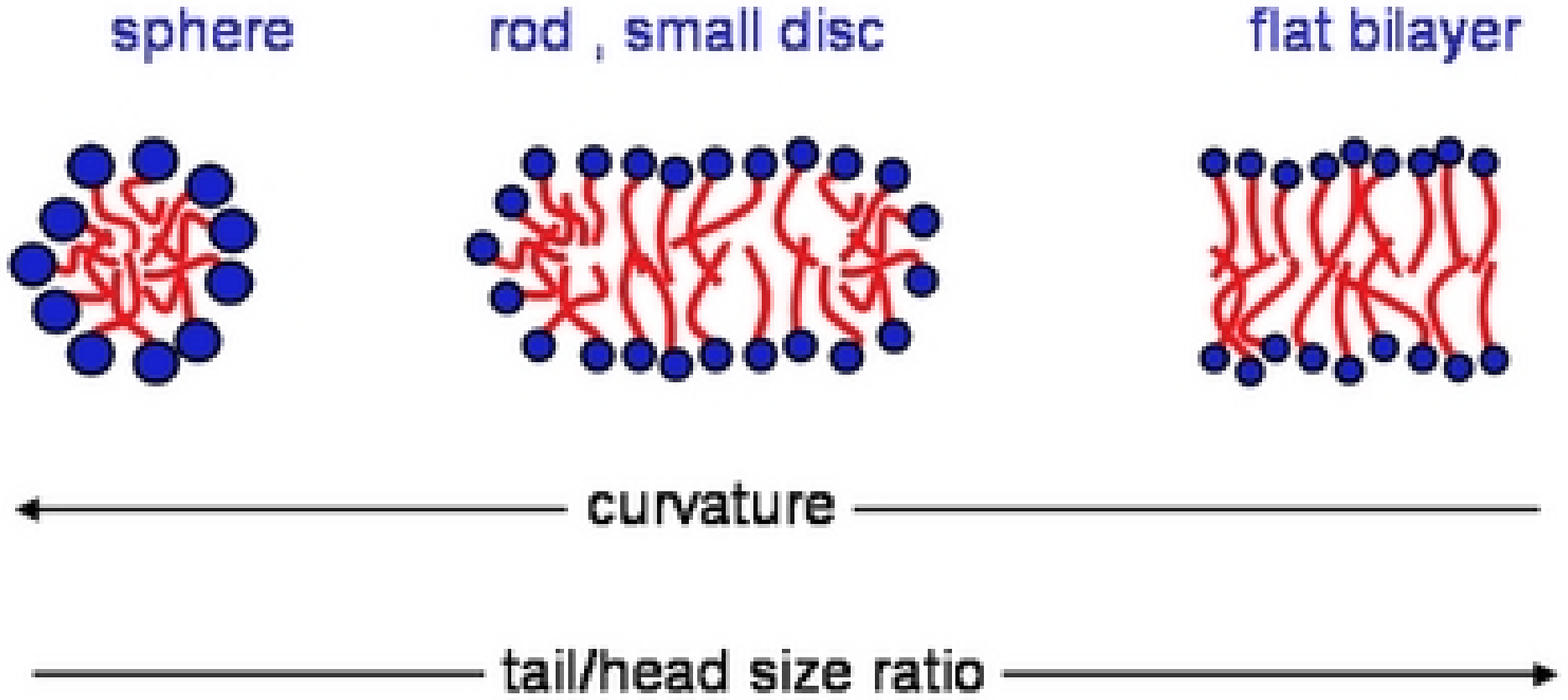}
\includegraphics[width=4.0 in]{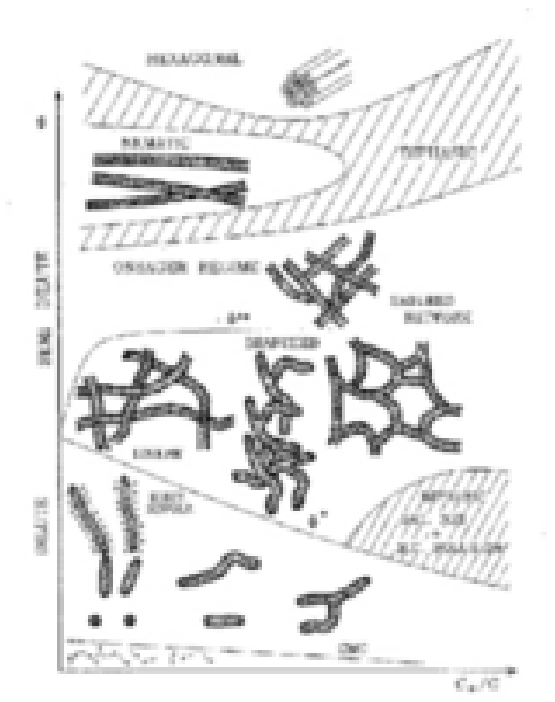}
\caption{Upper panel: Schematic view of aqueous surfactant self-assembly. Lower panel: Schematic phase diagram for self assembly of ionic amphiphiles into giant micelles and related structures. The vertical axis represents volume fraction $\Phi$ of amphiphile; the horizontal is the ratio $C_s/C$ of added salt to amphiphile concentrations. Figure in lower panel from F. Lequeux and S. J. Candau, in Ref.\cite{herb}, reprinted with permission.} \label{phasefig}
\end{figure}

An amphiphilic molecule is one that combines a water-loving (hydrophilic) part or `head group' with a with a water-hating (hydrophobic) part or `tail'. The head-group can be ionic, so that the molecule becomes charged by dissociation in aqueous solution; or nonionic (but highly polar, favouring a water environment), in which case the amphiphile remains uncharged. Zwitterionic head-groups, with two charges of opposite sign, are also common. The hydrophobic tail is almost always a short hydrocarbon (though fluorocarbons can also be used); in some cases (such as biological lipids) there are two tails. The most important property of amphiphilic molecules, from the viewpoint of theoretical physics at least, is their tendency to self-assemble by aggregating reversibly into larger objects. The simplest of these is a spherical aggregate called a `micelle' which in water has the hydrophobic tails sequestered at the centre, coated by a layer of headgroups; see Fig.\ref{phasefig}. (In a nonaqueous solvent, the structure can be inverted to create a `reverse micelle'.) 

For geometrical reasons, a spherical micelle is self-limiting in size: unless the amphiphilic solution contains a third molecular component (an oil) that can fill any hole in the middle, the radius of a micelle cannot be more than about twice the length of the amphiphile.
To avoid exposing tails to water, it also cannot be much less than this; the resulting `quorum' of a few tens of molecules for creation of a stable micelle leads to a sharp minimum in free energy as a function of aggregation number.
This collective aspect to micelle formation causes the transition from a molecularly dispersed solution to one of micelles to be rather sharp; micelles proliferate abruptly when the concentration is raised above the `critical micelle concentration' or CMC \cite{israelachvili}. 

In a spherical micelle the volume ratio of head- and tail-rich regions is also essentially fixed: such micelles are favoured by amphiphiles with relatively large size ratios between head and tail. Suppose this size ratio is smoothly decreased, for instance by adding salt to an ionic micellar solution (effectively reducing the head-group size by screening the coulombic repulsions).  The most stable local packing then evolves from the spherical micelle towards a cylinder; Fig.\ref{phasefig}. (Proceeding further, it becomes a flat bilayer; systems in which this happens are not addressed here.) Allowing for entropy, the transition from spheres to cylinders is not sudden, but proceeds via short cylindrical micelles with hemispherical end-caps. The bodies of these cylinders smoothly increase in length as the packing energy of the body falls relative to the caps; the micelles eventually become extremely long. The law of mass action, which favours larger aggregates, means that in suitable systems the same sequence can be observed by increasing concentration at fixed head/tail size ratio (fixed ionic strength). 

Since the organization of amphiphiles within the  cylindrical body is (in most cases) fluid-like, the resulting `giant micelles' soon exceed the so-called persistence length, at which thermal motion overcomes the local rigity and the micelle resembles a flexible polymer chain. This crossover may or may not precede the `overlap threshold' at which the volume occupied by a micelle -- the smallest sphere that contains it -- overlaps with other such volumes. Beyond this threshold, the chainlike objects soon become entangled but (unless extremely stiff) remain in an isotropic phase with no long-range orientational order. At very high concentrations, such ordering can arise, as can 
positional order, giving for example a hexagonal
array of near-infinite straight cylinders. Branching of micelles can also be important in both isotropic and ordered phases; a schematic phase diagram, applicable to many but not all systems containing giant micelles, is shown in Fig.\ref{phasefig}.

This article addresses primarily the isotropic phase of giant micelles, in a concentration range from somewhat below, to well above, the overlap threshold. 
This is the region where viscoelastic (but isotropic) solutions are observed. This regime merits detailed attention for two reasons. The first is that micellar viscoelasticity forms the basis of many applications, ranging from personal care products (shampoos) to specialist drilling fluids for oil recovery \cite{herb}. The second, is that as emphasised first by Rehage and Hoffmann \cite{rehagehoffmann}, viscoelastic micelles provide a uniquely convenient laboratory for the study of generic issues in nonlinear flow behaviour. This is partly because, unlike polymer solutions (which they otherwise resemble), the self-assembling character of micellar solutions causes them to self-repair after even the most violently nonlinear experiment. (In contrast, strong shearing of conventional polymers causes permanent degradation of the chains.) Our focus throughout this review is on rheology, which is the science of flow behaviour. 
Although we will refer in many places to experimental data, we make no attempt at a comprehensive survey of the experimental side of the subject, nor do we describe applications areas in any detail. For an up-to-date overview of both these topics, the reader is referred to a recent book \cite{GMB}, of which a shortened version of this article forms one Chapter.

We shall address the theoretical rheology of giant micelles at two levels. The first (in Section \ref{rrm}) is microscopic modelling, in which one seeks a mechanistic understanding of rheological behaviour in terms of the explicit dynamics ---primarily entanglement and reversible self-assembly--- of the giant micelles themselves. 
This yields so-called `constititive equations' which relate the stress in a material to its deformation history. Solution of these equations for simple experimental flow protocols presents major insights into the fascinating flow properties of viscoelastic surfactant solutions, including near-Maxwellian behaviour (exponential relaxation) in the linear regime, and drastic shear-thinning at higher stresses. These successes mainly concern the strongly entangled region where the micellar solution is viscoelastic at rest; in this regime, strong shear-thinning is usually seen. There are however equally strange phenomena occurring at lower concentrations where the quiescent solution is almost inviscid, but becomes highly viscoelastic after a period of shearing. These will also be discussed (Section \ref{thickening}) althought they remain, for the present, much 
less well understood.

Microscopic models of giant micelles under flow generally treat the micelles as structureless, flexible, polymer-like objects, albeit (crucially!) ones whose individual identities are not sustained indefinitely over time. This neglect of chemical detail follows a very successful precedent set in the field of polymer dynamics \cite{degennes,doiedwards}. There, models that contain only four static parameters (persistence length, an excluded volume parameter, the concentration of chains, and the degree of polymerization or chain length) and two more dynamic ones (a friction constant or solvent viscosity, and the so-called `tube diameter') can explain almost all the observed features of polymeric flows. Indeed, microscopic models of polymer rheology arguably represent one of the major intellectual triumphs of 20th century statistical physics \cite{degennesnobel}.

However, at least when extended to micelles, these microscopic constitutive models remain too complicated to solve in general flows, particularly when flow instabilities are present. (Such instabilities are sometimes seen in conventional polymer solutions, but appear far more prevalent in micellar systems.) Moreover, they omit a lot of the important physics, particularly couplings to orientational fields and concentration fluctuations, relevant to these instabilities. Therefore we also describe in Section \ref{macro} some purely macroscopic constitutive models, whose inspiration stems from the microscopic ones but which can go much further in addressing the complex nonlinear flow phenomena seen in giant micelles. These phenomena include for example ``rheochaos'', in which a steady shear deformation gives chaotically varying stress or vice versa. Our discussion of macroscopic modelling will take us to the edge of current understanding of these exotic rheological phenomena.

Prior to discussing rheology, we give in Section \ref{statics} a brief survey of the equilibrium statistical mechanics of micellar self-assembly. More detailed discussions of many of the static equilibrium properties of micelles can be found in \cite{GMB}; we focus only on those aspects needed for the subsequent discussion of rheology. Another key component for rheological modelling is the kinetics of micellar `reactions' whereby micelles fragment and/or recombine. These reactions are of course already present in the absence of flow, and represent the kinetic pathway whereby equilibrium (for quantities such as the micellar chain-length distribution) is actually reached. We review their properties also in Section \ref{statics}.

\label{secondary}
In developing the equilibrium statistical mechanics and kinetic theory for giant micelles (Section \ref{statics}), we should keep in mind both the successes and limitations of the rheological theories that come later. Such theories, since their first proposal by one of us in 1987 \cite{rrm} have had considerable success in predicting the basic features of linear viscoelastic relaxation spectra observed in experiments, and in inter-relating these, for any particular chosen system, with nonlinear behaviour such as the steady-state dependence of stress on strain rate. 
These dynamical models take as input the micellar size distribution, stiffness (or persistence length) and the rate constants for various kinetic processes that cause changes in micellar length and topology. Such inputs are theoretically well defined, but harder to measure in experiment. Nonetheless, there are a number of `primary' predictions (such as the shape of the relexation spectrum, and the inter-relation of linear and nonlinear rheological functions; see Section \ref{predictions}) for which the unknown parameters can either be fully quantified, or else eliminated. 
As an aid to experimental comparison, it is of course useful to ask how the rheological properties should depend on thermodynamic variables such as surfactant concentration, temperature, and salt-levels in the micellar system (Section \ref{secondarypredictions}). But in addressing these `secondary' issues, the dynamical models can only be as good as our understanding of how those thermodynamic variables control the equilibrium micellar size distribution, persistence length, and rate constants, as inputs to the dynamical theory. In many cases this understanding is only qualitative, so that these `secondary' experimental tests should not be taken as definitive evidence for or against the basic model.

\section{Statistical Mechanics of Micelles in Equilibrium}
\label{statics}

In line with the above remarks, we focus mainly on those aspects of equilibrium self-assembly that can affect primary rheological predictions. Most of the thermodynamic modelling can be addressed within mean-field-theory approaches (Sections \ref{mfpol}--\ref{rings}), although more advanced treatments show various subtleties that still await experimental clarification (Section \ref{bmf}). In Section \ref{kinetics} we turn to the kinetic question of how micelles exchange material with one another within the thermal equilibrium state.

\subsection{Mean Field Theory: Living Polymers}
\label{mfpol}

In typical giant micellar systems the critical micelle concentration (CMC) is low -- of order $10^{-4}$ molar for CTAB/KBr, for example.
(CTAB, cetyltrimethylammonium bromide, is a widely studied amphiphile. In what follows, we do not expand the acronyms for this or other such materials as their chemical formulas are rarely of interest in our context. KBr is, as usual, potassium bromide, added to alter the head-group interactions.) As concentration is raised above the CMC, uniaxial elogation occurs and soon micelles become longer than their persistence length $l_p$. This is the length over which appreciable bending occurs \cite{degennes}; once longer than this, micelles resemble flexible polymers.  Persistence lengths of order 10 - 20 nm are commonplace, though much larger values are possible in highly charged micelles at low ionic strength. 

As concentration is increased, there is an onset of viscoelastic behaviour at a volume fraction  $\tilde C$ usually identified with an `overlap' concentration $C^*$ for the polymers. (For problems with this identification, see \ref{rings}
below.) Above $C^*$, the wormlike micelles are in the so-called `semidilute' range of concentrations \cite{degennes} -- overlapped and entangled at large distances, but well separated from one another at scales below $\xi$, the correlation length or `mesh size'. In ordinary polymer solutions in good solvents, the behaviour at scales less than $\xi$ is not mean-field-like but described by a scaling theory with anomalous exponents \cite{degennes}. We return to this in Section \ref{bmf}, but note that these scaling corrections become small when the persistence length of a micellar cylinder is much larger than its diameter, giving modest values for a dimensionless `excluded volume parameter' $w$ \cite{degennes,doiedwards}. Therefore, a mean-field approach -- in which excluded volume interactions are averaged across the whole system rather than treated locally -- captures the main phenomena of interest, particularly in the regime of strong viscoelasticity at $C \ge \tilde C$.

The simplest mean field theory \cite{scott,mukerjee} assumes that no branch-points and no closed rings are present (rectified in Sections \ref{branch}, \ref{rings}), and ascribes a free energy $E/2$ to each hemispherical endcap of a micelle relative to the free energy of the same amount of amphiphilic material residing in the cylindrical body. Denoting by $c(N)$ the number density of aggregates containing $N$ amphiphiles or `monomers', the mean field free energy density obeys 
\begin{equation}
\beta F = \sum_Nc(N)[\ln c(N) + \beta E] + F_0(\phi)
\label{meanfield}
\end{equation}
Here $\beta = 1/k_BT$; the term in $E$ counts two end-caps per chain, and the $c\ln c$ piece comes from the entropy of mixing of micelles of different lengths. Within a mean-field calculation, these are the {\em only} terms sensitive to the size distribution $c(N)$ of the micelles; the free energy (including configurational entropy) of the cylindrical sections, alongside their excluded-volume interactions and all solvent terms, give the additive piece $F_0(\phi)$ which depends only on total volume fraction $\phi$. (It may also depend on ionic strength and related factors.) The volume fraction obeys
\begin{equation}
\phi = v_0C = v_0\sum_N N c(N)
\label{phidef}
\end{equation}
where $v_0$ is the molecular volume of the amphiphiles and $C$ their total concentration.

Minimizing (\ref{meanfield}) at fixed $\phi$ gives an exponential size distribution
\begin{equation}
c(N) \propto \exp[-N/\overline N] \;\;\; ; \;\;\;
\overline N \simeq \phi^{1/2}\exp[\beta E/2]
\label{size}
\end{equation}
The exponential form in each case is a robust result of mean field theory. The $\phi$-dependence in the second equation is also robust (it follows from mass action), but can be treated separately from the much stronger exponential factor only so long as parameters like $E$ and $v_0$ are themselves independent of concentration. (In ionic systems this is a strong and questionable assumption.) 
The formula for $\overline N$ as written in (\ref{size}) suppresses prefactoral dependences on $v_0,l_p$ and $a_0$, where $a_0$ is the cross-sectional area of the micellar cylinders; these are absorbed into our definition of $E$. So long as $a_0$ is constant, then exactly the same functional forms as in (\ref{size}) control $c(L)$ and $\bar L$, where $L\propto N$ is the contour length of a micelle. Within mean field, $L$ in turn controls the typical geometric size $R$ (usually chosen as either the end-to-end distance, or the radius of gyration) of a micelle via $R^2\simeq Ll_p$. This is the well-known result for gaussian, random-walk chain configurations \cite{degennes}. 

We can now work out, within our mean-field approach, the overlap concentration $C^*$, or overlap volume fraction $\phi^* = C^*v_0$. For a micelle of the typical contour length $\overline L$ we have $R\simeq n^{1/2}l_p$ where $n = \overline L/l_p$ is the number of persistence length it contains; this obeys $nl_pa_0/v_0 = \overline N$. The total volume of amphiphile within the region spanned by this micelle is  $\overline N v_0$ and the volume fraction within it therefore $\phi\simeq \overline Nv_0/R^3$. At the threshold of overlap, this $\phi$ equates to the true value $\phi^*$; then eliminating $\overline N$ via (\ref{size}) gives
\begin{equation}
C^*v_0 = \phi^* \simeq  (a_0/v_0^{1/3}l_p)^{6/5}e^{-\beta E/5}  
\label{phistar}
\end{equation}
For typical cases the dimensionless pre-exponential factor is smaller than unity, but nonetheless a fairly large $E$ is required if $\phi^*$ is to be below, say $1\%$. The regime of long, entangled micelles usually entails scission energies $E$ of around $10-20k_BT$; in practice, experimental estimates of $\phi^*$ (best determined by light scattering) are often in the range 0.05--5\% \cite{catescandau}. The scission energy $E$ of course depends on the detailed chemistry of the surfactant molecules and this (alongside micellar stiffness or persistence length) is one of the main points at which such details enter the theory. Very crudely, one can argue that doubling the mean curvature of a micellar cylinder to create an end-cap must cost about $k_BT$ per molecule in the end-cap region. (If packing energies were much higher than this, one would expect a crystalline rather than fluid packing on the cylinder, which is not typically observed, at least at at room temperature.) This gives $E\sim nk_BT$ where $n$ is the number of molecules in two endcaps. Within a factor two, this broadly concurs with the range $10-20k_BT$ stated above. More precise theoretical estimates also concur with this range, although values well outside of it are also possible for atypical molecular geometries, e.g. fluorosurfactants \cite{oelschlagermemory}.

The region around $\phi^*$ is where spectacular shear-thickening rheology occurs (see Section \ref{thickening}). In ionic micellar systems without excess of salt, the strong dependence of $l_p, E$ and other parameters on $\phi$ itself in this region means that the simple calculations leading to (\ref{size}), and hence the estimate (\ref{phistar}), are at their least reliable. More detailed theories, which treat electrostatic interactions explicitly, give a far stronger dependence of $\overline L$ on $\phi$ and also a narrower size distribution for the micelles \cite{mackintosh}. The overlap threshold $\phi^*$ itself moves to higher concentration due to the electrostatic tendency to stabilise short micelles.

\subsection{Role of Branching: Living Networks}
\label{branch}
The above assumes no branching of micelles. A mean-field theory can in principle be formulated to deal with self-assembled micellar networks having arbitrary free energies for both end caps and branch points \cite{drye}. This is, however, somewhat intractable for the general case. Fortunately things simplify considerably in the branching-dominated limit; that is, when there are many branch-points per end-cap.
For branching via $z$-fold `crosslinks' (each of energy $E_b$) one has, replacing (\ref{meanfield}), the following mean-field result \cite{drye}:
\begin{equation}
\beta F = \sum_Nc(N)[\ln c(N) + \beta E_bz^{-1}] + 2(z^{-1}-1)C\ln(2C)+
F_0(\phi)
\label{dryemeanfield}
\end{equation}
where $C$ is as defined in (\ref{phidef}), and  
$c(N)$ is now the concentration of network strands containing $N$ amphiphiles. 
To understand this result, note that the first logarithmic term is the translational entropy of a set of disconnected network strands. The second such term estimates the entropy loss on gathering the ends of these strands locally to form $z$-fold junction points. The term in $E_b$ counts the energy of these junctions and $F_0(\phi)$ has the same meaning as in
(\ref{meanfield}). 
The value of $z$ most relevant to micelles is $z = 3$, since for a system whose optimal local packing is a cylinder, a three-fold junction costs less in packing energy than $z>3$. Low $z$ is also favoured entropically: to create a four-fold junction one must fuse two three-fold ones with consequent loss of translational entropy along the network \cite{drye}. 

Minimizing (\ref{dryemeanfield}) to find the equilibrium strand length distribution, one finds this again to be exponential, with mean strand length $\overline L \sim\phi^{1-z/2}\exp[\beta E_b]$.
This result applies whenever the geometric distance between crosslinks, $\Lambda \simeq (\overline L l_p)^{1/2}$ greatly exceeds the geometric mesh size $\xi$, which within mean field theory obeys $\xi \sim a_0/\phi l_p$. This situation of $\Lambda \gg \xi$ is called an `unsaturated network' \cite{drye} and arises at high enough concentrations ($\phi \gg \phi^{sat}\simeq v_0^{-1}\exp[-\beta E/(3-z/2)]$). For $\phi \le \phi^{sat}$ one has a `saturated network' with $\Lambda\simeq\xi$. At low enough $\phi$ this saturated network can show a miscibility gap, where excess solvent is expelled from the system rather than increase $\xi$ which would sacrifice branch point entropy   \cite{drye}.

The rheology of living networks (see Section \ref{networks}) should differ strongly from the unbranched micellar case. Such a regime has been identified in several systems, primarily cationic surfactants at relatively high ionic strength \cite{lequeux,networkrefs,portesatnet}.
These accord with the expected trend for curvature packing energies: adding salt in these systems stabilizes negatively curved branch-points relative to positively curved end-caps \cite{networkrefs}.

\subsection{Role of Loop Closure: Living Rings}
\label{rings}
We have assumed in (\ref{meanfield}) that rings do not arise. A priori, however, there is nothing to stop micelles forming closed rings. Moreover, for unbreakable polymers (at least) the effect on rheology of closing chains to form rings is thought to be quite drastic \cite{ringpolymers}, so this assumption merits detailed scrutiny. It turns out to be satisfactory only when $E$ is not too large, so that $\phi^*$ in (\ref{phistar}) lies well above a certain volume fraction $\phi_{r}^{max}$, defined below, which signifies a maximal role for ring-like micelles. 

From (\ref{phistar}), as $E\to\infty$, the overlap threshold $\phi^*$ for open micelles tends to zero. In this limit, there is formally just a single micelle, of macroscopic length. This corresponds to an untenable sacrifice of translational entropy which is easily regained by ring formation. To study this, let us set $E\to\infty$ so that no open chains remain, but allow rings with concentration $c_r(N)$. Then, to replace (\ref{meanfield}), one has \cite{ppw}
\begin{equation}
\beta F = \sum_Nc_r(N)[\ln c_r(N) + \beta f_{r}(N)] + F_0(\phi)
\label{ringmeanfield}
\end{equation}
where $f_r(N) = -k_BT\ln(Z_r)$, and $Z_r$ is the configurational free energy cost of ring closure. Put differently, $E-f_r(N)$ is the total free energy cost of hypothetically opening a ring, creating two new endcaps but gaining an entropy $k_B\ln Z_r$. The latter stems both from the number of places such a cut could occur, and the extra configurations made available by allowing the chain ends to move apart.

For gaussian (mean-field-like) chains in three dimensions, it is easly shown that $Z_r = \lambda N^{-5/2}$ \cite{degennes}, where $\lambda$ is a dimensionless combination (as yet unknown \cite{catescandau}) of $a_0,v_0,l_p$. Minimizing (\ref{ringmeanfield}) at fixed $\phi$ then gives
\begin{equation}
c_r(N) = \lambda N^{-5/2}e^{\tilde\mu N}
\label{cring}
\end{equation}
where $\tilde\mu$ is a chemical-potential like quantity. Interestingly, this size distribution for rings shows a condensation transition. That is, for $\tilde \mu > 0$ the volume fraction $\phi_r = v_0\sum_NNc_r(N)$ is divergent, whereas for $\tilde \mu \le 0$ it can apparently be no greater than 
\begin{equation}
\phi_r^{max} = \lambda \sum_{N=N_{min}}^\infty N^{-3/2}
\label{maxring}
\end{equation}
This limiting value of $\phi_r$ depends not only on $\lambda$ but on $N_{min}$, which denotes the smallest number of amphiphiles that can make a ring-shaped micelle without prohibitive bending cost. (Such a micelle must presumably have contour length of a few times $l_p$.) 

This mathematical situation, in which there is an apparently unphysical `ceiling' $\phi_r^{max} < 1$ on the total volume fraction of rings a system can contain, is reminiscent of Bose condensation \cite{bose}. It represents the following physical picture, valid in the $E\to\infty$ limit. For $\phi \le \phi_r^{\max}$ one has the power law distribution of ring sizes in (\ref{cring}), cut off at large $N$ by an exponential multiplier (resulting from small negative $\tilde\mu$). For $\phi > \phi_r^{max}$, one has a pure power law distribution of rings, in which total volume fraction $\phi_r^{max}$ resides; plus an excess volume fraction $\phi-\phi_r^{max}$ which exists as a single `giant ring' of macroscopic length. This giant ring is called the condensate; its sudden formation at $\phi = \phi_r^{max}$ represents a true phase transition.

Obviously, an infinite ring is possible only because the limit $E\to\infty$ was taken; for any finite $E$, all rings with $N>\overline N$ as defined in (\ref{size}), including the condensate,
will break up into pieces (roughly of size $\overline N$). Indeed, if $E$ is finite one can, within mean-field theory, simply add the
chain and ring free energy contributions as
\begin{equation}
\beta F = \sum_Nc(N)[\ln c(N) + \beta E]+\sum_Nc_r(N)[\ln c_r(N) + \beta f_{r}(N)] + F_0(\phi)
\label{chainringmeanfield}
\end{equation}
From this one can prove that the condensation transition is smoothed out for any $E<\infty$ \cite{ringstatpaps}.
Nonetheless, if $E$ is large enough that the overlap threshold $\phi^*(E)$ obeying (\ref{phistar}) falls below $\phi_r^{max}$ obeying (\ref{maxring}), then the condensation transition of rings, though somewhat rounded, should still have detectable experimental consequences. These should mainly affect a (roughly) factor-two window in concentration either side of $\phi_r^{max}$. For $\phi \ll \phi_r^{max}$ there are no very large rings and hence limited opportunities for viscoelasticity. For $\phi \gg \phi_r^{max}$ the volume fraction $\phi-\phi_r$ of long chains exceeds that of rings, and the chains dominate.

Because of uncertainty over the values of $\lambda$ and $N_{min}$ in (\ref{maxring}) and how these might depend on chain stiffness, ionic strength, {\em etc.},  $\phi_r^{max}$ is one of the least well-charactarized of all static quantities for
giant micelles. In fact, there is relatively little (but some \cite{in}) experimental evidence for a ring-dominated regime in any micellar system, suggesting perhaps that, for reasons as yet unclear, $\phi_r^{max}$ lies well below the range of $\phi$ accessed in most experiments. However, as outlined in Section \ref{thickening} below, a ring-dominated regime might explain some of the strangest of all the rheological data in the shear-thickening regime just below $\tilde C$ \cite{rings}. 

Note that in an earlier review (Ref.~\cite{catescandau}) an impression was perhaps given that ring-formation matters only within scaling theories (discussed next) but not at the mean-field level. This is true only if $\phi_r^{max}$ is indeed small; in that case rings will only matter for very large $E$, and micelles are of sufficient size that excluded volume effects, even if locally weak, are likely to give scaling corrections to mean-field. However, rings are not purely a scaling phenomenon: even in a strictly mean-field picture, $\phi_r^{max}$ is a well-defined quantity at which a ring-condensation phase transition arises in the limit of large $E$.

\subsection{Beyond Mean Field Theory}
As with conventional polymers \cite{degennes}, 
micelles can exhibit non-gaussian statistics, induced by excluded-volume interactions arising from the inability of two different sections of the micelle to occupy the same spatial position.  In principle, this gives scaling corrections to all the preceding mean-field results, altering the various power law exponents that appear in equations such as (\ref{size},\ref{cring}).
As mentioned previously, however, micelles often have a persistence length $l_p$ large compared to their cross-section and therefore tend to have relatively weak excluded volume interactions. Therefore, giant micellar systems can often be expected to lie in a messy crossover region between mean-field and the scaling theory. For completeless we outline the scaling results here (see \cite{catescandau,drye,catesjphysique} for more details), but without attempting to track dependences on parameters like $l_p,a_0,v_0$.

First, consider a system with no branches or rings. In such a system, the excluded volume exponent $\nu \sim 0.588$ governs the non-gaussian behaviour of a self-avoiding chain; $R\sim L^{\nu}$ \cite{degennes}. This gives $\xi\sim\phi^{\nu/(1-\nu d)} \sim \phi^{-0.77}$ where $d = 3$ (the dimension of space). This leads  to a scaling of the transient elastic modulus $G_0$ (defined in Section \ref{rrm} below): $\beta G_0\sim\xi^{-d}\sim \phi^{2.3}$, which is proportional to the osmotic pressure $\Pi$ \cite{degennes}. This differs from the simplest mean-field-type estimate which has $G_0\sim \Pi\sim\phi^2$. Second, one finds in place of (\ref{size})
\begin{equation}
c(L) \propto \exp[-L/\overline L] \;\;\; ; \;\;\;
\overline L \simeq \phi^{y}\exp[\beta E/2]
\label{resize}
\end{equation}
where $y = [1+(\gamma-1)/(\nu d - 1)]/2 \simeq 0.6$; here $\gamma \simeq 1.2$ is another standard polymer exponent \cite{degennes}.
In practice this is rather a modest shift from 
the result in (\ref{size}).

In the presence of branch points, the important case remains $z = 3$. Here the mean field result for the mean network strand length
$\overline L \sim\phi^{-1/2}\exp[\beta E]$ becomes $\overline L \sim\phi^{-\Delta}\exp[\beta E]$ with an exponent $\Delta\sim 0.74$; the expression for $\Delta$ in terms of standard polymer exponents is given in Ref.~\cite{drye}. Similarly the mean-field result $\phi^{sat}\sim e^{-2\beta E/3}$ becomes $\phi^{sat}\sim e^{-\beta E/y}$ with $y\sim 0.56$ \cite{drye}.
Note that there is still the possibility of phase separation between a saturated network and excess solvent, even under good solvent conditions -- and there is some experimental evidence for exactly this phenomenon \cite{portesatnet}.

Scaling theories in the presence of rings become even more complicated \cite{ppw}; even the exponent $\nu$, governing the local chain geometry, is slightly different in the ring dominated regime \cite{catesjphysique}. Near $\phi_r^{max}$ (which has the same meaning as before, but no longer obeys (\ref{maxring})) there is a power-law cascade of rings, controlled by a distribution similar to (\ref{cring}), but with a somewhat larger exponent ($2.74$ instead of $5/2$) \cite{catesjphysique}. This cascade of rings creates `power law screening' of excluded volume interactions \cite{catesjpl}, causing $\nu$ to shift very slightly downward \cite{ppw}. Perhaps fortunately in view of these complications, the ring-dominated regime, if it exists at all, is poorly enough quantified experimentally that comparison with mean field theory is all that can be attempted at present.

\label{bmf}

\subsection{Reaction Kinetics in Equilibrium}
\label{kinetics}

Alongside the micellar length distributions addressed above, a key ingredient into rheological modelling is the presence of reversible aggregation and disaggregation processes (which we shall call micellar `reactions'), allowing micelles to exchange material. We will treat these reactions at the mean-field level, in which micelles are `well-mixed' at all times (so there is no correlation between one reaction and the next). Our excuse for this simplification is that, although deviations from mean-field theory are undoubtedly important in some circumstances and have been worked out in detail theoretically \cite{beno}, there is so far rather little evidence that this matters in the strongly entangled regime ($C\ge \tilde C$) where viscoelasticity is primarily seen. And, although in the shear thickening region ($C\simeq \tilde C$) it is quite possible that non-mean-field kinetic effects become important, there is so much else that we do not understand
about this regime (Section \ref{thickening}) that 
a detailed discussion of correlated reaction effects would appear premature.

\begin{figure}[tbp]
\centering
\includegraphics[width=3.0 in]{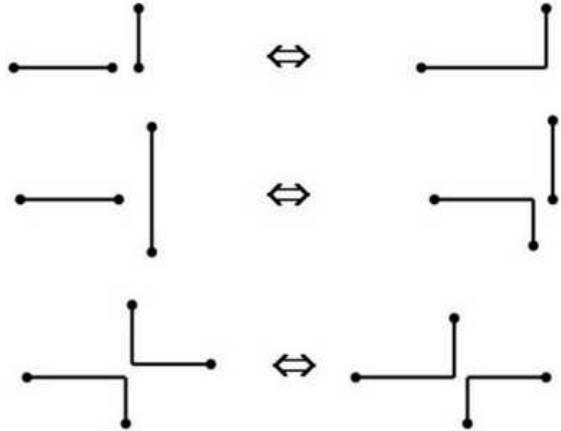}
\caption{The three main types of micellar reaction: Top, reversible scission; middle, end interchange; bottom, bond interchange} \label{reactionfig}
\end{figure}

We neglect branching and ring-formation in the first instance, and also distinguish reactions that change the aggregation number of a particular micelle $N$ by a small increment, $\Delta N\simeq 1$, from those which create changes $\Delta N$ of order $N$ itself. The former reactions can of course lead to significant changes in micellar size over time, but as $N$ increases, the timescale required for this gets longer and longer \cite{marquesstep}. Unless the reaction rates for all reactions of the second ($\Delta N\simeq N$) type are extremely slow, these latter will dominate for large aggregates. From now on, we consider only reactions with $\Delta N \simeq N$, of which there are three basic types: reversible scission, end-interchange, and bond-interchange, as shown in Figure \ref{reactionfig}.

In reversible scission, a chain of length $L$ breaks spontaneously into two fragments of size $L'$ and $L''=L-L'$. (Note that the conservation law $N''+N' = N$ really applies to $N$ and not $L$
as written here; but if we ignore the minor corrections to $L$ represented by the presence of end-caps, the sum of micellar lengths is also conserved.)
In thermal equilibrium the reverse process (end-to-end fusion) happens with exactly equal frequency; this follows from the principle of detailed balance \cite{db}. 
If, for simplicity, we assume that the fusion rate of chains of lengths $L$ and $L'$ is directly proportional to the product of their concentrations, then the fact that detailed balance holds for the equilibrium distribution (\ref{size}) can be used to reduce the full kinetic equations (as detailed, e.g., in Ref.~\cite{rrm}) to a single rate constant $k_{rs}$. This is the rate of scission per unit length of micelle, and is independent of both the position within, and the length of, the micelle involved \cite{rrm}. 
More relevant physically is
\begin{equation}
\tau_{rs} = (k_{rs}\overline L)^{-1}
\label{rs}
\end{equation}
which is the time taken for a chain of the mean length to break into two pieces by a reversible scission process. Note that, by detailed balance, the lifetime of a chain-end before recombination is also $\tau_{rs}$ \cite{rrm}. Moreover, solution of the full mean-field kinetic equations \cite{turnerjump} shows that if $\overline L$ is perturbed from equilibrium, for example by a small jump in temperature (T-jump), $\overline L$ relaxes monoexponentially to equilibrium with a decay time $\tau_{rs}/2$. 
(In fact this applies not only to $\overline L$ but to the entire perturbation to the micellar size distribution, which for this form of disturbance is an eigenmode of the kinetic equations \cite{turnerjump}.)
The response to a nonlinear, large amplitude jump is also calculable \cite{marquesjump}. These results allow $\tau_{rs}$ to be estimated from T-jump data \cite{kern}, providing an important constraint on the rheological models of Section \ref{rrm}.

Turning to end-interchange, this is the process where a `reactive' chain-end bites into another micelle, carrying away part of it (Figure \ref{reactionfig}). Assuming all ends to be equally reactive, and applying detailed balance, one finds that all points on all micelles are equally likely to be attacked in this way. There is, once again, a single rate constant $k_{ei}$, but now the lifetime of any individual chain end is $1/k_{ei}\phi$, since the availability of places to bite into is proportional to $\phi$. The lifetime of a micelle of the average length, before it is involved in an end-interchange reaction of some sort, is \cite{turnerjump}
\begin{equation}
\tau_{ei} = (4k_{ei}\phi)^{-1}
\label{ei}
\end{equation}
In contrast to the reversible scission case, analysis of the full mean-field kinetic equations \cite{turnerjump} shows that end-interchange is {\em invisible} in T-jump: for the specific form of perturbation that arises there, no relaxation whatever occurs by this mechanism. Beyond mean-field kinetics this would no longer hold, but there remains an important limitation to end-interchange in bringing the system to equilibrium. Specifically, {\em end-interchange conserves the total number of micelles}. Accordingly if a disturbance, whether rheological or thermal in origin, is applied that perturbs the total chain number $\sum_Nc(N)$ away from equilibrium, this will not fully relax until the time-scale $\tau_{rs}$ is attained, even if this is much larger than $\tau_{ei}$ \cite{turnerjump}. In the mean time, the end-interchange process relaxes the size distribution $c(N)$ towards the exponential form $c(N) \propto \exp[-N/\overline N]$ of (\ref{size}); but with a nonequilibrium value of $\overline N$.
This separation of time scales may lie at the origin of strange `structural memory' effects seen in certain systems (Section \ref{memory} below) \cite{rings}. 

Note that since in our simple models the micellar energy is fixed by the number of end caps, conservation of micellar number in end-interchange reactions (and also bond-interchange, below) is tantamount to conservation of the total energy stored in such end caps. An energy-conserving processes cannot, unaided, relax a system after a jump in temperature. However, since $E$ is really a free energy and the dynamics is not microcanonical, conservation of micellar number is perhaps the more fundamental concept in distinguishing interchange from reversible scission kinetics; in subsequent discussions, we take this view.

Finally we turn to the bond-interchange process \cite{bint} in which micelles transiently fuse to form a four-fold link before splitting again into differently connected components (Figure \ref{reactionfig}). This process, like end-interchange, conserves chain number. Indeed it does not even alter the identity of chain ends. Since, in entangled polymeric systems, stress relaxation occurs primarily at the chain ends, bond-interchange is far less effective than reversible scission or end-interchange in speeding up the disentanglement of micelles (see Section \ref{rrm}). In fact, although a breaking time $\tau_{bi} = (k_{bi}\overline L \phi)^{-1}$ can be defined, this enters the rheological models differently from $\tau_{rs}$ or $\tau_{bi}$ (Section \ref{bint} below).
Bond interchange also allows chains to effectively pass through one another by decay of the four-fold intermediate, creating a somewhat different relaxation channel for chain disentanglement and stress relaxation \cite{rehagebi}. However (as previously discussed in Section \ref{branch}) a transient four-fold link
is likely to dissociate rapidly into two three-fold links. Such three-fold links are in turn the transition states of the end-interchange process. If these links disconnect rapidly, then the end-interchange process (which their decay represents) is probably dominant over bond interchange. If they do not decay rapidly, then it is likely that their existence cannot be ignored for static purposes; one has a branched system in equilibrium (see Section \ref{branch}).

The reaction kinetics in branched micellar networks is far from easy to cast in terms of simple mean-field reaction equations, as studied, {\em e.g.}, in Ref. \cite{turnerjump} for unbranched chains. However, within such a network, alongside any bond-interchange reactions that are present, structural relaxation can still occur by reversible scission or end-interchange of a section of the micellar network between junctions. Time-scales $\tau_{rs}$ or $\tau_{ei}$ can then be defined as the lifetime of a typical network strand before destruction by such a process. In the reversible scission case 
(\ref{rs}) still holds, now with $\overline L$ the mean strand length in the network \cite{lequeux}.

In the presence of rings, the three reaction schemes of Figure \ref{reactionfig} remain applicable in principle. It is then notable that the chain number $\sum_Nc(N)$, though not the ring number $\sum_Nc_r(N)$, is still conserved by the two interchange processes. Whenever open chains are present, reversible scission is needed for them to reach full thermal equilibrium \cite{rings}.

\subsection{Parameter Variations}

As stated previously, the static mean-field theories given above (in Sections \ref{mfpol} -- \ref{rings}) take as their parameters $E,l_p,a_0,v_0$. Also relevant is the excluded volume parameter $w$ \cite{degennes, doiedwards}. This controls the strength of repulsions between sections of micelle; for hard core interactions this is a function of $l_p$ and $a_0$, but in general $w$ also depends on all local interaction forces between sections of micelles. Nonetheless, within mean-field, this parameter only affects the purely $\phi$-dependent term $F_0(\phi)$ in (\ref{meanfield}, \ref{dryemeanfield}, \ref{ringmeanfield}) and hence has no effect on the mean micellar length $\overline L$ or the size distribution $c(L)$. (The most important role of $w$ is, in fact, to control the crossover to the scaling results discussed in Section \ref{bmf} above.)
All of the parameters $E,l_p,a_0,w$ in principle can have explicit dependence on the volume fraction $\phi$. This certainly occurs in ionic micellar systems at low added salt, where the ionic strength depends strongly on $\phi$ itself and modulates directly parameters such as $E$ and $l_p$. Ion-binding and similar effects can also be strongly temperature dependent. 
Similar remarks apply to the reaction rate constants $k_{rs},k_{ei}$ considered in Section \ref{kinetics} above, and hence also to their activation energies $E_A \equiv -\partial \ln k/\partial \beta$. The rheological consequences of these parameter variations are discussed in Section \ref{secondarypredictions}.

\section{Theoretical Rheology}
Since microscopic models for giant micelle rheology draw strongly from earlier progress in modelling conventional polymers, we review that progress briefly here. (See \cite{doiedwards} for a definitive account.) In doing so we can also establish some of the concepts and terms used in rheology -- a field which remains regrettably foreign to the majority of physics graduates.

\subsection{Basic Ideas}

\begin{figure}[hbtp]
	\centering
	\includegraphics[width=4.truecm]{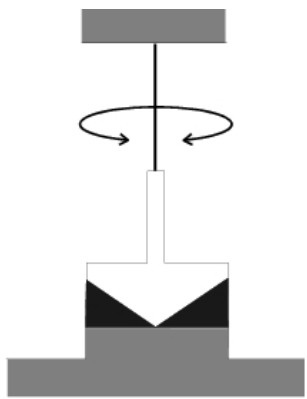}
	\caption[]{A cone-plate rheometer. The sample (black) sits between a rotating cone (white) and a solid plate (grey).}
	\label{cone}
\end{figure}

Rheology is the measurement and prediction of flow behaviour. The basic experimental tool is a rheometer -- a device for applying a controlled stress to a sample and measuring its deformation, or vice versa. However, in recent years a variety of rheophysical probes, which allow simultaneous microscopic characterisation or imaging, have been developed \cite{pine,callaghan}. For the complex flows that can arise in giant micelles, these enhanced probes offer important additional information about how microstructure and deformation interact.
Many rheometers use a Couette cell, comprising two concentric cylinders, of radius $r$ and $r+h$ with the inner one rotating. (See Figure \ref{fig:couette} below for an illustration of this geometry.) Others use a cone-plate cell (Figure \ref{cone}) where a rotating cone contacts a stationary plate at its apex, 
with opening angle $\theta$. In the limit of small $h/r$ or small $\theta$, each device results in a uniform stress in steady state; in each case, the shear stress can be measured from the torque. Some cone-and plate devices can also measure `normal stress differences' defined below.

\subsubsection{Statistical Mechanics of Stress}
\label{stressstat}

We shall use suffix notation, with roman indices and the usual summation convention, for vectors and tensors; letters $a...w$ can therefore stand for any of the three cartesian directions $x,y,z$. Greek indices will be reserved for labels of other kinds.

Consider a surface element of area $dA$ with unit normal vector $n_i$. Denote by $dF_i$ the force acting on the interior of the surface element caused by what is outside. If $n_i$ is reversed (switching the definitions of interior and exterior), then so is $F_i$; this accords with Newton's third law. Writing the usual vectorial area element as $dS_i = n_idA$, we have
\begin{equation}
dF_i = \sigma_{ij} dS_j 
\label{stressdef}
\end{equation}
which defines the stress tensor $\sigma_{ij}$. This tensor is symmetric.
The hydrostatic pressure is {\em defined} via the trace of the stress tensor, as $p = - \sigma_{ii}/3$; what matters in rheology the (traceless) `deviatoric' stress
$
\sigma^{\mbox{\small dev}}_{ij} = \sigma_{ij} + p\delta_{ij}
$.
This includes all shear stresses, and also two combinations of the diagonal elements, usually chosen as the two normal stress differences, 
\begin{equation}
N_1 = \sigma_{xx} - \sigma_{yy}\;\;\;\;\;;\;\;\;\;\;N_2 = \sigma_{yy}-\sigma_{zz}\;
\label{normalstresses}
\end{equation} 

\begin{figure}[hbtp]
	\centering
	\includegraphics[width=6.truecm]{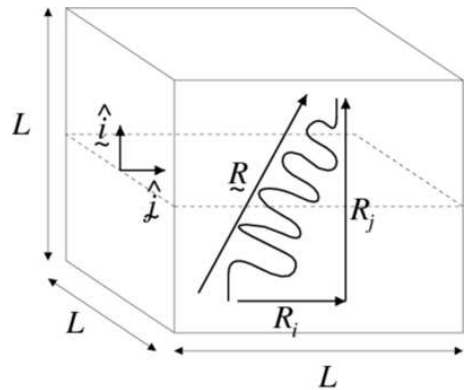}
	\caption[]{Contribution from a polymer `subchain' to the stress tensor. The endpoints of the chain can be viewed as two particles, with the chain in between supplying a `spring force' between them.}
	\label{figure3}
\end{figure}

For simplicity we assume pairwise interactions between particles. (The choice of what we define as a particle is clarified later.) The force $f_i^{\alpha\beta}$ exerted by particle $\alpha$ on particle $\beta$ then depends on their relative coordinate $r_i^{\alpha\beta}$ (measured by convention from $\alpha$ to $\beta$). But this pair of particles contributes to the force $dF_i$ acting across a surface element $dS_i$ only if the surface divides one particle from the other. 
The probability of this happening is $dS_ir_i^{\alpha\beta}/V$ where $V$ is the volume of the system. (This is easiest seen for a cubic box of side $L$ with a planar dividing surface of area $A = L^2$ with normal $n_i$ along a symmetry axis; Figure \ref{figure3}). The separation of the particles normal to the surface is clearly $\ell = r_i^{\alpha\beta}n_i$, and the probability of their lying one either side of it is then just $\ell/L$, which can be written as $A r_i^{\alpha\beta}n_i/V$. 
Accordingly, the total force across a surface element $dS_i$ is
$
dF_i = -\sum_{\alpha\beta}dA (r_j^{\alpha\beta}n_j) f_i^{\alpha\beta}/V
$
which by definition acts outward (hence the minus sign). Bearing in mind (\ref{stressdef}), this gives
\begin{equation}
\sigma_{ij} =  
-V^{-1}\sum_{\alpha\beta}r_i^{\alpha\beta} f_j^{\alpha\beta} = -\rho^2V\langle r_if_j\rangle
\label{statmechstress2}
\end{equation}
where the average is taken over pairs and $\rho$ is the mean particle density.

An example is shown in Figure \ref{figure3}, where a polymer `subchain' is shown crossing the surface. At a microscopic level, one could choose the individual monomers as the particles, and their covalent,  van der Waals, and other  interactions as the forces in (\ref{statmechstress2}). (This is often done in computer simulation \cite{allen,kremer}.) But so long as the force $f_i^{\alpha\beta}$ is suitably redefined as an effective, coarse grained force that includes entropic contributions, we can equally well consider a polymer chain as a sparse string of `beads' connected by `springs'. At this larger scale, the interaction force $f_i^{\alpha\beta}$ has a universal and simple dependence on $r_i^{\alpha\beta}$, deriving from an `entropic potential' $U(r_i) = (3k_BT/2b^2) r_ir_i$, where $b^2 \equiv \langle r_ir_i \rangle$. This is a consequence of the well-known gaussian distribution law for random walks, of which the polymer, at this level of description, is an example. 
The entropic potential is defined so that the probability distribution for the end-to-end vector of the subchain obeys $P(r_i) \propto \exp[-U(r_i)/k_BT]$; this form identifies $U$ as the free energy. 
The force now obeys
\begin{equation}
f_i^{\alpha\beta} = -dU (r_i^{\alpha\beta})/dr_i^{\alpha\beta} = -(3k_BT/b^2) r_i^{\alpha\beta}
\label{forcederiv}
\end{equation}
which gives, using (\ref{statmechstress2}), the polymeric contribution to the stress tensor:
\begin{equation}
\sigma_{ij}^{{\mbox{\small pol}}} = \frac{N_{{\mbox{\small spr}}}}{V}\frac{3k_BT}{b^2}\langle r_i r_j\rangle
\label{polymerstress}
\end{equation}
Here the average is over the probability distribution $P(r_i)$ for the end-to-end vectors of our polymeric subchains (or springs); $N_{{\mbox{\small spr}}}/V$ is the number of these per unit volume.
In polymer melts, contributions such as the one we just calculated completely dominate the deviatoric stress. In solutions there may also be a significant contribution from local viscous dissipation in the solvent. In this case, although a formula such as (\ref{statmechstress2}) still holds in principle, it is more convenient to work with (\ref{polymerstress}) and add a separate solvent contribution directly to the stress tensor. For a Newtonian solvent, the additional contribution is $\sigma_{ij}^{{\mbox{\small sol}}} = \eta^{{\mbox{\small sol}}} (K_{ij}+K_{ji})$, where $K_{ij}$ is the velocity gradient tensor introduced below.
 
\subsubsection{Strain and Strain Rate}

Consider a uniform, but possibly large, deformation of a material to a strained from an unstrained state. The position vector $r_i$ of a material point is thereby transformed into  $r_i'$; the deformation tensor $E_{ij}$ is defined by
$
r_i' = E_{ij} r_j .
$
For small deformations, one can write this 
as $E_{ij} = \delta_{ij}+e_{ij}$ so that the displacement $u_i = r'_i - r_i$ obeys $u_i = e_{ij}r_j$. Alternatively we may write this as $e_{ij} = \nabla_j u_i$.
Consider now a time-dependent strain, for which $v_i \equiv \dot u_i$ defines the fluid velocity, which depends on the position $r_i$. We define the velocity gradient tensor
$
K_{ij} = \nabla_j v_i = \dot e_{ij}
$;
this is also sometimes known as the `rate of strain tensor' or `deformation rate tensor'. If we now consider a small strain increment, $\dot e_{ij} \delta t$,
\begin{equation}
r_i(t+\delta t) = (\delta_{ij} + \dot e_{ij}\delta t)\, r_j(t)
\end{equation}
The left hand side of this is, by definition, $E_{ij}(t+\delta t)r_i(0)$ where the time-dependent deformation tensor $E_{ij}(t)$ connects coordinates at time zero with those at time $t$. Inserting also $r_j(t) = E_{jk}(t) r_k(0)$ we obtain $\dot E_{ij} = K_{ik}E_{kj}$, or equivalently
\begin{equation}
\partial E_{ij}/\partial t = \dot e_{ik} E_{kj} 
\label{rateofstrain}
\end{equation}

An important example is simple shear. Consider a shear rate $\dot\gamma$ with flow velocity along $x$ and its gradient along $y$: then $v_i = \dot\gamma y\delta_{ix}$. The velocity gradient tensor is $K_{ij} = \dot \gamma \delta_{ix}\delta_{jy}$, that is, $K_{ij}$ is a matrix with $\dot\gamma$ in the $xy$ position and all other entries zero. Solving (\ref{rateofstrain}) for arbitrary $\dot\gamma(t)$ then gives
$E_{ij}^{tt'} = \delta_{ij} + \gamma(t,t') \delta_{ix}\delta_{jy}
$
where $E_{ij}^{tt'}$ is defined as the deformation tensor connecting vectors at time $t$ to those at time $t'$, and $\gamma(t,t') = \int_t^{t'}\dot\gamma(t'') dt''$ is the total strain between these two times. 

\if{
A second example is uniaxial extension (pulling chewing gum), with stretching along the $x$ direction at strain rate $\dot \epsilon$, accompanied by shrinking in the other two directions as required to conserve volume: 
$
K_{ij} = \dot\epsilon\left(\delta_{ix}\delta_{jx} - \delta_{iy}\delta_{jy}/2-\delta_{iz}\delta_{jz}/2 \right).$
From this one finds that $E_{ij}^{tt'}$ also takes a similar (diagonal) form, with $E_{xx} = \exp \epsilon(t,t')$, $E_{yy} = E_{zz} = \exp(-\epsilon(t,t')/2)$, and $\epsilon(t,t') = \int_t^{t'} \dot \epsilon(t'') dt''$. Note that for steady strain rate, this flow gives exponential stretching of embedded vectors, but the same is not true of simple shear.
}\fi

\subsection{Linear Rheology}
\label{linrheol}
Linear rheology addresses the response of systems to small stresses. 
\label{linstepstrain}
Imagine an undeformed block of material which is suddenly subjected, at time $t_1$, to a small shear strain $\gamma$. Taking the displacement along $x$ and its gradient along $y$, we then have for the resulting deformation tensor $E_{ij} = \delta_{ij} + \gamma \delta_{ix} \delta_{jy}$. Suppose we measure the corresponding stress tensor $\sigma_{ij}(t)$. Linearity, combined with time-translational invariance of material properties, requires that 
\begin{equation}
\sigma_{yx} =\sigma_{xy} = G(t-t_1)\gamma
\label{lingt}
\label{stepstrain}
\end{equation}
and that all other deviatoric components of $\sigma_{ij}$ vanish, at linear order in $\gamma$, by symmetry. (For example, $N_1(\gamma) = N_1(-\gamma)$, which requires $N_1  = O(\gamma^2)$.) 
This defines the linear step-strain response function $G(t)$. This function is zero for $t<0$; it is discontinuous at $t=0$, jumping to an initial value which is very large (on a scale set by $G_0$, defined below). This largeness reflects the role of microscopic degrees of freedom; there follows a very rapid decay to a more modest level arising from mesoscopic (polymeric) degrees of freedom. In most cases this level persists for a while, making it useful to identify it as $G_0$, the transient elastic modulus. (In models that ignore microscopics, one can identify $G_0 = G(t\to 0+)$.) On the timescale of mesoscopic relaxations, which are responsible for viscoelasticity, $G(t)$ then falls further. 

Now suppose we apply a time-dependent, but small, shear strain $\gamma(t)$. By linearity, we can decompose this into a series of infinitesimal steps of magnitude $\dot\gamma(t')dt'$; the response to such a step is $d\sigma_{xy}(t) = G(t-t')\dot\gamma(t')dt'$. We may sum these incremental responses, giving 
\begin{equation}
\sigma_{xy}(t) = \int_{-\infty}^tG(t-t')\dot\gamma(t')dt'
\label{linconstit}
\end{equation} 
where, to allow for any displacements that took place before $t=0$, we have extended the integral into the indefinite past. Hence $G(t)$ is the memory kernel giving the linear stress response to an arbitrary shear rate history. This is an example of a {\em constitutive equation}. However, the constitutive equation for nonlinear flows is far more complicated.

In steady shear $\dot\gamma(t)$ is constant; therefore from (\ref{linconstit}) one has $\sigma_{xy}(t) = \dot\gamma\int_{-\infty}^t G(t-t') dt'$. However, the definition of a fluid's linear viscosity (its `zero-shear viscosity', $\eta$) is the ratio of shear stress to strain rate in a steady measurement when both are small; hence 
\begin{equation}
\eta = \int_0^{\infty}G(t)dt = \lim_{\omega\to 0}\left[G^*(\omega)/i\omega\right] \;.
\label{viscodef}
\end{equation}
This is finite so long as $G(t)$ decays to zero faster than $1/t$ at late times, which
is true in all viscoelastic liquids (as opposed to solid-like materials), including giant micelles.

\subsubsection{Oscillatory Flow; Linear Creep}
\begin{figure}[hbtp]
	\centering
	\includegraphics[width=9.truecm]{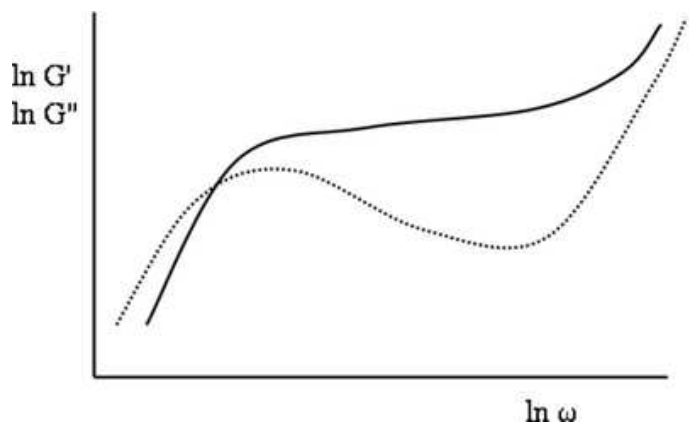}
	\caption[]{Artist's impression of the viscoelastic spectrum for a typical polymeric material; the storage and loss moduli $G'(\omega), G''(\omega)$ are solid and dotted lines respectively.}
	\label{figure4}
\end{figure}

The case of an oscillatory flow is often studied. We write $\gamma(t) = \gamma_0 e^{i\omega t}$ (taking the real part whenever appropriate); substituting in (\ref{linconstit}) gives after trivial manipulation
\begin{equation}
\sigma_{xy}(t) = \gamma_0 e^{i\omega t} G^*(\omega)
\label{gstardef}
\end{equation}
where $G^*(\omega) \equiv i\omega \int_0^\infty G(t) e^{-i\omega t} dt$; this is called the complex modulus. 
The complex modulus, or `viscoelastic spectrum', is conventionally written $G^*(\omega) = G'(\omega) + iG''(\omega)$ where the real quantities $G'$ and $G''$ are respectively the in-phase or elastic response, and the out-of-phase or dissipative response. (These are called the `storage modulus' and the `loss modulus' respectively.)  
Many polymeric fluids exhibit a `longest relaxation time' $\tau$ in the sense that for large enough $t$, the relaxation modulus $G(t)$ falls off asymptotically like $\exp[-t/\tau]$. In this case one has at low frequencies $G' \sim \omega^2$ and $G''\sim \omega$. For polymer melts and concentrated solutions, as frequency is raised $G'$ passes through a plateau whereas $G''$ starts to fall; eventually at high frequencies both rise again. This is sketched in Figure \ref{figure4} where, as is common practice, a double logarithmic scale is used to plot the viscoelastic spectra. 

One can also study the steady-state flow response to an oscillatory stress. This defines a frequency-dependent complex compliance $J^*(\omega)$; however, within the linear response regime this is just the reciprocal of $G^*(\omega)$.
Suppose, instead of applying a step strain as was used to define $G(t)$ in (\ref{stepstrain}), we apply a small step in shear stress of magnitude $\sigma_0$ and measure the strain response $\gamma(t)$. This defines a compliance function  $\gamma(t)/\sigma_0 =J(t)$ which is the functional inverse of $G(t)$ (that is, $\int J(t)G(t-t') dt = \delta(t')$). To see this, one can repeat the derivation of (\ref{gstardef}) with stress and strain interchanged, to find that $J^*(\omega) = i\omega\int J(t) e^{-i\omega t} dt$.
For a viscoelastic liquid $\gamma(t)$ rises smoothly from zero, and the system eventually asymptotes to a steady flow: $\gamma(t\to\infty) = \sigma_0 (t/\eta +  J_e^{(0)})$. The offset $J_e^{(0)}$, measured by extrapolating the asymptote back to the origin, 
is called the steady-state compliance. It can be written as
$J_e^{(0)} = \int_0^\infty t\,G(t)dt/\eta^2$
and is therefore more sensitive to the late-time part of $G(t)$ than the viscosity $\eta = \int_0^\infty G(t)dt$.

\subsubsection{The Linear Maxwell Model}
\label{linmaxsec}
\begin{figure}[hbtp]
	\centering
	\includegraphics[width=9.truecm]{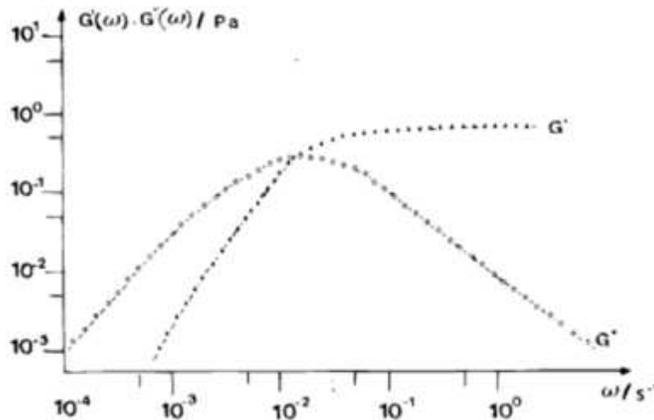}
	\caption[]{Viscoelastic spectrum for a system of entangled micelles: arguably nature's closest approach to the linear Maxwell model, for which the peak in $G''$ is perfectly symmetric and $G'$ crosses through this peak at the maximum. Figure reprinted with permission from Ref.\cite{rehagefigure}.}
	\label{figure5}
\end{figure}

The simplest imaginable $G(t)$ takes the form $G(t)= G_0\exp(-t/\tau_M)$ for all $t>0$ and is called the linear Maxwell model, after its inventor James Clerk Maxwell. $G_0$ is a transient elastic modulus and $\tau_M$ a relaxation time (in this model, it is the {\em only} such time) called the Maxwell time. The viscosity is $\eta = G_0\tau_M$; note that a Newtonian fluid is recovered by taking $G_0 \to \infty$ and $\tau_M \to 0$ at fixed $\eta$. In nature, nothing exists that is quite as simple as the Maxwell model: but the low-frequency linear viscoelasticity of certain giant micellar systems is remarkably close to it (Figure \ref{figure5}). The viscoelastic spectrum of the Maxwell model is $G^*(\omega) = G_0 i\omega\tau_M/(1+i\omega\tau_M)$ whose real and imaginary parts closely resemble Figure \ref{figure5}: a symmetric maximum in $G''$ on log-log through which $G'$ passes as it rises towards a plateau. This is distinct from ordinary polymers, where the peak is lopsided (with slope closer to $-1/2$ on the high $\omega$ side), with $G'$ not passing through the maximum (Figure \ref{figure4}).
Understanding the near-Maxwellian behaviour of giant micelles in linear rheology is one of the main
achievements of the `reptation-reaction' models outlined in Section \ref{rrm} below.

\if{
In an obvious extension of the Maxwell model one can add several Maxwell-like `modes' to give $G(t)=\sum_i G_i \exp[-t/\tau_i]$. This is generally curve fitting but can also be motivated microscopically in some cases (see for example Eq.\ref{mucalc} below).
}\fi

\subsection{Linear Viscoelasticity of Polymers: Tube Models}
\label{tubes}
\begin{figure}[hbtp]
	\centering
	\includegraphics[width=12.truecm]{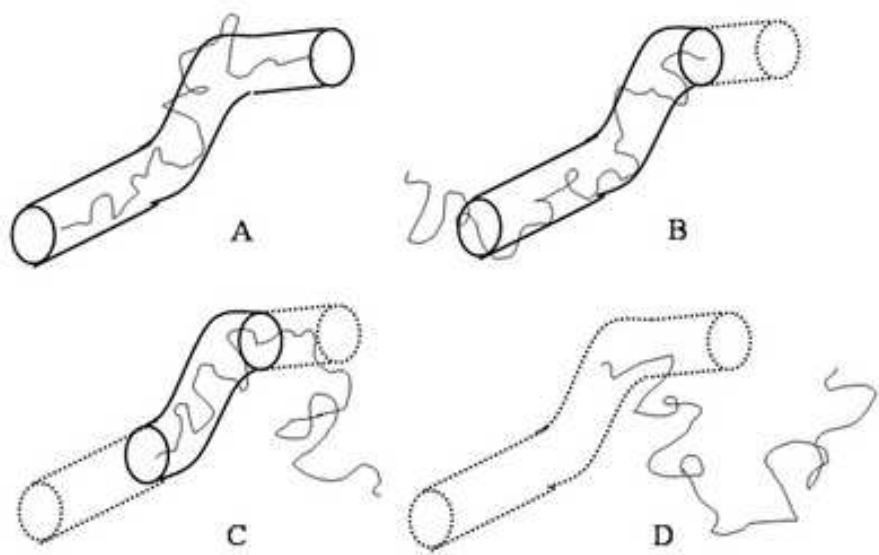}
	\caption[]{A polymer chain (light line). Surrounding chains present obstacles that the chain cannot cross. These can be modelled by a tube (heavy line). The stress relaxation response after step strain is controlled by the fraction of the intial tube still occupied by the chain at time $t$. 
Frames A-D show the state of the tube at four consecutive times, with vacated regions shown dotted.
Although the emerging chain creates new tube around itself (not shown) this part is assumed to be created in the strained environment, and hence carries no shear stress.}
	\label{figure6}
\end{figure}

Figure \ref{figure6} shows a flexible polymer. The chain conformation is a random walk; its end-to-end vector is gaussian distributed. 
In both polymeric and micellar systems there are  corrections to gaussian statistics arising from excluded volume effects at length scales smaller than the static correlation length $\xi$. These effects are screened out at larger distances \cite{degennes}, and their effects in micelles anyway limited (see Section \ref{bmf}); we ignore them here.

Dense polymers are somewhat like an entangled mass of spaghetti, lubricated by Brownian motion. The presence of other chains strongly impedes the thermal motion of any particular chain. Suppose for a moment that the ends of that chain are held fixed. In this case, the effect of the obstacles can be represented as a tube (Figure \ref{figure6} A). Because it wraps around a random walk, the tube is also a random walk; its number of steps $N_T$ and step-length $b$ (comparable to the tube diameter) must obey the usual relation $\langle R^2\rangle = N_Tb^2$ where $R$ is the end-to-end distance of both the tube and the chain. This distance can be measured by scattering with selected labelling, as can, in effect, the tube diameter (or step length), by looking at fluctuations in chain position on short enough timescales that the chain ends don't move much. However, there is no fundamental theory that can predict $b$; in what follows it is a parameter. It is quite large, so that chains smaller than a few hundred monomers do not feel the tube at all. (The largeness of $b$ remains an active topic of research \cite{mcleish02}.) In what follows we will address strongly entangled materials for which $N_T \gg 1$, ignoring many subtle questions that arise when $N_T$ is of order unity.

Suppose we now take a dense polymer system and perform a sudden step-strain with shear strain $\gamma$. The chain will instantaneously deform with the applied strain. Since a deformed random walk is not maximally random, but biased, this causes a drop in its configurational entropy. Quite rapidly, though, degrees of freedom at short scales (within the tube) can relax by Brownian motion. Once this has happened, the only remaining bias is at the scale of the tube: the residual entropy change $\Delta S$ of the chain is effectively that of the tube in which it resides. A calculation \cite{doiedwards} of the entropy of deformed random walks gives a resulting free energy change
\begin{equation}
\Delta F = - T \Delta S = \frac{1}{2} G_0 \gamma^2
\label{godef}
\end{equation}
where we identify $G_0$ as the transient elastic modulus; this comes out as $G_0 = 4k_BTn/5$ where $n$ is the number of tube segments per unit volume. Hence the elastic modulus is close to, but not exactly, $k_BT$ per tube segment.

What happens next? The chain continues to move by Brownian motion, as do its neighbours. Although the individual constraints  may come and go to some extent, the primary effect is as if the chain remains hemmed in by its tube (Figure \ref{figure6}). Therefore it can diffuse only along the axis of the tube (curvilinear diffusion). The curvilinear diffusion constant $D_c$ is inversely proportional to chain length $L$ \cite{degennes}. 
Curvilinear diffusion allows a chain to escape through the ends of the tube. When it does so, the chain encounters new obstacles and, in effect, creates new tube around itself. However, we assume that this new tube, which is created at random {\em after} the original strain was applied, is undeformed. This turns out to be a very good approximation, mainly since $b$ is so large: the deformation at the tube scale leads to a local, molecular level alignment that is very small indeed. (Such an alignment might `steer' the emerging chain end so that new tube was correlated with the old; this effect is measurable \cite{tirrell}, but negligible for our purposes.) This causes the stored free energy $\Delta F$ to decay away as $\Delta F = G(t)\gamma^2/2$ where
\begin{equation}
G(t) = G_0 \mu(t)
\end{equation}
Here we identify $\mu(t)$ as the fraction of the original tube (created at time zero) which is still occupied, by any part of the chain, at time $t$. (In Figure \ref{figure6}, this part of the tube is shown with the solid line in each time frame; the remaining, vacated, regions are shown dotted.)
The problem of finding $\mu(t)$ can be recast \cite{degennes} as the problem of finding the survival probability up to time $t$ of a particle of diffusivity $D_c$ which lives on a line segment $(0,L)$, with absorbing boundary conditions at each end; the particle is placed at random on the line segment at time zero. To understand this, choose a random segment of the initial tube and paint it red; then go into a frame where the {\em chain} is stationary and the tube is moving. The red tube segment, which started at a random place, diffuses relative to the chain and is lost when it meets a chain end. This tube segment is our particle, and its survival probability defines $\mu(t)$.  

It is remarkable that the tube concept simplifies our dynamics from a complicated many-chain problem, first into a one-chain (+tube) problem, and then into a one-particle problem. The result of this calculation, a good revision exercise in eigenfunction analysis \cite{doiedwards}, is:
\begin{equation}
\mu(t) = \sum_{n=\mbox{\small odd}} \frac{8}{n^2\pi^2}\exp\left[-n^2t/\tau_R\right]
\label{mucalc}  
\end{equation}
where $\tau_R = L^2 \pi^{-2}/D_c$. This parameter is called the `reptation time' (`reptate' means to move like a snake through long grass), and sets the basic timescale for escape from the tube. The calculated $\mu(t)$ is dominated by the slowest decaying term -- hence it is not that far from the Maxwell model, though clearly different from it, and resembles the left part of Figure \ref{figure4}. (To understand the upturn at the right hand side of that figure, one needs to include intra-tube modes; see \cite{doiedwards}.)
From this form of $\mu(t)$ follow several results: for example the viscosity is $\eta = \int G(t) dt= G_0 \tau_r \pi^2/12$ and the steady-state compliance obeys $J_e^{(0)}G_0 = 6/5$. Thus the tube model gives quantitative inter-relations between observable quantities, and the number of these relations significantly exceeds the number of free parameters in the theory --- which can be chosen, in effect, as $G_0$ and a diffusivity parameter $\tilde D_c = D_cL$. 

The model predicts that $\eta = G_0 L^3/(12\tilde D_c)$; since $G_0$ is independent of molecular weight, $\eta$ at fixed $\phi$ varies as $L^3$ for long chains. The experiments lie closer to $\eta \sim L^{3.4}$, at least for modest $L$, but with a prefactor such that the observations lie {\em below} the tube model's prediction until extremely large $L$ is attained (at which point, in fact, the data bend over towards $L^3$). This viscosity deficit at intermediate chain lengths has, in recent years, been successfully accounted for by studying more closely the role of intra-tube fluctuation modes and their effects on other chains; see \cite{milner}.

\subsection{Nonlinear Rheology}
Nonlinear rheology addresses the response of a system to finite or large stresses.
In the absence of a superposition principle, such as the one that holds for linear response, the range of independent measurements is much wider. Nonlinear versions exist of the step-strain and step-stress response measurements discussed in Section \ref{linrheol}, and of oscillatory measurements in which either stress or strain oscillate sinusoidally (though in the nonlinear regime, the induced strain or stress will have a more complicated waveform). 

\label{nonlinstepstrain}
In nonlinear step strain, a deformation $E_{ij} = \delta_{ij} + \gamma \delta_{ix} \delta_{jy}$ is suddenly applied at time $t_1$, just as in Section \ref{linstepstrain}, but now $\gamma$ need not be small. Analogous to (\ref{lingt}) we define
\begin{equation}
\sigma_{xy} = G(t-t_1;\gamma)\gamma
\end{equation}
where a factor of $\gamma$ ensures that $G(t-t_1;0)=G(t-t_1)$ (so the small-strain limit coincides with the linear modulus defined previously). A system is called `factorable' if $G(t-t_1;\gamma) = G(t-t_1)h(\gamma)$, but this is not the general case.
Whereas at linear order all other deviatoric components of $\sigma_{ij}$ vanished by symmetry, in the nonlinear regime one can expect to measure finite normal stress differences $N_1,N_2$, as defined in (\ref{normalstresses}). In some cases, including many systems containing giant micelles, these quantities greatly exceed the shear stress $\sigma_{xy}$ \cite{rehagehoffmann}.

Another key experiment in the nonlinear shear regime is to measure the `flow curve', that is, the relationship $\sigma(\dot\gamma)$ in steady state. For a Newtonian fluid this is a straight line of slope $\eta$; upward curvature is called shear-thickening and downward curvature shear-thinning. Flow curves can also exhibit vertical or horizontal discontinuities: these are usually associated with an underlying instability to an inhomogeneous flow, to which we return in Section \ref{macro}.

\subsubsection{Nonlinear Step Strain for Polymers}
Imagine a dense polymer system to which a {\em finite} strain is suddenly applied. We are thinking mainly of shear, but can equally consider an arbitrary strain tensor $E_{ij}$. As previously discussed, the random walk comprising the tube, which describes the slow degrees of freedom, becomes non-random. If we define the tube as a string of vectors $b u_i^{\alpha}$ (where $\alpha$ labels the tube segment) then the initial $u_i^{\alpha}$ are random unit vectors. After deformation
\begin{equation}
u_i^\alpha \to E_{ij} u_j^\alpha
\end{equation}
where it is a simple matter to prove \cite{doiedwards} that the average length of the vector has gone up:  $\langle| E_{ij}u_j^\alpha|\rangle_\alpha \equiv \chi >1$, where the average so defined is over the initial, isotropic distribution. The length increment is of order $\gamma^2$ (for the usual reasons of symmetry; strains $\gamma$ and $-\gamma$ must be equivalent, macroscopically) but for large strains cannot be neglected.

This increase in the length of the tube is rapidly relaxed by a `breathing' motion \cite{doiedwards} of the free ends (one of the intra-tube modes mentioned previously). This rapid retraction kills off a fraction $1-
1/\chi$ of the tube segments, so that in effect $n \to n/\chi$. Retraction also relaxes the magnitude, but not the direction, of the mean spring force in a tube segment back to the equilibrium value. The resulting force according to (\ref{forcederiv}) is $f_i^\alpha = (3k_BT/b) E_{ij}u_j^\alpha /|E_{ij}u_j^\alpha|$, while the corresponding end-to-end vector of the segment is $bE_{ij}u_j^\alpha$. Substituting these results in (\ref{polymerstress}) gives
\begin{equation}
\sigma_{ij}(t>t_1) = \frac{3 n k_BT}{\langle| E_{ij}u_j^\alpha|\rangle_\alpha} \left\langle\frac{E_{ik}u^\alpha_k E_{jl}u^\alpha_l}{|E_{im}u^\alpha_m|} \right\rangle_\alpha\; \mu(t-t_1) \;  
\label{nonlinpolstep}
\end{equation}
Here the final $\mu(t-t_1)$ is inserted on the grounds that, after retraction is over, the dynamics proceeds exactly as discussed previously for escape of a chain from a tube.

This stress relaxation is of factorable form (now choosing $t_1 = 0$): 
\begin{equation}
\sigma_{ij}(t) = 3 n k_BT Q_{ij}(E_{mn}) \mu(t)
\label{relax}
\end{equation}
which defines a tensor $Q_{ij}$ as a function of the step deformation $E_{mn}$. Computing $Q_{ij}$ involves only angular integrations over a sphere, since the $\alpha$ average in (\ref{nonlinpolstep}) is over isotropic unit vectors \cite{doiedwards}. 
Expanding the result in $\gamma$ for simple shear gives $Q_{xy} = 4\gamma/15 + 0(\gamma^2)$; this confirms the value of the transient modulus $G_0$ quoted after (\ref{godef}) above. In finite amplitude shear, $Q_{ij}$ is sublinear in deformation: this is called `strain-softening' and the same physics is responsible for shear-thinning in polymers under steady flow. 

There are two ways to explain this sublinearity. One is retraction, leading to loss of tube segments. The other is `overalignment': a randomly oriented ensemble of tube segments will, if strained too far, all point along the flow direction. Hence none will cross a plane transverse to the flow as required to give a shear stress (Figure \ref{figure3}). But the second argument is fallacious unless retraction also occurs (the number of chains crossing the given plane is otherwise conserved) and indeed crosslinked polymer networks, where retraction cannot happen because of permanent connections, do not strain-soften.
Like many other predictions of the tube model, these ones are quantitative to 10 or 15 percent. Note that the factorability stems from the separation of timescales between slow (reptation) modes and the faster ones (breathing) causing retraction; close experimental examination shows that the factorisation fails at short times.

\subsubsection{Constitutive Equation for Polymers}
Alongside shear thinning, polymeric fluids exhibit several exotic phenomena under strong flows; these go by the names of rod-climbing, recoil, the tubeless syphon, {\em etc.} \cite{larson}. Because the behaviour of a viscoelastic material cannot be summarised by a few linear or nonlinear tests, the goal of serious theoretical rheology is to obtain for each material studied a {\em constitutive equation}: a functional relationship between the stress at time $t$ and the deformation applied at all previous times (or vice versa). The tube model, in its simplest form (involving a further simplification called the `independent alignment approximation, or IAA') has the following constitutive equation, due (like so much above) to Doi and Edwards \cite{doiedwards}:
\begin{equation}
\sigma_{ij}^{\mbox{\small pol}} (t) = G_0 \int_{-\infty}^t \dot\mu(t-t')Q_{ij}(E_{mn}^{tt'}) \; dt'
\label{polymerce}
\end{equation}
where $Q_{ij}(E_{mn})$ is as defined in (\ref{relax}) and $E_{mn}^{tt'}$ denotes the deformation tensor connecting the shape of the sample at time $t$ to that at an earlier time $t'$. (Recall this is found by solving (\ref{rateofstrain}) with initial condition $E_{mn}^{t't'} = \delta_{mn}$, so it is fully determined by the strain rate history.) This is the deformation seen by tube segments that were created at time $t'$; $G_0 Q_{ij}$ gives the corresponding stress contribution. The factor $\dot\mu(t-t')$ (with $\mu(t)$ obeying (\ref{mucalc})) is the probability that a tube segment, still alive at time $t$, was created at the earlier time $t'$. 

We see that, despite its tensorial complexity, the constitutive equation for the tube model (within the IAA approximation, at least) has a relatively simple structure in terms of an underlying `birth and death' dynamics of tube segments. The Doi-Edwards constitutive equation (\ref{polymerce}) has formed the basis of a series of further advances in which not only IAA but several other simplifications of the tube model have been improved upon -- see Section \ref{gen2tube} and the review by McLeish \cite{mcleish02}. Often these more careful theories add no further parameters to the model; it is remarkable that, in almost every case, agreement with experiment gets better rather than worse when such changes are made. This is a very strong indication that the basic concept of the tube mode is very nearly correct -- something far from obvious when (\ref{polymerce}) was first written down in 1978 \cite{graessley}. Among its `unforced triumphs'  were $J_e^{(0)}$ independent of molecular weight; $J_e^{(0)}G_0$ a constant not far above unity; zero-shear viscosity $\eta\sim L^3$ (not far from the experimental exponent); and factorability in step strain with roughly the right strain dependence \cite{graessley}.

\subsection{Upper Convected Maxwell Model and Oldroyd B Model}
\label{sec:ucmm}

For some macroscopic purposes (Section \ref{macro} below), constitutive models like (\ref{polymerce}), and the analogues presented in Section \ref{rrm} for giant micelles, are rather too complicated. Most of the macroscopic studies start instead from simpler models which (thanks to various adjustable parameters) can be tuned to mimic the micellar problem to some extent.
\label{dumbell}
Some of these simpified models can in turn be motivated by the so-called dumb-bell picture, which in fact predated the tube model by many years.

A polymer dumb-bell is defined as two beads connected by a gaussian spring. We forget now about entanglements, and represent each polymer by a single dumb-bell, whose end-to-end vector is $R_i$. The force in the spring is $f_i = -\lambda R_i$. (Hence $\lambda = 3k_BT/N_mb_m^2$ where $N_m$ is the number of monomers in the underlying chain and $b_m$ is the bond length; but this does not actually matter once we adopt the dumb-bell picture.) In thermal equilibrium, it follows that $\langle R_i R_j \rangle_e = k_BT\delta_{ij}/\lambda$ and we can write (\ref{polymerstress}) as
$\sigma_{ij}^{\mbox{\small pol}} = n_D \lambda \langle R_iR_j\rangle
$,
where $n_D = N_D/V$ is the number of dumb-bells per unit volume.
The dumb-bell model assumes that the two beads undergo independent diffusion subject to (a) the spring force, and (b) the advection of the beads by the fluid in which they are suspended. These ingredients can be combined to give a relatively simple equation of motion for $\sigma_{ij}^{\mbox{\small pol}}$, as follows.

First, consider diffusion alone. This would give
$d\langle R_iR_j\rangle/dT = 4k_BT\delta_{ij}/\zeta
$. This equation says that the separation vector evolves through the sum of two independent diffusion processes, each of diffusivity $D=k_BT/\zeta$, and hence with combined diffusivity $2D$; $\zeta$ is the friction factor (or inverse mobility) of a bead.
Next, add the spring force: this gives a diffusive regression towards the equilibrium value of  $\langle R_iR_j\rangle_e = k_BT\delta_{ij}/\lambda$, that is:
$
d\langle R_iR_j\rangle/dt = ({4k_BT}/{\zeta}) \left(\delta_{ij}-\lambda\langle R_iR_j\rangle/k_BT\right)
$. Finally, we allow for advection of beads by the flow; on its own this would give $\dot R_i = K_{ij}R_j$, from which it follows that
$d\langle R_iR_j\rangle/dt|_ {\mbox{\small flow}} = K_{il}\langle R_lR_j\rangle +\langle R_iR_l\rangle K_{jl}.
$
Combining these elements yields
\begin{equation}
\frac{d}{dt}\langle R_iR_j\rangle = K_{il}\langle R_lR_j\rangle +\langle R_iR_l\rangle K_{jl}+\frac{4k_BT}{\zeta} \left(\delta_{ij}-\lambda\langle R_iR_j\rangle/k_BT\right)
\end{equation}
which is equivalent to
\begin{equation}
\frac{d}{dt}\sigma^{\mbox{\small pol}}_{ij} = K_{il}\sigma^{\mbox{\small pol}}_{lj}+\sigma^{\mbox{\small pol}}_{il} K_{jl}+\tau^{-1} \left(G_0\delta_{ij}-\sigma^{\mbox{\small pol}}_{ij}\right)
\label{ucmm}
\end{equation}
where $\tau = \zeta/(4k_BT\lambda)$ is the relaxation time, and $G_0 = N_D\lambda/V$ is the transient modulus, of the system. This is a differential constitutive equation, which can also be cast into an integral form resembling (\ref{polymerce}); it is called the `upper convected Maxwell model' \cite{larson}. 

The equations above consider only the polymeric contribution to the stress. To this can be added a standard, Newtonian contribution from the 
solvent (see Section \ref{stressstat} above)
\begin{equation}
\sigma_{ij} = \sigma^{\mbox{\small pol}}_{ij} + \eta^{\mbox{\small sol}}(K_{ij}+K_{ji}) \label{jsmodel}
\end{equation}
Eq.\ref{jsmodel} defines the so-called Oldroyd B fluid.
This model is the most natural extension to nonlinear flows of the linear Maxwell model of Section \ref{linmaxsec}, and so its adoption for macroscopic flow modelling in micellar systems, which are nearly Maxwellian in the linear regime, is highly appealing. However, this is not enough -- in particular it cannot describe the spectacular shear-thinning behaviour, and related flow instabilities, seen in these systems. The simplest model capable of this is called the Johnson-Segalman model, which will be presented in Section \ref{macro}; it reduces to Oldroyd B in a certain limit, but has additional parameters allowing a much closer approach to micellar rheology. 

The Oldroyd B fluid is also closely related \cite{larson} to the Giesekus model which has sometimes been advocated as a versatile modelling tool for macroscopic micellar rheology \cite{giesekusadv}. Caution is needed however: this can easily become pure curve-fitting if, for instance, the model assumes homogeneous uniform flow when (as explored in Section \ref{macro}) the experimental flow curve in fact represents an average over what is an unsteady or inhomogeneous situation.

\section{Microscopic Constitutive Modelling of Giant Micelles}\label{rrm}

In 1987 one of us \cite{rrm} proposed an extension of the tube model of polymer viscoelasticity that allows incorporation
of micellar reactions. This led to a predictive constitutive model for viscoelastic surfactant solutions. Here we review the model (Section \ref{rrmint}), outline its main rheological predictions
(Section \ref{predictions}) and briefly overview the extent to which these have been experimentally verified. There follow discussions of complexities arising from ionicity effects and branching in entangled micelles (Section \ref{salt}). 

Although generally successful in the highly entangled region, the microscopic approach initiated by Ref.\cite{rrm} has not proved easily generalizable to the shear-thickening window around the viscoelastic onset threshold $\tilde\phi$. Here, one has a system which apparently becomes entangled only as a result of structural buildup upon shearing; in some cases there is also evidence of nematic or other ordering within the resulting shear-induced viscoelastic structure.
Attempts to model these phenomena  are briefly outlined in Section \ref{thickening}.
Then, we address in Section \ref{memory} various `structural memory' effects, in which material properties of micellar systems can evolve on time scales much longer than the Maxwell time.

\subsection{ Slow Reaction Limit}
\label{unbreakable}

Consider first a system of (linear, unbranched) giant micelles for which the kinetic timescales $\tau_{rs},\tau_{ei},\tau_{bi}$ governing reversible scission and interchange reactions are exceedingly long. After waiting this long time, the system will achieve equilibrium with size distribution obeying (\ref{size}). (Below we assume mean-field theory is appropriate unless otherwise stated.) This creates an exponentially polydisperse system, but, if the micelles are entangled, as we from now on assume, the system is otherwise equivalent to a set of unbreakable polymers. This is because the identity of any individual chain is preserved on the time scale of its stress relaxation. Hence, for the purposes of calculating the stress relaxation function $\mu(t)$, defined previously by writing $G(t) = G_0\mu(t)$ as the step-strain response in (\ref{stepstrain}), one has a pure polymer problem.
Recall that $G_0$, the plateau modulus, depends only on the micellar contour length per unit volume; if parameters such as $l_p,a_0$ are held constant, $G_0\sim \phi^2$ in mean-field (or $\phi^{2.3}$ in a scaling approach).

The simplest approach then is to write the overall stress relaxation function as the length-weighted average over all the chains present in the system
\cite{rrm,catescandau}:
\begin{equation}
\mu(t) = \sum_L Lc(L)\mu_L(t)/\sum_L L c(L)
\label{polydisperse}
\end{equation}
Here $\mu_L(t)$ is the function defined in (\ref{mucalc}) appropriate to the given chain length $L$, which controls the reptation time $\tau_R$ in that expression via $\pi^2\tau_R = L^3/\tilde D_c$. (Recall that $\tilde
D_c$, the curvilinear diffusivity, is $L$-independent, though it does depend on $\phi,l_p,a_0$ and the solvent viscosity $\eta^{\mbox{\small sol}}$ which controls the local drag on a chain.) An estimate of (\ref{polydisperse}) gives \cite{rrm}
\begin{equation}
\mu(t) \simeq \exp[-A(t/\tau_R)^{1/4}]
\label{stretch}    
\end{equation}
This relaxation function has a characteristic relaxation time given by the reptation time for a chain of the average length (we abbreviate $\tau_R \equiv \tau_R(\overline L)$) but, in contrast to the result (\ref{mucalc}) for monodisperse chains (let alone the linear Maxwell model of Section \ref{linmaxsec}) it represents an extremely nonexponential decay.  

The above crude result for $G(t)$  can doubtless be much improved by applying
modern `dynamic dilution' concepts which account for the removal of
constraints comprising the tube around a long chain, on the time scale
of reptation of the shorter ones \cite{mcleish02}. (We
have not seen this worked out for the specific case of exponential polydispersity; but see \cite{recentpoly} for an indication of the state of the art.) A strongly nonexponential relaxation would nonetheless remain; indeed such
decays are a well-known experimental signature of polydispersity. The experimental observation of a near monoexponential
relaxation (e.g., Figure \ref{figure5}) in many viscoelastic micellar
systems thus proves the presence of a different relaxation mechanism
from simple reptation \cite{catescandau}. On the other hand, there are
a number of experimental systems where results similar to
(\ref{stretch}) are observed \cite{rehagehoffmann}; this can be taken
as evidence that $\tau_R(\overline L) \ll \tau_{rs},
\tau_{ei},\tau_{bi}$, so that micellar reactions have negligible
direct effect on stress relaxation, in those systems.

\subsection{Reptation-Reaction Model}
\label{rrmint}

We now define $\tau_{b}$, the mean breaking time for a micelle, as the lesser of $\tau_{rs}$ in (\ref{rs}) for reversible scission, or $\tau_{ei}$ in (\ref{ei}) for end interchange. (Bond interchange is dealt with separately  below.) This is the lifetime of a chain before breaking, and also, to within a factor 2 in the case of end-interchange, the lifetime of an end before recombination. We have assumed that whichever reaction is faster, dominates. (It would be a slight improvement to add both channels in parallel but, given that they will have different activation energies, one probably has to fine-tune the system to make them have comparable rates.)
We also define a parameter $\zeta = \tau_{b}/\tau_R$, the ratio of breaking and reptation times.  When this is large, one recovers the results of the preceding section. 

\begin{figure}[tbp]
\centering
\includegraphics[width=3.0 in]{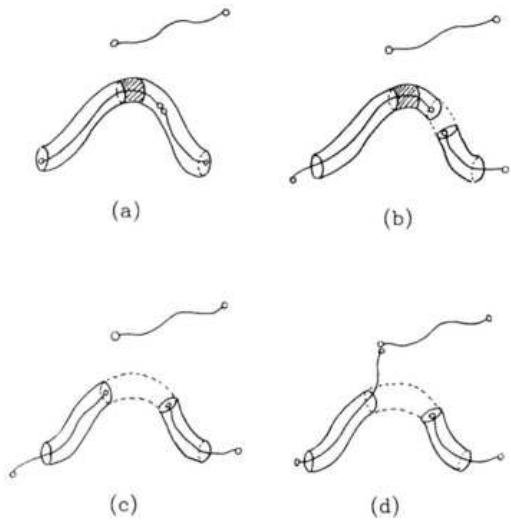}
\caption{The stress relaxation process in the reptation-reaction model with reversible scission. For end interchange, the new end is instead created by attack of another chain end (which connects to the right hand fragment) and destroyed by its own reaction with the central part of another chain. Figure adapted from Ref.\cite{rrm}.} \label{rrmfig}
\end{figure}

Ref.\cite{rrm} proposes that, when $\zeta$ is small, the dominant mode of stress relaxation is as shown in Figure \ref{rrmfig}. The stress relaxation function $\mu(t)$ is, just as for unbreakable chains (Section \ref{tubes}), the probability that a randomly-chosen tube segment, present at time zero, survives to time $t$ without a chain end passing through it. However, the original chain ends do not survive long enough for ordinary reptation to occur; instead, each tube segment has to wait for a break to occur close enough to it, that the new chain end can pass through the given tube segment before disappearing again. The distance $l$ an end can move by reptation during its lifetime $\tau_b$ obeys $\tilde D_c(\overline L) l^2 \simeq \tau_b$; hence $(l/ \overline L)^2 \simeq
\tau_b/\tau_R$ where $\tau_R=\tau_R(\overline L)$ as defined previously. The waiting time $\tau$ for a new end to appear within $l$ is $\tau_b \overline L/l$. This gives, for $\zeta \ll 1$, a characteristic stress relaxation time
\begin{equation}
\tau \simeq (\tau_b\tau_R)^{1/2}
\label{squareroot}
\end{equation}
which is the geometric mean of the timescales for breaking and for (unbreakable) reptation. Moreover, if we also define $\overline \zeta  = \tau_b/\tau$, then in the limit $\overline\zeta \ll 1$ of rapid breaking, both the length of the particular micelle seen by a tube segment, and the position of the tube segment within that micelle, are randomized by the reaction kinetics of order $\overline \zeta^{-1}$ times during the stress relaxation process itself. This causes a rapid averaging to occur, so that all tube segments experience near-identical probabilities for stress relaxation; there is no dispersion in relaxation rates and accordingly, in this limit, the resulting relaxation function $\mu(t)= \exp[-t/\tau]$ is a purely Maxwellian, mono-exponential decay \cite{rrm}.

For modest $\overline \zeta$, deviations from the Maxwellian form are of course expected; these have been studied numerically using a modification of the stochastic diffusion process described prior to (\ref{mucalc}). (Some results for $\overline \zeta$ of order unity are presented in the next section.) Note also that as $\zeta$ falls below $1/N_T$, where $N_T$ is the number of tube segments on the average chain (a quantity that is often of order 10-50 in micelles) there is a crossover to a new regime. In this very rapid breaking regime, the dominant motion of a chain end during its lifetime is not curvilinear diffusion, but 
a more complex motion called `breathing' in which the length of the chain in its tube has fluctuations. This motion is well understood for unbreakable chains \cite{doiedwards} and gives instead of (\ref{squareroot}) \cite{rrm}
\begin{equation}
\tau \simeq \tau_b^{3/4}\tau_R^{1/4}N_T^{1/4}
\label{breathing}
\end{equation}
The deviations from a pure Maxwellian relaxation spectrum in this regime have also been studied \cite{granek}, within a model that  allows partial access to the high-frequency regime, in which stress relaxation involves not only breathing but the intra-tube `Rouse modes' \cite{doiedwards}. This gives a high-frequency turnup in the viscoelastic spectra $G'(\omega), G''(\omega)$, as already depicted schematically for unbreakable polymers in Figure \ref{figure4}. Such a turnup does occur in micelles \cite{granek} although it can lie beyond the experimental window (as is the case in Figure \ref{figure5}).

\subsubsection{Bond Interchange}
\label{bint}
As mentioned in Section \ref{kinetics}, bond-interchange does not directly create or destroy chain ends and so is less effective than reversible scission or end-interchange at causing stress relaxation. Nonetheless, enhancement of relaxation does occur, because bond interchange will occasionally bring what was a distant chain end very close to a given tube segment, allowing relaxation to proceed faster than on a chain undergoing no reactions. This requires the interchange event to create a chain end no further away than $l$ (which was defined as the curvilinear distance a chain can move during its lifetime); the waiting time for this is not $\tau_b = (k_{bi}\overline L \phi)^{-1}$ but $\tau_b\overline L/l$. As shown in \cite{bint} the result is to replace (\ref{squareroot}) with
\begin{equation}
\tau = \tau_R^{2/3}\tau_b^{1/3}
\label{binttau}
\end{equation}
There is a second effect, discussed already in Section \ref{kinetics}, which is the `evaporation' of the tube caused by those bond interchange processes whose effect is to pass one chain through another. Closer inspection shows that this does not affect the regime just described in which chain ends are reptating on the time-scale of their survival, but alters the analogue of (\ref{breathing}) where they move by breathing modes on this time scale. For details, see \cite{bint}.

\subsubsection{Constitutive Equation for Giant Micelles}
We now turn from the stress relaxation function $\mu(t)$ (which, alongside $G_0$, is enough to determine all linear viscoelastic properties) to the nonlinear constitutive equation of the 
reptation-reaction model. This was first worked out in \cite{jphyschem}. We assume $N_T^{-1}\ll\zeta\ll 1$ so that the linear response behaviour is Maxwellian with relaxation time obeying (\ref{squareroot}), and, more importantly, all tube segments are governed by the same relaxation dynamics. We also assume that the rates of micellar reactions are unperturbed by shear; in the highly entangled regime, interesting rheology arises even at modest shear rates, so this is a reasonable approximation. (We will revisit it for barely-entangled systems  in Section \ref{thickening}.)

As calculated in Ref.\cite{jphyschem}, the constitutive equation for giant micelles is
written in terms of the deviatoric part of the polymer stress as
\begin{eqnarray}
\sigma_{ij}^{\mbox{\small pol,dev}} &=& \frac{15}{4}G_0[W_{ij}-\delta_{ij}/3]
\label{ceone}\\
W_{ij}(t) &=& \int_{-\infty}^t{\cal B}(v(t'))\exp\left[-\int_{t'}^t{\cal D}(v(t''))\right]\tilde Q_{ij}(E_{mn}^{tt'}) dt'
\label{cetwo}\\
v(t) &=& W_{ij}(t)K_{ij}(t)
\label{cethree}\\
\tilde Q_{ij}(E_{mn}) &=& \left\langle\frac{E_{ik} u_k E_{jl} u_l}{|E_{im}u_m|}\right\rangle_0
\label{cefour}
\end{eqnarray}
In (\ref{ceone}), $W_{ij}(t) = \langle u_iu_j\rangle$ is the second moment of the distribution at time $t$ of unit vector orientations $u_i$ for tube segments; this controls the polymeric stress in a manner similar to that discussed in Section \ref{nonlinstepstrain} for unbreakable chains. (Indeed, this equation represents essentially the deviatoric part of (\ref{polymerstress}).) The central constitutive equation is (\ref{cetwo}), in which
${\cal B}$ and ${\cal D}$ are birth and death rates for tube segments.  These rates obey complicated equations \cite{jphyschem} but are well approximated for $v>0$ by ${\cal D} = 1/\tau + v$, ${\cal B} = 1/\tau$; and for $v<0$ by ${\cal D} = 1/\tau$, ${\cal B} = 1/\tau - v$. Here $v(t)$, which obeys (\ref{cethree}), is the rate of destruction of tube segments by the same retraction process as was outlined for unbreakable chains in Section \ref{nonlinstepstrain}; this is positive in most flows but can become negative if a flow is suddenly reversed \cite{doiedwards}, in which case chains instead spill out of their tubes causing an addition to the birth rate ${\cal B}$.
In the absence of flow (retraction rate $v = 0$), the birth and death rates are of course equal (by detailed balance), and both are given by $1/\tau$, which is the death rate of a tube segment by the reptation-reaction process shown in Figure \ref{rrmfig}.

The physical content of (\ref{cetwo}) is that the stress in the system at any time $t$ is found by integrating over past times $t'$ the creation rate ${\cal B}(v(t'))$ for tube segments, multiplied by an exponential factor which is the survival probablity of these segments up to time $t$, times $\tilde Q_{ij}$ which is the stress contribution of such a surviving segment (allowing for both its elongation and orientation by the intervening deformation $E_{mn}^{tt'}$). The quantity $\tilde Q_{ij}$ is in turn calculated in (\ref{cefour}) where the average is over an isotropic distribution of initial tangent vectors $u_i$. This is a close relative (but not identical) to $Q_{ij}$ defined for unbreakable chains in (\ref{polymerce}); the difference lies essentially in the independent alignment approximation, IAA, which we do not require here \cite{jphyschem}.
We see that, in the limit of rapid breaking $\overline \zeta \ll 1$, the coupling of reactions into the reptation mechanism produces a constitutive equation that is hardly more complicated than the standard one (\ref{polymerce}) for entangled polymers and, because IAA
is not required, achieves this with slightly fewer approximations than in the case of unbreakable chains. Moreover there is no worry about polydispersity (which plagues comparison of (\ref{polymerce}) with experiment) since this is erased by the fast averaging process. Giant micelles in the fast-reaction regime therefore offer a very interesting arena for testing quantitatively the fundamental ingredients of the tube model and related concepts in polymer dynamics \cite{jphyschem}. 

Note that the reptation-reaction model, as embodied in (\ref{ceone}--\ref{cefour}), describes entanglements at the level of `first generation' tube models (as surveyed in \cite{doiedwards}). Since these equations were first proposed for micelles in 1990, there has been significant work on developing `second generation' tube models which give predictions for unbreakable polymers that are significantly closer to experimental observations. The relevance of these `second generation' tube models to micelles is discussed in Section \ref{gen2tube}.

In the linear viscoelastic limit, Eqs. (\ref{ceone}--\ref{cefour}) reduce to  the linear Maxwell model, which obeys (\ref{linconstit}) with $G(t) = G_0\exp[-t/\tau]$.
Accordingly all linear viscoelastic quantities reduce to those of a pure Maxwellian fluid with $\tau$ obeying (\ref{squareroot}).  However, the full constitutive model is, in both structure and content, quite unlike (\ref{ucmm}) for the upper convected Maxwell model, which would be ---to an empirical rheologist--- a `natural' nonlinear extension of the linear Maxwell model. (There are other nonlinear extensions which do show such effects, including the so-called `co-rotational' Maxwell model. However the way stress elements move with the fluid in these models is not natural for polymers \cite{larson}.)  

\subsection{Primary Rheological Predictions}
\label{predictions}
As discussed previously, we distinguish primary predictions, which do not require much input in terms of how parameters such as $E,k_{rs},l_p$ depend on concentration or temperature, from secondary predictions, which do require this information. (For the latter, see Section \ref{secondarypredictions}.)

\subsubsection{Linear Spectra: Cole-Cole Plots}

The reptation-reaction model predicts that, in the rapid breaking regime, $\mu(t)$ approaches an exponential form as described above. Such behaviour has by now been reported dozens of times in the literature (see e.g. \cite{rehagehoffmann,majid,hoffmannrev,rehagefigure}, and many more recent papers). Its prediction from a microscopic model is nontrivial: while many simple models, such as that of Maxwell himself, `predict' mono-exponential decay, this is actually an assumption, not a result, of such models. In contrast, the reptation-reaction mechanism quantitatively explains the monoexponential behaviour, in terms of the rapid averaging of tube-segments over chain length and position within the chain on timescale $\tau_b = \overline \zeta \tau$. 

\begin{figure}[tbp]
\centering
\includegraphics[width=4.0 in]{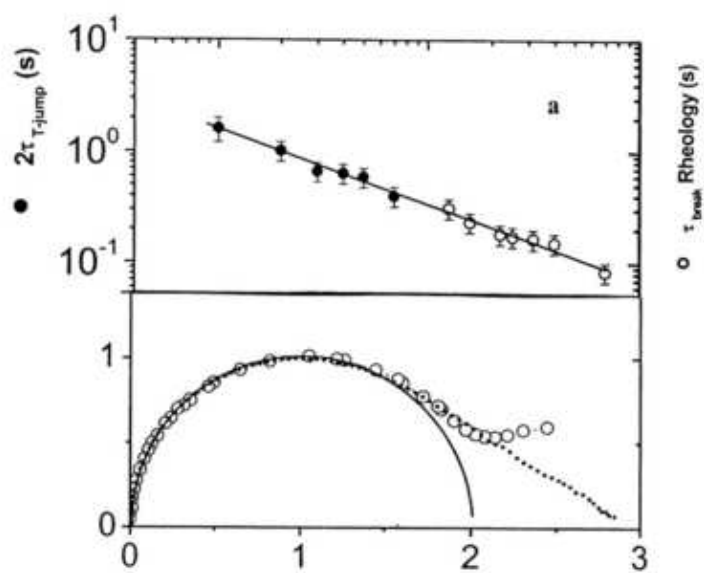}
\caption{Upper panel: comparison of $\tau_{rs}$ measured by T-jump and rheology in a CePyCl/NaSal (2:1) system on a log plot. The horizontal axis is log concentration with data spanning 0.001-0.07 g/cm$^3$.
Lower panel: fit of Cole-Cole data, viz $G''(\omega)$ vs. $G'(\omega)$ on a parametric plot (open symbols) to the numerical model of \cite{turnerlang91} (dots). Data is normalized by the radius of the osculating semicircle. Note that the turnup, caused by Rouse modes \cite{granek} is not part of the fitted data. Figures adapted from  \cite{oelschlagertjump2}, courtesy C. Oelschlaeger.} \label{oelschlagerfig}
\end{figure}

By numerical methods \cite{rrm,turnerlang91,granek}, the model can also predict systematic deviations from the Maxwellian form for $\zeta$ of order unity. In an experimental system where the rapid breaking regime does not cover the entire phase diagram but can be left by varying a parameter (salt concentration, for example), these predictions
offer additional tests of the model, particularly in cases where $\tau_b$ can be estimated independently \cite{kern}.
Such deviations are rendered most visible in the so-called Cole-Cole representation \cite{turnerlang91,kern}
whereby $G''(\omega)$ is plotted against $G''(\omega)$; for exponential $\mu(t)$ the result should be a perfect semicircle. For systems well within the rapid-breaking regime, this has been confirmed with remarkable accuracy in a large number of cases (see, {\em e.g.}, \cite{catescandau,kern,rehagebi,hoffmannrev,rehagefigure,berretcole,zanacole}). Significant deviations are predicted \cite{turnerlang91} at $\overline \zeta \ge 0.4$. In at least one case, where $\tau_b$ is independently determined by temperature jump on the same system, a fit between the experimental and theoretical Cole-Cole plots gives good agreement on the value of $\overline\zeta = \tau_b/\tau$ \cite{kern}. 
In more recent measurements, the breaking times determined by T-jump at volume fractions below and around $\tilde\phi$
were found to extrapolate with good coherence to those measured by rheology at higher concentrations \cite{oelschlagertjump}; see Figure \ref{oelschlagerfig}. This suggests, not only that the reptation-reaction model is correct for this class of systems, but that breakage in it takes place by reversible scission, not end-interchange (see Section \ref{kinetics} above).
In some systems ({\em e.g.}, \cite{berretcole}) deviations from the semicircle involve a sharp upturn at high $\omega$; this can be understood qualitatively in terms of the intervention of intra-tube `Rouse modes' \cite{granek}.
For systems where the deviation from the semicircle has a region of negative slope prior to the upturn (as in Figure \ref{oelschlagerfig}), the location of the minimum can be used \cite{granek} to estimate the mean micellar chain length $\overline L$. This was done for the system of \cite{kern}, and  a physically reasonable trend for $\overline L(\phi)$ found \cite{granek}.

In summary, the linear viscoelastic behaviour of entangled micellar systems, across a wide range of different chemical types (see, {\em e.g.}, \cite{majid}) shows a regime of
strongly Maxwellian relaxation in accord with the reptation-reaction model. The leading shape corrections to the relaxation spectrum as one departs this regime are also well accounted for, in many systems, by that model. These statements involve no knowledge about how the kinetic and structural inputs to the model vary with concentration, ionic strength, or temperature.

\subsubsection{Nonlinear Rheology: Shear-Banding}

The micellar constitutive equation (\ref{cetwo}) can be solved for nonlinear step strain; results show generally similar features to unbreakable polymer rheology \cite{jphyschem}. In practice  comparison to these predictions are often complicated by 
strain-induced instabilities and/or strain-history effects \cite{rehagehoffmann}, the latter perhaps related to those discussed below in Section \ref{memory}. The overall similarity between the micellar constitutive equation and that for unbreakable polymers nonetheless means that most of the nonlinear viscoelastic functions (normal stress coefficients $\Psi_{1,2} \equiv \lim_{\dot\gamma \to 0}(N_{1,2}/\dot\gamma^2)$, etc.), should be broadly similar to those for unbreakable chains \cite{jphyschem}. 

The main arena for comparing nonlinear predictions of the microscopic model with experiments on micelles has involved `steady' flow. This flow is not always literally steady, however; see Section \ref{macro}.  Steady flow, at least if homogeneous, is fully characterized by the `flow curve' $\sigma(\dot\gamma)$, which relates shear stress to strain rate in steady state, and the normal stress difference curves $N_1(\dot\gamma)$ and $N_2(\dot\gamma)$.
Until the early 1990s, studies of the flow curve for giant micelles simply {\em assumed} homogeneity of the flow. Checking for this has since become much easier with a variety of modern techniques \cite{pine,callaghan,nmr,nmr2,nmr3,birefringence}. 

\begin{figure}[tbp]
\centering
\includegraphics[width=4.0 in]{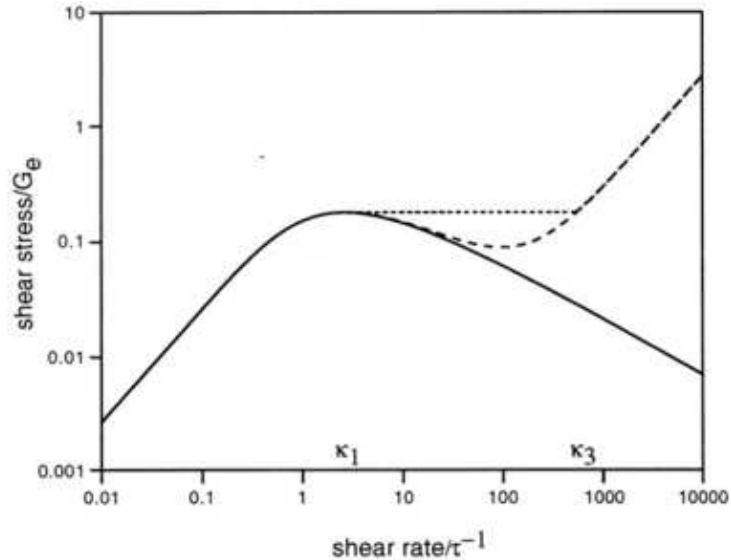}
\caption{Flow curves for reptation-reaction model: solid line, by solution of (\ref{ceone} -- \ref{cefour}); dashed line, with additional quasi-newtonian stress calculated (with one fit parameter) as per \cite{cmm}; dotted-line, top-jumping shear-banded solution (unaffected by that parameter). Figure courtesy N. Spenley \cite{spenleythesis}.} \label{spenley0fig}
\end{figure}

A very striking observation, reported first in a CPySal/NaSal system (compare Figure \ref{figure5}) \cite{rehagehoffmann,rehagefigure}, was that above a certain strain rate $\dot\gamma_p$, the shear stress $\sigma$ attains a plateau value $\sigma=\sigma_p$, remaining at this level for at least two decades in $\dot\gamma\ge\dot\gamma_p$. At the same time the normal stress difference $N_1$ continues to increase almost linearly. This represents shear thinning of a quite drastic kind. It is exploited in technologies such as hand lotions and shampoos, allowing a highly viscous liquid to be pumped or squeezed out of the bottle through a narrow nozzle; in a non-shear-thinning fluid of equal viscosity the bottle would be likely to break first \cite{catesherb}. 
Such behaviour in strong shear flows is as ubiquitous a feature of the nonlinear rheology of giant micelles
as is the Maxwellian spectrum in the linear rheology.
Its explanation came in \cite{spenley}, where the reptation-reaction constitutive equations (\ref{ceone}--\ref{cefour}) were solved in steady state (Figure \ref{spenley0fig}). It was found that the shear stress $\sigma$ has, as a function of $\dot\gamma$, a maximum at $(\dot\gamma,\sigma) = (2.6/\tau, 0.67 G_0)$. Such a nonmonotonic flow curve is known to be unstable \cite{spenley,spenleythesis,banding}; but a steady flow can be recovered by developing `shear-bands'.

\begin{figure}[tbp]
\centering
\includegraphics[width=4.0 in]{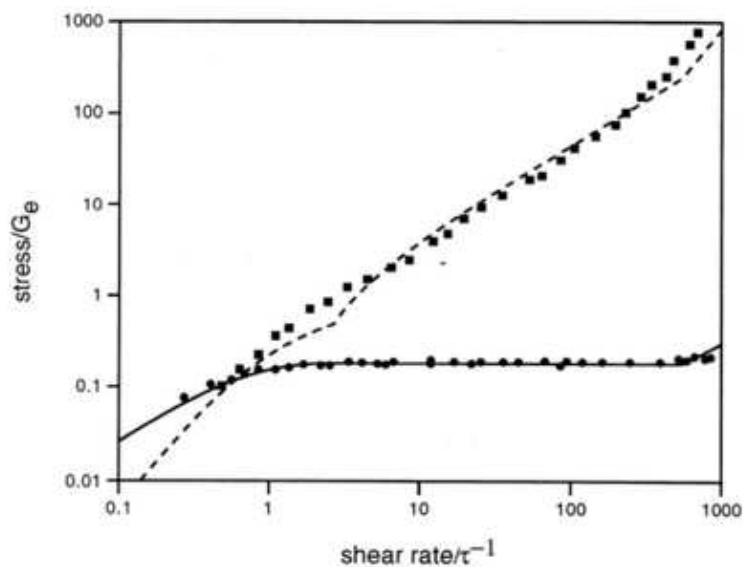}
\caption{Comparison of reptation-reaction prediction for shear banded flows \cite{spenley} with observations of \cite{rehagehoffmann}. Dots are experiments for shear stress $\sigma$ (below at right) and normal stress difference $N_1$. Solid curve is shear stress in the reptation-reaction model assuming top-jumping but with no adjustable parameters. The dashed curve for normal stress has one fitting parameter (as per Figure \ref{spenley0fig}). Figure courtesy N. Spenley \cite{spenleythesis}.} \label{spenleyfig}
\end{figure}

For a shear-thinning system such as this, the shear bands comprise layers of fluid with unequal strain rates but equal stress, their layer normals in the velocity gradient direction \cite{spenley}. In this way, the decreasing part of the flow curve is bypassed by coexistence between two bands, one at low $\dot\gamma=\dot\gamma_1$ and one at high $\dot\gamma_2$, each of which is on an increasing (hence normally stable) part of the curve. The relative amounts of the two bands arrange themselves to match the macroscopically averaged imposed strain rate. Assuming the nature of the coexisting states does not vary as their amounts change (an assumption that ignores coupling to concentration fields, see Section \ref{macro}), this gives a horizontal stress plateau as observed in \cite{rehagehoffmann}. The value of the plateau in CpySal/NaSal could be reproduced by assuming `top jumping', in which one of the coexisting states is at the maximum of the flow curve described previously (giving $\sigma_p = 0.67 G_0$). However, it is now known that the mechanism for selecting coexisting stresses is more complicated than this \cite{olmstedluball}; top-jumping gives an upper estimate for $\sigma_p$,
and lower values have since been seen in many other materials ({\em e.g.}, \cite{grand,berret}).

The theory of \cite{spenley} assumes that the high shear branch (which in the system of \cite{rehagehoffmann} is at $\dot\gamma_2\ge 1000$s$^{-1}$) is not purely Newtonian, as it would be if the micelles were smashed to pieces. Instead this phase is taken to contain highly aligned micelles, which can contribute a very large  $N_1$ while maintaining a small shear stress $\sigma$. (This requires an improvement to the tube model at high stresses \cite{cmm}, giving a linear $\sigma(\dot\gamma)$ at high shear rates, Figure \ref{spenley0fig}.) This theory accounts for the observation in \cite{rehagehoffmann} that $N_1$ continues to increase almost linearly with $\dot\gamma$ throughout the shear-banding plateau at constant $\sigma(\dot\gamma) = \sigma_p$ (Figure \ref{spenleyfig}). 

Since its first prediction in \cite{spenley}, the evidence for shear-banding in viscoelastic micellar solutions has become overwhelming \cite{pine,nmr,nmr2,nmr3,birefringence,berretband}. Some authors have preferred interpretations involving a shear-induced phase transition to a nematic state \cite{berret,berretband}. However, this is not excluded by the above arguments, which make negotiable assumptions about the physics of the high-shear branch at $\dot\gamma = \dot\gamma_2$. Such `semi-thermodynamic' mechanisms of shear-banding can also explain of the `unfolding' of the nonmonotonic flow curve on varying $\phi$ or other conditions, whereby the shear-banding region gets narrower and eventually disappears \cite{berret}; but a crossover from the Maxwellian rapid-breaking regime into the unbreakable polymer limit can explain this too (see below). Even more complicated identifications of the high shear branch are not ruled out; these might include a long-lived gel phase similar to that induced by flow in the shear-thickening region (see Section \ref{thickening}). In all cases, there can also be a concentration differential between the two bands. This is not addressed in the reptation-reaction model, and causes a ramp \cite{olmstedramp} rather than a plateau, in shear stress (Sections \ref{sec:onset}, \ref{sec:steady}). Such couplings to nematic and/or concentration fields take us beyond the limit of practical calculability within such a model; these phenomena can instead be addressed with the macroscopic approaches of Section \ref{macro}.

Note again that the `natural' nonlinear generalization of the linear Maxwell model, namely the upper convected Maxwell model of (\ref{ucmm}), does not predict shear banding. The widespread observation of this phenomenon  \cite{rehagehoffmann,nmr,nmr2,nmr3,birefringence} is thus not implicit in the linear response of the system, but confirms a quite separate primary prediction of the reptation-reaction model.

\subsubsection{Second Generation Tube Models: Convective Constraint Release}
\label{gen2tube}

One might ask whether such banding phenomena can be expected in ordinary, unbreakable polymers. Though seemingly robust for micelles,  there is only very limited (and controversial) evidence for shear banding in polymers \cite{polymerband}; the consensus view is that, even for monodisperse chains, shear banding does not occur in the unbreakable case. Certainly there is no experimental evidence for shear banding in even moderately {\em polydisperse} polymers (let alone the exponential polydispersity represented by (\ref{size})). Since this physics should be recovered in the unbreakable limit, the reptation-reaction model includes among its robust primary predictions that the shear banding instability, seen in the Maxwellian regime, should disappear as one crosses into the highly non-Maxwellian regime at $\zeta = \tau_b/\tau_R \gg 1$. 

Intriguingly however, the `first generation' tube models, which underly the reptation-reaction constitutive equations (\ref{ceone}--\ref{cefour}), do actually predict a nonmonotonic flow curve for monodisperse unbreakable chains in steady shear. This has formed a large part of the inspiration for the development of `second generation' tube-model constitutive equations. These deploy the concept of `convective constraint release' (CCR) whereby the nonlinear stretching and retraction of chains in strong flows reduces the strength of entanglements \cite{CCR,milner2,mcleish02}. Most of these models entail at least one phenomenological parameter describing the strength of the CCR effect; in \cite{milner2}, for example, the relevant parameter is called $c_\nu$. 

It is important to ask whether the CCR concept, if consistently applied to the case of entangled micelles, restores monotonicity of the flow curve in that case also. If it did, this would create a direct conflict between the modelling of unbreakable chains (which require CCR to restore monotonicity and avoid shear bands) and micelles (which require nonmonotonicity to explain shear bands). This issue is addressed directly in \cite{milner2}. It is found that, on varying the parameter $c_\nu$ (which vanishes in the first-generation theories) a significant range exists where the flow curve for unbreakable chains is monotonic but that for micelles is not. The requirement that the model admit shear-banding in micelles is then used by the authors of \cite{milner2} to limit $c_\nu$ to that window. 

The second-generation constitutive model for micelles derived in \cite{milner2} directly combines the physics of CCR with the reptation-reaction picture that led to (\ref{ceone}--\ref{cethree}). The resulting equation set is less transparent, however; we refer the reader to \cite{milner2} for details. In brief, retaining (\ref{ceone}), these authors derive a differential equation for a quantity $f_{ij}(x)$ whose integral over $x$ plays the role of $W_{ij}$ in (\ref{ceone}--\ref{cethree}). This differential equation involves a convection term related to (\ref{cethree}), a stretching/orientation term related to (\ref{cefour}), and a new CCR term ($\propto c_\nu$) which has a relatively elaborate dependence on the flow history. Presumably, although it is not proved explicitly in \cite{milner2}, this formulation reduces to (\ref{ceone}--\ref{cethree}) when the CCR term is switched off. 

Although a careful numerical comparison has not yet been made for anything other than steady shear, it seems likely that for modest values of $c_\nu$ (chosen to retain the nonmonotonicity of the flow curve) the primary rheological predictions of this second-generation micellar model broadly coincide with those found from the original reptation-reaction model Eqs.~(\ref{ceone}--\ref{cefour}). This expectation is based on more extensive comparisons that have been done between first- and second-generation models in the case of unbreakable chains \cite{milner2}, where rheological features unconnected with flow-curve monotonicity are not strongly altered.
 
\subsection{Secondary Predictions: Concentration and Temperature Dependence}
\label{secondarypredictions}
If $E$, $k_{rs}$ (or $k_{ei},k_{bi}$, as applicable) and $l_p$ are independent of concentration, a number of secondary rheological predictions can be obtained for the concentration dependence of $\tau$ (and hence viscosity $\eta = G_0\tau$) and related quantities. However, these depend on: (a) whether the scaling or mean-field theory applies; (b) whether reversible-scission, end-interchange, or bond-interchange dominates; (c) whether one is in the pure reptation-reaction regime governed by (\ref{squareroot}), or in the breathing-reaction regime obeying (\ref{breathing}). Accordingly such laws for concentration trends are at best a rough guide, from which only rather strong deviations can be taken as firm evidence against the underlying model. 

Nonethless, we present in Table \ref{tableone} `best estimate' scaling laws for an exponent $g$ relating the Maxwell time to concentration ($\tau\sim\phi^g$) for different reaction schemes in the three different regimes governed by unbreakable chains, reptation-reaction, and breathing-reaction processes. These best estimates, reproduced from \cite{bint}, make use of scaling exponents in the absence of rings (see Section \ref{bmf} above) for both static properties and the effective tube diameter. (In the scaling perspective, the latter is the mesh size, $\xi$ \cite{degennes}.) However, the results would not be very different if mean-field estimates were applied consistently instead. To convert from $\tau$ to the viscosity, one multiplies by $G_0\sim\phi^{2.3}$.

\begin{table}[t]
\label{tableone}
\caption{The exponent $g$ (where $\tau \sim \phi^g$ and $\eta \sim \phi^{g+2.3}$) for various regimes.}
\begin{center}
\begin{tabular}{crrrr}
\hline
Reaction &
$\zeta \gg 1$ & 
$1\gg \zeta\gg N_T^{-1}$ & 
$N_T^{-1} \ll \zeta$
\\\hline
reversible scission & 3.4 & 1.4 & 0.9\\
end interchange & 3.4 & 1.2 & 0.6 \\
bond interchange & 3.4 & 1.7 & 0.3 \\
\hline
\end{tabular}
\end{center}
\end{table}

We note that $g$ is always positive; in no regime does $\tau$ fall with $\phi$. This is reasonably robust, but could be altered if for some reason the reaction rate for the relevant reaction scheme ($k_{rs}$, say) was strongly increasing with $\phi$. Observation of small positive exponents is not disproof of the model, but might be evidence that a system is in the breathing regime with interchange kinetics. Thirdly, even if kinetic parameters such as $k_{rs,ei,bi}$ and static ones such as $E,l_p$ are $\phi$-independent as assumed, and breaking is rapid, there can be crossovers between six different regimes (Table \ref{tableone}); these might normally prevent the experimental verification of simple power laws of any kind. Even more complex trends can be rationalized, {\em a posteriori}, once ionicity effects are expressly considered \cite{majid}; see Section \ref{salt}. Although perhaps more informative in being broken than obeyed, the scaling laws in Table \ref{tableone} do give reasonable agreement with some of the simpler experimental systems (e.g., \cite{catescandau,berretcole,canscale}).


One can make further interesting predictions by assuming that $E$ is independent of $T$ and also that the activation energy $E_A$ (for whichever reaction is dominant) is likewise $T$-independent. Note that $E_A$ and $E$ are not the same, even for reversible scission (unless there happens to be no activation barrier for fusion of chain ends). Nonetheless, since $E$ and $E_A$ are likely to be much bigger than the Arrhenius energy $E_S$ for the solvent viscosity (which controls the curvilinear friction parameter $\tilde D_c$), Equation (\ref{squareroot}) for reversible scission (say) predicts that $\tau \propto \exp[\beta E']$ with $E' \simeq (E_A+E)/2$. Since $E_A$ is measurable in temperature jump, this allows a check on the equilibrium scission energy $E$; reasonable values (around $20k_BT$) have been obtained in this way, as have reasonable trends for dependence on ionic strength and other factors \cite{catescandau,oelschlagertjump,oelschlagermemory}.
In one recent study \cite{oelschlagermemory}, strong evidence is given for the switching off of the reversible scission process at low ionic strength, at least in the specific classes of ionic surfactants studied there. This can be rationalized in terms of the electrostatic and bending contributions to the activation barriers for micellar reactions. (See \cite{leng} for a recent use of such ideas in a different context.) At low salt, the micellar breaking time $\tau_b$ is then controlled by an interchange process, invisible to T-jump, and the activation energy comparison with T-jump data described above is inapplicable.

\subsection{ Role of Branching: Ionicity Effects}
\label{salt}
In ionic micellar systems, and also zwitterionic ones, the overall ionic strength and/or the degree of specific counterion binding can strongly influence $E, E_A, l_p$ and other parameters that were assumed constant in the above analysis. While naive application of scalings such as those in Table \ref{tableone} is thereby precluded, it is possible in many cases to make sense of the trends, at least {\em a posteriori}, within the physical precepts of the reptation-reaction picture, so long as branching is allowed for. Much work along these lines is reviewed in \cite{majid}.

\label{networks}

As described for equilibrium statistics in Section \ref{branch}, the presence of branching in micellar chains can alter  both static and rheological predictions for their properties. One crucial observation is that micellar branch-points are {\em labile}; they are always free to slide along the length of a micelle. This is quite different from crosslinks in conventional polymers and, counterintuitively, means that branching generically decreases the viscosity of the system, rather than increasing it. This was first recognized in an important paper by Lequeux \cite{lequeux}, who showed that, if branch points are present in an entangled micellar network, the curvilinear diffusion constant of a chain end is $D_c = \tilde D_c/\overline L$, where $\overline L$ now denotes (roughly speaking) the distance to the nearest other chain end {\em or junction point} in the network. This is because a branched network, at scales above the distance between junctions, offers an efficient reservoir for micellar contour length. (This can be stored or borrowed by sharing it with other network strands, rather than having to move it all the way to the other end of a given, unbranched chain.) Accordingly some of the main rheological predictions of the reptation-reaction model can be retained, so long as $\overline L$ carries this new interpretation. 
Once branching is widespread, as shown in Section \ref{branch} one then has $\overline L \sim \phi^{-1/2}$ (within a mean-field picture); for rheological purposes, the system behaves {\em as if} micelles were becoming shorter with concentration.
This can lead to a rheological relaxation time exponent $g$ that, for reversible scission reactions, is barely larger than zero ($\tau\sim\phi^g = \phi^{0.15}$ in a scaling picture \cite{oelschlagertjump}). For interchange reactions $g$ can even be somewhat negative, in contrast with the  results of Table \ref{tableone}. 

In a careful recent study \cite{oelschlagertjump,oelschlagertjump2}, Oelschlaeger {\em et al}  correlate surfactant hydrophobicity, salt, and counterion binding efficiency  with the observed $g$ values. These authors find evidence that the unbranched reptation-reaction model is applicable at relatively low ionic strength, high hydrophobicity, and low counterion binding efficiency. 
They then argue that the model of Lequeux \cite{lequeux} explains the falling $g$ values seen when ionic strength is increased, surfactant hydrophobicity decreased, or counterion binding increased \cite{majid}. Each of these can be argued to favour the branching of micelles either through curvature or electrostatic energy effects \cite{oelschlagertjump}. 
Moving such parameters further in the same direction can lead to significantly negative $g$  and/or a complete collapse of viscoelasticity \cite{portesatnet}. The latter may signify the onset of network saturation (see Section \ref{branch}); by this point all potential entanglements have been replaced by labile branch points, and resistance to shear is very low. 

\subsection{ Shear Thickening}
\label{thickening}

Of equal fascination to the Maxwellian and shear-thinning behaviour described above, is the phenomenon of shear thickening \cite{oelschlagermemory,
hoffmannthicken,matthys,oda,pinethick,
berretmemory}. This is seen in a window of volume fraction around the onset of viscoelasticity at $\phi\simeq\tilde\phi$; an initially inviscid or barely viscoelastic system is found, after a period of prolonged shearing above a critical shear rate $\dot\gamma_c$, to convert into a much more viscous state. In some cases the viscosity increase is modest (a factor between two and ten) and the state relaxes quite rapidly to the previous one when shearing ceases \cite{hoffmannthicken}. In other cases, the new phase is a long-lived gel; shear banding is often implicated in the formation of such a gel \cite{pinethick}. The thickened phase contains micelles that are nearly fully aligned (the optical extinction angle is close to zero); its formation involves a nontrivial latency time which in some cases is geometry-dependent \cite{oelschlagermemory}.

At the time of writing there is no microscopic model capable of explaining these phenomena at anything like the level of a constitutive equation, nor indeed any consensus among theorists as to the mechanisms involved. Model-building attempts  have been made from time to time, based for example on shear-induced aggregation or polymerization of rodlike micelles \cite{aggregation}. There are also models that couple a gelation transition to shear bands \cite{goveas1,goveas2}, and some that ascribe the shear thickening directly to ionic or electrokinetic phenomena \cite{liu}. All these models have drawbacks; for example, aggregation models normally require very high shear rates for onset whereas experimental values are low; thickening is sometimes seen in nonionic micelles \cite{matthys}; {\em etc.}.
In the next Section we briefly review a  radical approach in which shear-thickening is speculatively attributed to the presence of micellar rings \cite{rings}. 

\subsection{ Role of Rings}
\label{ringstuff}
As discussed in Section \ref{rings}, the expected influence of micellar rings is maximal around $\phi_r^{max}$, the (sadly unknown!) volume fraction at which, in a system of high enough scission energy $E$,  a cascade of rings crosses over to a semidilute solution of open chains. The reptation-reaction model assumes strong entanglement and hence requires $\phi \gg \phi_r^{max}$; in this regime any remaining rings are smaller than the mesh size $\xi$ \cite{ppw} and hence of rather little rheological consequence.
Thus the role of rings is limited to the concentration range below the strongly entangled regime, where a cascade of rings is predicted (Section \ref{rings}).

In the putative cascade-of-rings phase, governed in mean-field by (\ref{cring}) and present in theory at large enough $E$, it is an open question \cite{rings} whether or not the rings interlink so as  to form a percolating linked network. For unbreakable ring polymers, formation of such a structure would create a permanently elastic solid 
known as an `olympian gel' \cite{degennes}.
Although in a micellar system the rings would have a finite breakage time, one still expects drastic changes to the flow behaviour under any conditions where significant concatenation can arise, as seems possible in the `cascade of rings' regime.
Suppose for now that the rings are indeed linked for $\phi\simeq\phi_r^{max}$. If so, then were micellar kinetics suddenly to be switched off, creating an olympian gel, the system would acquire some finite residual modulus $G_r$ (assumed small), indefinitely resisting attempts to impose a steady shear flow. Restoring a finite delinking time $\tau_l$, this gel becomes a viscoelastic fluid of viscosity $\eta _r = G_r\tau_l$. It is possible that, if $G_r$ is small enough, $\eta_r$ remains comparable to that of the solvent; the sample is only marginally viscoelastic and in that sense would be considered to have $\phi\simeq\tilde\phi$ (and identified as $\phi\simeq\phi^*$ if an assumption of linear chains were made). 
Notice now that the viscoelastic linked-ring fluid does not in fact require complete percolation of linked rings; it only requires that linked structures extend far enough that their configurational relaxation times exceed $\tau_l$. (If these large structures are few and far between, the better is the assumption of small $G_r$.)

If such a linked-ring fluid exists around $\phi\simeq\tilde\phi$, any shear rate $\dot\gamma \ge \tau_l^{-1}$ will cause strongly nonlinear effects. The orientating effect of elastic strains (of order $\dot\gamma\tau_l$) on the linked rings will alter the reaction rates for bond-interchange and (where present) other reactions. For example, pulling two interlocked rings in opposite directions could promote bond interchange at their contact point, increasing the creation rate for a larger ring and shifting the mean ring size upwards. This polymerization tendency could well cause shear thickening, as could tension-induced chain scission, by pushing the equilibrium in (\ref{chainringmeanfield}) towards polymerization. 

This scenario, though speculative, does explain several observed features of the shear-thickening process. One is a shear-rate threshold for the transition that is lower than any reasonable estimate of the reorientation time for marginally overlapped linear micelles \cite{rings}. Second, since the observed gel phase is fully aligned, a large total strain must be applied before it is fully formed. This implies a latency time at least of order $\tau_lN_T^{1/2}$, and potentially much longer if the shear-thickening transition is caused by only a slight rate inequality between formation and destruction of larger-than-average micelles \cite{rings}. The latter mechanism could also give a geometry-dependent latency time as reported experimentally \cite{oelschlagermemory}.
Third, any ring-dominated regime requires suppression of open chains in the quiescent state and hence large $E$. Factors favouring large $E$ include raising counterion lipophilicity and using gemini (twin-tailed) surfactants; such factors do seem broadly to cause a reduction in $\dot\gamma_c$ and enhancement of the viscosity jump.
Finally, this picture can tentatively explain some strange memory effects seen around the shear thickening transition, discussed next.

\subsection{ Structural Memory Effects}
\label{memory}

Structural memory is the presence in a system of internal degrees of freedom, other than the stress, which relax on a time scale $\tau_s$ that is at least comparable to the stress relaxation time $\tau$ itself (and in some cases vastly longer). For example, one can find among micellar systems some instances where $\tau$ itself is at most a few seconds; but the value one measures for $\tau$ in repeat experments depends in a complicated manner on the process history of the sample over, say, the preceding 24 hours. 
It is possible, though not yet certain, that structural memory plays a major role in the exotic rheological phenomena, such as rheochaos, considered in Section \ref{macro} below.
Though hinted at anecdotally from the earliest days of the subject \cite{hoffmannrev,hoffmannthicken}, it is only very recently that structural memory effects in micellar systems have been studied in systematic
detail \cite{oelschlagermemory,berretmemory}. 
\if{
These studies concern the neighbourhood of the shear-thickening transition, although the generic explanation of structural memory given below applies equally in the strongly entangled regime (under conditions when reversible scission 
is slow). It would therefore be interesting to look more carefully for these effects at higher concentrations.
}\fi

Among effects observed in \cite{oelschlagermemory,berretmemory} are the following. In some shear-thickening micellar systems subjected to steady shearing, there is an initial latency time $\tau_{lat}$ (seconds or minutes) for the thickening to occur. After this, however, the stress level continues to adjust slowly over time scales $\tau_s$ of order hours or days before finally achieving a steady state. If shearing is stopped, the stress relaxes quite rapidly but the {\em memory of having been sheared} persists for times of order $\tau_s$: if shearing is resumed within this period, a quite different $\tau_{lat}$ is measured. Moreover the stress level immediately after latency is closer to the ultimate steady-state value, and almost identical to it if the switch off period has been short compared to $\tau_s$. Finally, the latency time can also be raised or lowered by a prior incubation at elevated or reduced temperature \cite{oelschlagermemory,berretmemory}, even in a sample that has never been sheared.

These phenomena point to a robust structural property, perturbed by shear but also by temperature, as the carrier of structural memory in micellar systems. One such property immediately springs to mind, namely the micelle size distribution. However, as discussed in Section \ref{kinetics}, this relaxes rapidly to its equilibrium form (in mean field, this is the usual $c(L) \propto \exp[-L/\overline L]$)  when any kind of micellar reaction is present. On the other hand, as also discussed in Section \ref{kinetics}, the  
relaxation of the mean micelle length $\overline L$ is contingent on the presence of reversible scission reactions. This is because interchange reactions conserve the chain number $\sum_Lc(L)$; and, given the fixed shape of the distribution, $\overline L$ can only change if the chain number
 does so. As emphasized in Section \ref{kinetics}, chain number is conserved by interchange reactions, even when rings are present; but ring number is not itself conserved.
 
The structural memory effects reported in \cite{oelschlagermemory,berretmemory} close to the shear thickening transition can be explained in outline if one assumes that (a) both rings and open chains are present and (b) reversible scission reactions are very slow. The slow relaxation time is $\tau_s = \tau_{rs}$, the time scale for the chain number to reach equilibrium. Stress relaxation is not slow, since the faster rate for end interchange or bond interchange reactions will dominate, allowing fast relaxation of  all quantities other than $\overline L$, including stress.
A system which is sheared or thermally treated for a time long compared to $\tau_{rs}$ will acquire a steady-state chain number appropriate to those conditions; if conditions are now changed, it will take a time of order $\tau_{rs}$  to relax to the new value. Among things that can vary with chain number are, of course, the latency time in the thickening transition; and also the Maxwell time for stress relaxation in the quiescent state.

To understand how the structural memory time can become so long  \cite
{oelschlagermemory,berretmemory}, note that the reaction rates for the three reaction schemes (reversible scission, end-interchange, and bond-interchange) all involve different activation energies. For instance, end-interchange involves passing through a state which contains a threefold junction in place of a single end cap; this gives an estimate $E_b-E/2$ for the relevant activation energy, where $E_b$ is the energy of the branch point. The activation energy for scission, however, cannot be less than $E$, the energy to create two end-caps. (An activation energy cannot be less than the energy rise between initial and final states.) Using these estimates, we find $\tau_{rs}/\tau_{ei} \sim \exp[(3E/2-E_b)/k_BT]$. Hence it is indeed quite possible to have a structural relaxation time $\tau_s = \tau_{rs}$ (governing relaxation of the micellar size distribution) that exceeds by orders of magnitude the dominant kinetic timescale involved in stress relaxation. The required chemical conditions are those that favour large $E$ (small $\tilde\phi$) and modest $E_b$. (If the latter is too small, there will be significant branching of micelles at concentrations around $\tilde\phi$; this will further complicate the rheology, but not affect the basic time scale separation under discussion here.) These considerations are broadly consistent with the observed trends in $\tau_s$ on varying the choice of surfactant, ionic strength, and other factors \cite{oelschlagermemory,berretmemory,rings}.

Parts of above scenario could hold equally well for giant micelles in the fully entangled regime, where rings are not important. For a system showing slow reversible scission in T-jump \cite{oelschlagertjump,oelschlagermemory}, it would be interesting to look for the effects of nonlinear flow, and also thermal pre-treatment, on the Maxwell time. If the nonlinear flow creates a shear banding region, this might reveal whether the average chain length in the high-shear band is significantly perturbed by the flow.  
A more radical speculation is that at least in some cases, when a
shear-induced gel phase that forms below $\tilde\phi$ persists after
stress is removed (and does not retain nematic order
\cite{berretband}), this state is in disequilibrium {\em solely}
through having an enhanced $\overline L$ ( of lifetime $\tau_s =
\tau_{rs}$). If so, the shear-induced gel state is just another
instance of the entangled regime of giant micelles, to which the
reptation-reaction model can be applied. An open issue
would be how $\overline L$ acquires its nonequilibrium value during
the induction period of the shear-thickening transition.

\section{Macroscopic Constitutive Modelling}

\label{macro}

So far we have discussed microscopic constitutive modelling, which
aims to predict rheology from an understanding of the microscopic
dynamics of the polymer-like micelles themselves. We expressed the
stress tensor (\ref{polymerstress}) as a sum over micellar chain
segments, and derived the constitutive equation (\ref{cetwo}) for its
dynamics. Pursuing this approach further, particularly to address
nonstationary shear-banded flows, becomes prohibitively
complicated. Indeed, current microscopic models for micelles lack some of the
important physics of these flows, such as coupling between flow and
concentration or collective orientational fluctuations. To make
progress, we now turn to phenomenological models which, thanks to
various adjustable parameters, can be tuned to mimic the micellar
problem to some extent.

Before discussing individual models, we sketch the basic
features that are common to all of them. The stress $\sigmatot_{ij}$ is taken to comprise additive contributions from the polymer-like
micelles and from a Newtonian solvent (as per (\ref{jsmodel})):
\be
\label{eqn:contributions}
\sigmatot_{ij}=\sigmapol_{ij} + \etasol (K_{ij}+K_{ji})
\ee
The relevant instabilities are viscoelastic and not inertial in
origin, so we work throughout at zero Reynolds number. The stress then
obeys the force balance equation
\be
\label{eqn:forceBalance}
\nabla_i\sigmatot_{ij}-\nabla_j p = \nabla_i \sigmapol_{ij} + \etasol \nabla^2 v_j -\nabla_j p= 0
\ee
Here $p$ is an isotropic pressure, which maintains fluid
incompressibility:
\be
\label{eqn:incomp}
\nabla_iv_i=0
\ee
The viscoelastic stress $\sigmapol_{ij}$ is then
written as a function of some underlying microstructural
quantities, whose identities vary according to the system and
regime of interest. Common choices include the concentration $\phi$
and molecular deformation $W_{ij}$ of the polymeric component; the
orientation tensor $Q_{ij}$ in nematics; the micellar length
distribution $P(L)$, {\em etc.}:
\be
\label{eqn:basicStress}
\sigmapol_{ij}=\sigmapol_{ij}(W_{ij},\phi,\cdots)
\ee
Among these microstructural variables it is important to distinguish
``fast'' from ``slow'' variables. The former quickly relax to local
steady-state values determined by the latter, whereas each ``slow''
variable requires its own dynamical equation of motion.  Formally, the
slowest variables are the ``hydrodynamic'' ones, which relax at a
vanishing rate $\omega\propto k^\alpha$ ($\alpha>0$) for small
wavenumbers $k\to 0$, either because a long-wavelength distortion of
the variable costs very little free energy (broken symmetry), or
because the quantity is conserved and therefore obeys diffusive
dynamics ($\alpha=2$). In the formal limit $k\to 0$, all other
variables are fast in comparison to these hydrodynamic modes.  In
viscoelastic solutions however, variables that are not strictly
hydrodynamic nonetheless relax very slowly. An example is the
molecular deformation $W_{ij}$ governed by the Maxwell time $\tau$
(often seconds or minutes). It is therefore essential to add to the
list of ``slow'' variables $W_{ij}$ and any other slow but formally
non-hydrodynamic quantities~\cite{milner93}.

As discussed in Sec.~\ref{memory} above, many micellar systems show a
pronounced structural memory, with degrees of freedom that relax on a
time scale greater than the intrinsic
deformational relaxation time $\tau$. Indeed, there are instances in
which $\tau$ is itself at most a few seconds, but with a value that
varies between repeat experiments in a complicated (though
reproducible) manner that depends on the process history over the
proceeding day or so. This points to an underlying structural property
that evolves slowly, on a timescale of hours/days. One such candidate
is the mean micellar length, with a timescale $\tau_{\rm rs}\gg\tau$;
another is the micellar concentration.

Consider, then, a scenario in which the relevant dynamical variables
are the micellar deformation $W_{ij}$, the micellar concentration
$\phi$, and the mean micellar length $\bar{L}$, with respective
relative relaxation timescales $\tau\ll \Lambda^2/\mathcal{D}\ll \tau_{\rm rs}$, for a
micellar diffusion coefficient $\mathcal{D}$ and sample size $\Lambda$. The
concentration evolves with conserved dynamics of the form
\be
\label{eqn:basicConc}
\Dt{\phi} = \nabla M\nabla \mu( \nabla_n v_m, W_{nm},\phi)
\ee
with a mobility $M$, proportional to $\mathcal{D}$.  The derivative
$\Dt\equiv\partial_t+v_l\nabla_l$ denotes rate of change in a fluid
element convected with the flow field $v_i$. The direct counterpart of
(\ref{eqn:basicConc}) for nonconserved quantities such as $W_{ij}$
and $\bar{L}$ is an equation of the form
\be
\label{eqn:basicW}
\Dt{W_{ij}} = \frac{1}{\tau} G_{ij}( \nabla_nv_m, W_{nm},\phi)
\ee
Sometimes, however, this differential structure is replaced by an
integral form
\be
\label{eqn:basicL}
\bar{L}=\int_{-\infty}^t dt' \mathcal{M}(t-t')g(t')
\ee
where $g$ depends on time $t'$ via all relevant
quantities $W_{ij}(t')$, {\em etc.}.  Only in the case of an exponential
memory kernel, $\mathcal{M}(t)=\exp(-t/\tau)/\tau$, can
(\ref{eqn:basicL}) be recast into the  differential form of
(\ref{eqn:basicW}). The integral form of (\ref{eqn:basicL}) emphasises,
for long-lived variables such as $\bar{L}$, the structural memory at time $t$ to all earlier times $t'<t$.

Note that (\ref{eqn:basicStress}) above assumes the polymeric stress
to be instantaneously prescribed by any given configuration of the
underlying microscopic variables, some of which themselves obey
``slow'' dynamical equations. This approach retains the basic
principle that even macroscopic models should ultimately be motivated
by microscopics, however oversimplified the evolution equations
(\ref{eqn:basicConc}--\ref{eqn:basicL}) become. A simpler, and
very common, approach consists of directly prescribing the dynamics of
the polymeric stress itself by means of an autonomous viscoelastic
constitutive equation.  Fundamentally, though, this should still be
thought of in terms of the evolution of an underlying microstructural
quantity: for example, in Section~\ref{sec:ucmm}, the equation
(\ref{ucmm}) for stress evolution is actually an instance of
(\ref{eqn:basicW}).

The basic structure outlined above encompasses the microscopic models
discussed earlier. For example, the reptation-reaction model has
(\ref{ceone}) for (\ref{eqn:basicStress}); (\ref{cetwo} -- \ref{cefour}) which can be cast into the form of (\ref{eqn:basicW});
and assumes the concentration to remain uniform (so that
(\ref{eqn:basicConc}) is suppressed, with $\phi$ constant in
(\ref{eqn:basicStress})). In the macroscopic approach to constitutive
modelling, one instead arrives at equations of the form
(\ref{eqn:basicStress} -- \ref{eqn:basicW}) by ansatz, or by an
exact description of a simplified system (such as the dumb-bell model
of Section~\ref{sec:ucmm}).  Further guidance comes from the generic
constraints of translational and rotational invariance, and from Onsager
reciprocity.  The crucial advantage of the macroscopic approach is
that it allows coupling between the flow and microstructural
quantities such as $\phi$, $Q_{ij}$ and $P(L)$ to be incorporated in a
simple way. Another ingredient, almost always absent from microscopic
models (though see Ref.~\cite{dhont2003b}), is that operators such as
$G$ in (\ref{eqn:basicW}) should contain spatially nonlocal terms,
which are needed to correctly describe the structure of spatially
inhomogeneous ({\em e.g.}, banded) flows~\cite{olmstedluball}:
\be
\Dt{ W_{ij}} = G^{\mbox{\small local}}_{ij}(W_{nm}, \nabla_nv_m,\phi) +\mathcal{D} \nabla^2 W_{ij}
\ee

\subsection{Johnson-Segalman Models for Shear Thinning}

\label{sec:JS}

Within the framework just described, we now discuss some specific
models of shear thinning. These are
designed to reproduce, at the level of macroscopic modelling, the
nonmonotonic constitutive curve of the microscopic reptation-reaction
model (Sec.~\ref{predictions}), for which homogeneous flow is unstable
with respect to the formation of shear bands.
From now on, we reserve the
term ``constitutive curve'' for the underlying nonmonotonic relation
between stress and strain rate, and use ``steady state flow curve'' for
the actual stress/strain-rate relation measured in an experiment. In
the banding regime, the former is unstable and gives way to the
latter: the two only coincide when the flow remains homogeneous.

The most widely used model was originally devised by Johnson and
Segalman~\cite{johnson77}, and later extended by Olmsted {\em et
al}~\cite{olmsted99a} to include the spatially nonlocal terms needed
to describe the structure of the interface between the bands. Force
balance and incompressibility are given by (\ref{eqn:forceBalance})
and (\ref{eqn:incomp}). The viscoelastic stress of
(\ref{eqn:basicStress}) is assumed to depend linearly on the molecular
deformation tensor $W_{ij}$, and on the concentration $\phi$ via a
modulus $G$:
\be
\sigmapol_{ij}=G(\phi)W_{ij}
\ee
The deformation tensor $W_{ij}$ obeys diffusive Johnson-Segalman (dJS) dynamics as follows:
\be
\label{eqn:js}
D_t W_{ij} = a(D_{il}W_{lj} + W_{il}D_{lj}) + (W_{il}\Omega_{lj}-\Omega_{il}W_{lj}) + 2D_{ij} -\frac{1}{\tau(\phi)}W_{ij} + \mathcal{D}\nabla^2W_{ij}
\ee
in which $D_{ij}$ and $\Omega_{ij}$ are respectively the symmetric and
antisymmetric parts of the velocity gradient tensor
$K_{ij}=\nabla_jv_i$. (Note that the deformation tensor $W_{ij}$ used
here differs from that in Section~\ref{sec:ucmm} by a trivial
isotropic contribution.)  For the moment we assume that $\phi$ is
uniform so that $G(\phi)$ and $\tau(\phi)$ are constants in space and
time for any sample. 

\begin{figure}[tbp]
\centering
\includegraphics[width=3.5 in]{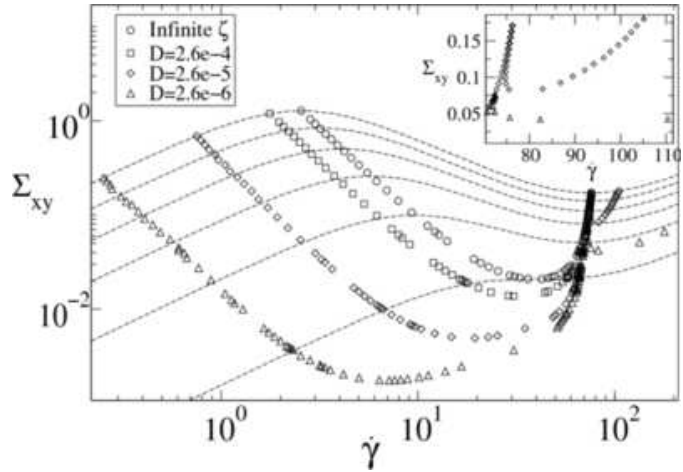}
\caption{Dashed lines: constitutive curves for the diffusive Johnson-Segalman model with $G\sim\phi^{2.2}$, $\tau\sim\phi^{1.1}$ and
various $\phi$.
Circles: limits of linear stability in the dJS model.  Squares:
corresponding limits for the full dJS$\phi$ model with an
experimentally realistic micellar diffusion coefficient. (Diamonds and
triangles are for an artificially reduced $\mathcal{D}$.)  Inset: zoom
on large $\gdot$. Figure reprinted with permission from Ref.~\cite{fielding2003c}.
\label{fig:spinodals_with_phi}}
\end{figure}

Setting $a=1$ and $\mathcal{D}=0$ in (\ref{eqn:js}) we recover Oldroyd B
dynamics, as derived in Section~\ref{sec:ucmm} by considering an
ensemble of relaxing dumb-bells undergoing affine deformation under shear. Although, as described there, Oldroyd B is
the most natural extension to nonlinear flows of the linear Maxwell
model of Section~\ref{linmaxsec}, its trivial constitutive curve
$\sigma_{xy}(\gdot)=G\gdot\tau+\etasol\gdot$ fails to capture the dramatic
shear thinning and related flow instabilities seen in wormlike
micelles.
To allow for shear thinning, the Johnson-Segalman model invokes a
`slip parameter' $a$, with $|a|\le 1$. When $|a|<1$, the dumb-bells no
longer deform affinely, but slip relative to the flow field. The
resulting constitutive curve is then
\be
\sigma_{xy}(\gdot)=\frac{G\gdot\tau}{1+(1-a^2)\gdot^2\tau^2}+\etasol\gdot
\ee
The viscoelastic contribution (first term) now shear-thins
dramatically and is nonmonotonic, increasing as $G\gdot\tau$ for small
$\gdot$ before decreasing towards zero at higher shear rates (the
maximum is at $\gdot\tau\sqrt{1-a^2}=1$). In contrast, the Newtonian
solvent stress always increases with $\gdot$. The overall shape of the
constitutive curve thus depends on the relative strength of these
contributions. For $\etasol>G\tau/8$ the solvent dominates the
viscoelastic stress to restore monotonicity (bottom dashed curve in
Fig.~\ref{fig:spinodals_with_phi}). For lower $\etasol$ the negative
slope survives over some range of shear rates (upper dashed curves in
Fig.~\ref{fig:spinodals_with_phi}). Assuming that the modulus
increases with concentration as $G\sim \phi^{2.2}$ and the relaxation
time as $\tau(\phi)\sim \phi^{1.1}$~\cite{bint}, one obtains the full
family of dashed curves $\sigma_{xy}(\gdot,\phi)$ of
Fig.~\ref{fig:spinodals_with_phi}~\cite{fielding2003c}. As discussed in
Sec.~\ref{sec:steady} below, the spatial gradient terms in
(\ref{eqn:js}) are needed to confer a finite interfacial width
$l\propto\sqrt{\mathcal{D}\tau}$ between the bands~\cite{olmstedluball}.

In work related to the above, Ref.~\cite{rossi}, Cook and Rossi considered the dynamics of an
ensemble of bead-spring dumb-bells of finite extensibility, subject to spring,
Brownian and hydrodynamic forces, allowing for slip. They thereby
derived a coupled set of equations of motion for the stress and the
number density of dumb-bells. When the finite extensibility
parameter tends to zero, these reduce to the Johnson-Segalman
model. They studied the predictions of this model, including shear
banding, in planar shear for a variety of boundary conditions. In
Ref.~\cite{rossi1}, this work was generalised to Taylor-Couette flow.

\subsubsection{Concentration Coupling}

Above we assumed the micellar concentration $\phi$ to remain
spatially uniform. Generically, however, one expects concentration
fluctuations to be important in sheared multi-component solutions when
different species have widely separated relaxation
times~\cite{brochdgen77,milner91,deGen76,Brochard83}, as seen
experimentally in
Refs.~\cite{WPD91,GerHigCla99,WheIzuFul96,KadEgm97}. This was first
explained by Helfand and Fredrickson~\cite{HelfFred89} in the context of polymer
solutions, as follows. Under shear, parts of a
stretched polymer chain (or micelle) in regions of low viscosity will,
on relaxing to equilibrium, move further than parts mired in regions
of high viscosity and concentration. A relaxing chain thus on average
moves towards the high concentration region. This provides a positive
feedback whereby chains migrate up their own concentration gradient,
leading to flow-enhanced concentration fluctuations.

In a remarkable paper, Schmitt {\em et al}~\cite{schmitt95} outlined
the implications of this feedback mechanism for the shear banding
transition. They predicted an enhanced tendency to form bands,
together with the existence of a concentration difference between the
bands in steady state. They futher noted that this difference would
lead to a slight upward ramp in the stress ``plateau'' of the steady
state flow curve. Subsequently, strongly enhanced concentration
fluctuations were seen in the early-time kinetics of shear band
formation~\cite{DecLerBer01}. Observations of a ramping stress plateau
are now widespread; for example, see Ref.~\cite{berretband}. In
Ref.~\cite{fielding2003c}, therefore, one of us proposed an extension to
the dJS model, by combining the constitutive equation (\ref{eqn:js})
with a two-fluid model for flow-concentration coupling.

The basic assumption of this two-fluid
approach~\cite{brochdgen77,milner91,deGen76,Brochard83} is a separate
force-balance for the micelles (velocity $\vm_i$, volume fraction
$\phi$) and solvent (velocity $\vs_i$, volume fraction $1-\phi$) in
any fluid element.  Any relative velocity $\vrel_i=\vm_i-\vs_i$
(implicitly assumed zero in the ordinary dJS model) can then give rise
to concentration fluctuations.  The forces and stresses acting on the
micelles are assumed as follows: (i) the usual viscoelastic stress
$G(\phi)W_{ij}$; (ii) an osmotic force $\phi
\nabla_i \delta F/\delta\phi$ derived from a free energy $F$, leading
to conventional micellar diffusion; (iii) a drag force
$\zeta(\phi)\vrel_i$ impeding motion relative to the solvent with a
drag coefficient $\zeta$; (iv) an additional Newtonian stress
$2\,\phi\,\etapol_{\rm m}\,\Dm_{ij}$ due to fast micellar relaxation
processes such as Rouse modes, where $\Dm_{ij}$ is the symmetric
traceless part of the micellar strain rate tensor; and (v) a
hydrostatic pressure. The solvent experiences (I) the usual Newtonian
viscous stress; (II) a drag force (equal and opposite to that on the
micelles) and (III) a hydrostatic pressure.
Now adding (i)--(v), one obtains a force balance equation for the micelles.
Adding (I)--(III) we get the corresponding equation for the
solvent. The sum of these two equations gives the force balance equation for the fluid element as a whole
\be
0=\nabla_iG(\phi)W_{ij} - \phi\nabla_j\frac{\delta F(\phi)}{\delta \phi} + 2\nabla_i\,\phi\,\etapol \,\Dm_{ij} + 2\nabla_i\,(1-\phi)\,\etasol\,\Ds_{ij} -\nabla_jp
\ee
which replaces (\ref{eqn:forceBalance}). (We
have redefined $\etasol$ slightly here by explicitly extracting the
prefactor $1-\phi$.)  Another combination of the two equations yields the relative velocity
$\vrel_i$, which governs the concentration fluctuations:
\bea
D_t\phi &=& \nabla_i\phi(1-\phi)\vrel_i\nonumber\\
                               &=& -\nabla_j\frac{\phi^2(1-\phi)^2}{\zeta(\phi)}\left[\frac{\nabla_iG(\phi)W_{ij}}{\phi}-\nabla_j\frac{\delta F}{\delta\phi} + \frac{2\nabla_i\phi\etapol\Dm_{ij}}{\phi}-\frac{2\nabla_i(1-\phi)\etasol\Ds_{ij}}{1-\phi}\right]\nonumber
\eea
The evolution of the molecular deformation tensor $W_{ij}$ is
prescribed by dJS dynamics, as before, but now with the velocity $v_i$
in (\ref{eqn:js}) re-interpreted as the micellar velocity $\vm_i$.

The constitutive curves of this ``dJS$\phi$ model'', which by
definition describe purely homogeneous flow, are the same as for the
original dJS model. (Dashed lines in
Fig.~\ref{fig:spinodals_with_phi}.) The relevance of the new coupling
is that any {\em heterogeneity} in the flow variables now affects the
concentration field, and vice versa. As we will show below, this
enhances the tendency to form shear bands (Sec.~\ref{sec:onset}) and
leads to a concentration difference between the bands in steady state
(Sec.~\ref{sec:steady}), as earlier predicted in Ref.~\cite{schmitt95}. In the
limit of large drag, $\zeta\to\infty$, concentration fluctuations are
suppressed, and we recover the original dJS model. A two-fluid model
of shear banding was independently developed by Yuan and Jupp in
Ref.~\cite{yuan2002}. Here it was claimed that gradient terms in the
free-energy functional for the concentration field are sufficient to
give unique stress selection (see Sec.~\ref{sec:steady} below) for the
banded state, without the need for gradient terms in the viscoelastic
constitutive equation. (This is at odds with Ref.~\cite{olmstedramp}; the
discrepancy deserves further investigation, but might be attributable to the particular numerical scheme employed in \cite{yuan2002}.)

\subsubsection{A Simplified Scalar Model}

The dJS model is the simplest tensorial model to capture the
negatively sloped constitutive curve of the full reptation-reaction picture.  An even simpler model~\cite{spenley96} neglects normal
stresses and considers only the shear stress $\sigma =\sigma_{xy}$ and
the shear component $\gdot=\nabla_yv_x$ of the strain rate tensor. It
further chooses units that equate both $\tau$ and $G$ to unity, so that
$\sigmapol_{xy}=W_{xy}$. The force balance and constitutive equations
are then
\be
\label{eqn:spenley1}
\sigma=\sigmap + \etasol\gdot
\ee
\be
\label{eqn:spenley2}
\tau\partial_t\sigmap = -\sigmap + g(\gdot\tau)+l^2\nabla^2\sigmap
\ee
in which the choice $g(x) \equiv x/(1+x^2)$  is made; this recovers a JS-like constitutive curve.

\subsection{Shear Banding Instability}
\label{sec:onset}

The above models each have a non-monotonic constitutive curve
$\sigma(\gdot)$. The negatively sloping part of this is well known to be mechanically
unstable~\cite{Yerushalmi70}. Under conditions of imposed shear rate,
the system can recover steady flow by separating into shear bands, one
on each of the stable, positively sloping branches (discussed fully in
Sec.~\ref{sec:steady} below). In this section, we discuss the initial
onset of banding in the unstable regime, by means of a linear
stability analysis.

For simplicity, we start by analyzing the simple scalarised model of
(\ref{eqn:spenley1}, \ref{eqn:spenley2}). Consider a homogeneous
initial state at an applied shear rate $\gdot_0$. The viscoelastic
shear stress is $\sigmapol = g(\gdot_0\tau)$ and the total shear
stress $\sigma = g(\gdot_0\tau)+\etasol\gdot_0$. Consider now a small
heterogeneous perturbation away from this state, decomposed into
Fourier modes with wavevectors confined for simplicity to the flow
gradient direction $y$:
\be
\label{eqn:lin0}
\sigmapol(y,t)=\sigmapol_0 + \sum_{n=0}^\infty \delta\sigmapol_n\cos\left(\frac{n\pi y}{\Lambda}\right)\exp(\omega_n t)
\ee
with a similar expression for $\gdot$. We assume flow between
parallel plates at $y=0$ and $y=\Lambda$ with plate conditions of zero-slip
for the velocity and zero-flux for the stress,
$\partial_y\sigmapol=0$. The $\omega_n$ are growth rates to be
determined. If $\omega_n<0 \;\forall\;n$, the flow is
stable; $\omega_n>0$ signifies instability and the onset of banding.
Linearising (\ref{eqn:spenley1}) and (\ref{eqn:spenley2}) in these
perturbations, and recognising that the total stress must remain
uniform by force balance, we obtain
\be
\label{eqn:lin1}
0=\delta \sigmapol_n + \etasol\delta\gdot_n
\ee
\be
\label{eqn:lin2}
\omega\tau\delta\sigmapol_n = - \delta\sigmapol_n + g'(\gdot_0\tau)\delta\gdot_n - l^2k_n^2\delta\sigmapol_n 
\ee
in which the wavevector is $k_n=n\pi/\Lambda$, and the function $g(x)$ and parameter $l$ are as defined in (\ref{eqn:spenley2}). Combining these,
\be
\label{eqn:growthRate}
\omega_n\tau=-\frac{1}{\etasol}S'(\gdot_0\tau)-l^2k_n^2
\ee
in which $S'=g'(\gdot_0\tau)+\etasol$ is the slope of the underlying
constitutive curve. When this is negative we
have positive growth rates $\omega_n>0$ for wavevectors $k_n$ less
than a large cutoff (scaling as $1/l$). This signifies linear
instability and the onset of shear banding. A related analysis, for the Doi-Edwards model for unbreakable polymers, can be found in Ref.\cite{DE4}.

The assumption above is that the system starts on the underlying
homogeneous constitutive curve. In practice, it is impossible to
prepare an initially homogeneous state within the unstable
region. Instead, one performs a shear startup experiment at the
desired shear rate $\gdot_0$.  The linearisation must therefore now be
done about this time-dependent startup flow, complicating the analysis
somewhat~\cite{fielding2003a}. However, the basic stability properties
turn out to be unaltered from the simplified calculation presented
above.

In the scalarised model, there is
only one dynamical variable, $\sigmapol$. What about
the dJS model of (\ref{eqn:js})? In principle, tensoriality now
confers three dynamical stress components,
dependent on three components of the strain rate, $\nabla_i v_j$, with
$ij=xx,xy,yy$. (We have
suppressed concentration and set $\tau(\phi)=1$ and $G(\phi)=1$, so
$\sigmapol_{ij}=W_{ij}$.) However, if we allow spatial variation only
in the flow-gradient direction $y$, incompressibility confines the
velocity to the flow direction $x$, and only one component
$\nabla_yv_x\equiv
\gdot$ of the strain-rate tensor is relevant. It further turns out
that, of two linear combinations $Y,Z$ of $\sigmapol_{xx}$ and
$\sigmapol_{yy}$, only $Z$ is relevant: $Y$ decouples and
decays. The governing equations then reduce
to the form
\be
\label{eqn:structJS1}
\sigma_{xy} = \sigmapol_{xy} + \etasol\gdot
\ee
\be
\tau\partial_t \sigmapol_{xy} = f(\gdot,\sigmapol_{xy},Z)+l^2\nabla^2\sigmapol_{xy}
\ee
\be
\label{eqn:structJS3}
\tau\partial_t Z = g(\gdot,\sigmapol_{xy},Z)+l^2\nabla^2 Z
\ee

For stability purposes the
$\nabla^2$ terms can now be neglected since they act only to cut off
any instability at high wavevectors ($kl\gg 1$) as seen above.
Linearising these equations in small perturbations about a state of
homogeneous flow on the underlying constitutive curve, we then obtain a
quadratic equation for the growth rate: $\omega_n^2+b\omega_n+c=0$.
For this to have at least one unstable root, $\omega_n>0$, we need
$b<0$ or $c<0$. For this model it can be shown that $c=\tilde{c}S'$,
where $S'$ is the slope of the constitutive curve, and $b$ and
$\tilde{c}$ are positive at all shear rates.  Therefore, the condition
for instability is $S'<0$ as before: homogeneous flow is unstable in
the intermediate range of shear rates where the constitutive curve has negative slope (circles in
Fig.~\ref{fig:spinodals_with_phi}).

\if{In Sec.~\ref{sec:temporal} below, we introduce a model defined by
equations of the same basic structure as (\ref{eqn:structJS1}) to
(\ref{eqn:structJS3}), with constitutive curves of same shape as in
Fig.~\ref{fig:spinodals_with_phi}.  As we
will see, this model shares the instability described above in the
region of negative constitutive slope. It also displays a new
instability at high shear rates, where $S'>0$, now as a result of
$b<0$. This inhibits the formation of a stable high shear band and
leads to unsteady chaotic dynamics of the ultimate banded state, as
discussed in Sec.~\ref{sec:spatioTemporal}. Coupling the dJS model to
concentration also extends the regime of linear instability (see Fig.~\ref{fig:spinodals_with_phi}), though
not sufficiently to destabilise the
high shear band in the manner just described.
For the moment, however, we assume that the high shear
branch remains stable, and next discuss steady shear-banded states in which structural variation occurs only in one spatial dimension, normal to the interface between the bands.
}\fi

\subsection{Steady Shear Bands}
\label{sec:steady}

\begin{figure}[tbp]
\centering
\includegraphics[width=6.0 in]{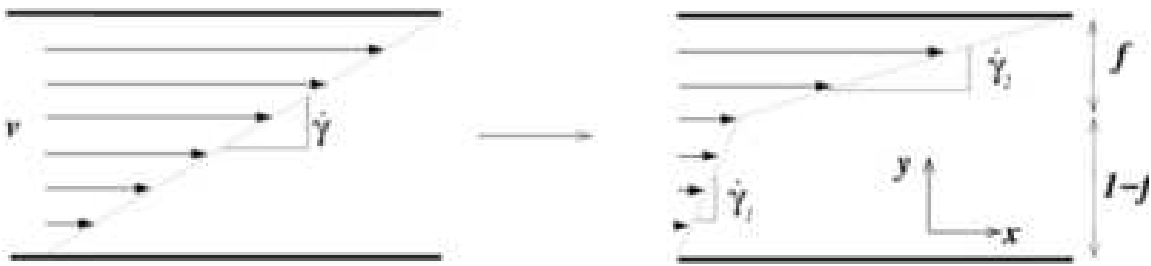}
\caption{Left: homogeneous flow. Right: shear banded flow.
\label{fig:bands}}
\end{figure}

\begin{figure}[tbh]
\centering
\includegraphics[width=3.0 in]{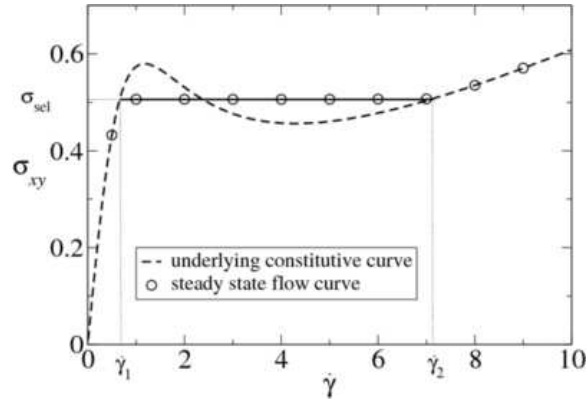}
\caption{Constitutive curve (dashed line) and steady state flow curve (solid where different) in the dJS model ($a=0.3$, $\etasol=0.05$, $G=\tau=1$). Figure reprinted with permission from Ref.~\cite{fielding2005}.
\label{fig:flowCurvea0.3}}
\end{figure}

For an applied shear rate in the regime of decreasing stress, we have
seen above that homogeneous flow is unstable. For a constitutive curve
of the shape shown in Fig.~\ref{fig:spinodals_with_phi} (or indeed
Fig.~\ref{spenley0fig} above for the reptation-reaction model, or
Fig.\ref{fig:flowCurvea0.3} below for dJS), this instability triggers
formation of two bands of shear rates $\gdot_1$ and $\gdot_2$, one on
each of the stable branches, with a flat interface between bands whose
normal is oriented in the flow-gradient direction $y$
(Fig.~\ref{fig:bands}). The relative volume fractions ($f, 1-f$) of
the bands arrange themselves to match the spatially averaged shear
rate $\gdot$ imposed on the cell as a whole. (It is this averaged
quantity, namely the ratio of the velocity difference between plates
to the width of the gap, that now appears on the abscissa of the
experimental flow curve, $\sigma(\gdot)$.) As $\gdot$ increases, the
width $f$ of the high shear band increases at the expense of the low
shear band. Force balance demands that the shear stress
$\sigmatot_{xy}$ is common to both the bands.  Assuming that the
nature of each band does not vary as their amounts change ({\em i.e.},
neglecting concentration coupling), this gives a plateau in the
observed flow curve $\sigma(\gdot)$.

\begin{figure}[tbh]
\centering
\includegraphics[width=3.5 in]{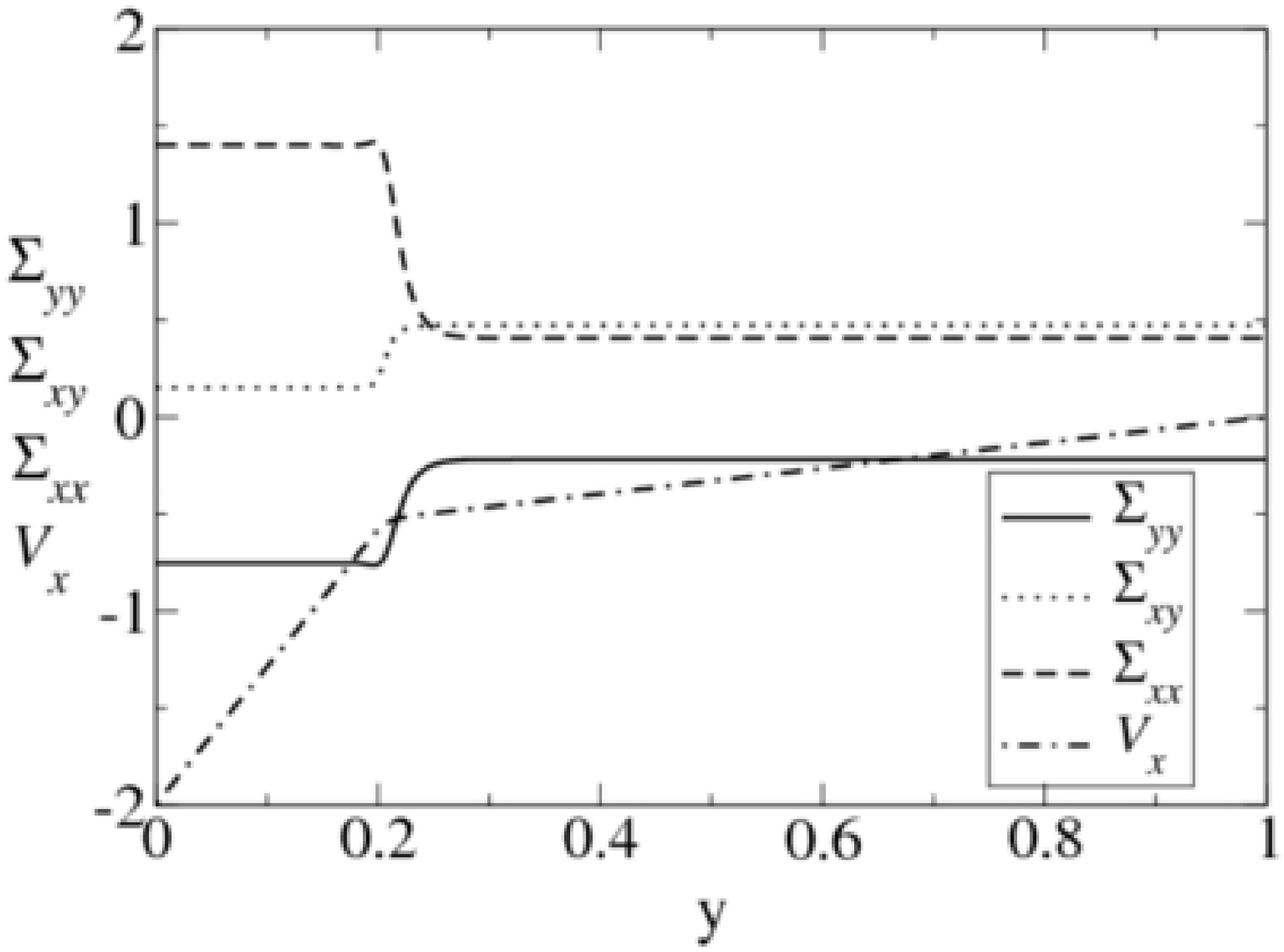}
\caption{Shear banded profile predicted by evolving the dJS equations in one spatial dimension, $y$. The imposed shear rate $\gdot=2.0$, towards the left of the stress plateau in Fig.~\ref{fig:flowCurvea0.3}. $G=\tau=1$. Figure reprinted with permission from Ref.~\cite{fielding2005}.
\label{fig:profile}}
\end{figure}

The scenario just described, and first proposed for micelles
in~\cite{spenley}, was confirmed by explicit numerical calculation
within the dJS model~\cite{fielding2005}. The resulting steady state
flow curve is shown in Fig.~\ref{fig:flowCurvea0.3}, and indeed
comprises two homogeneous branches ($\gdot<\gdot_1$ and
$\gdot>\gdot_2$) connected by a plateau across the banding regime
$\gdot_1<\gdot<\gdot_2$. A typical flow profile in the banding regime
is shown in Fig.~\ref{fig:profile}. Note the smooth variation across
the interface, which has a width $l\propto
\sqrt{\mathcal{D}\tau}$ set by the spatially non local (stress diffusion) term in (\ref{eqn:js}).  Without this term, the interface would be an
sharp discontinuity whose position is not reproducible between
different startup runs ~\cite{olmsted99a}. The nonlocal term confers a
smooth interface and a robust, reproducibly selected value
$\sigma_{\rm sel}$ of the plateau stress. The selected stress is then
the only one at which a stationary front can exist between the
bands~\cite{olmstedluball}. We note that an alternative nonlocal JS model
incorporating a higher order gradient of the deformation-rate tensor
was offered by Yuan in~\cite{Yuan99} and solved numerically in
two dimensions (the flow/flow-gradient plane). The model was shown to give a uniquely
selected stress, although the model was not fully resolved in the sense that the width of the interface
between the bands depended on the fineness of the numerical mesh.

\begin{figure}[tbp]
\centering
\includegraphics[width=3.5 in]{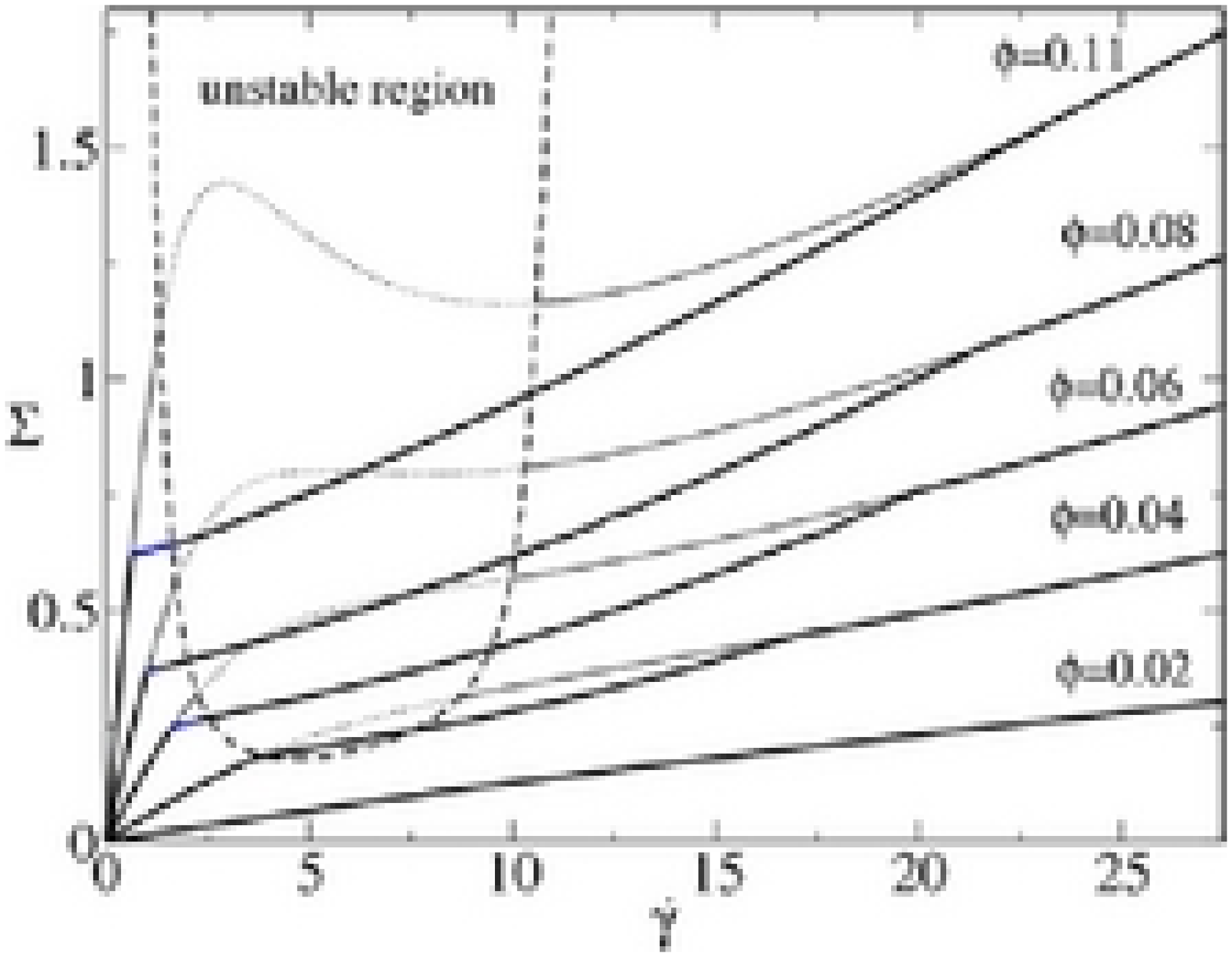}
\caption{Steady state flow curves in the dJS$\phi$ model for different values of the average concentration $\phi$. Concentration coupling now confers an upward slope in the banding regime. Figure reprinted with permission from Ref.~\cite{olmstedramp}.
\label{fig:fc_alpha1p0e-3_gamma0p0_new}}
\end{figure}
\begin{figure}[tbp]
\centering
\includegraphics[width=3.0 in]{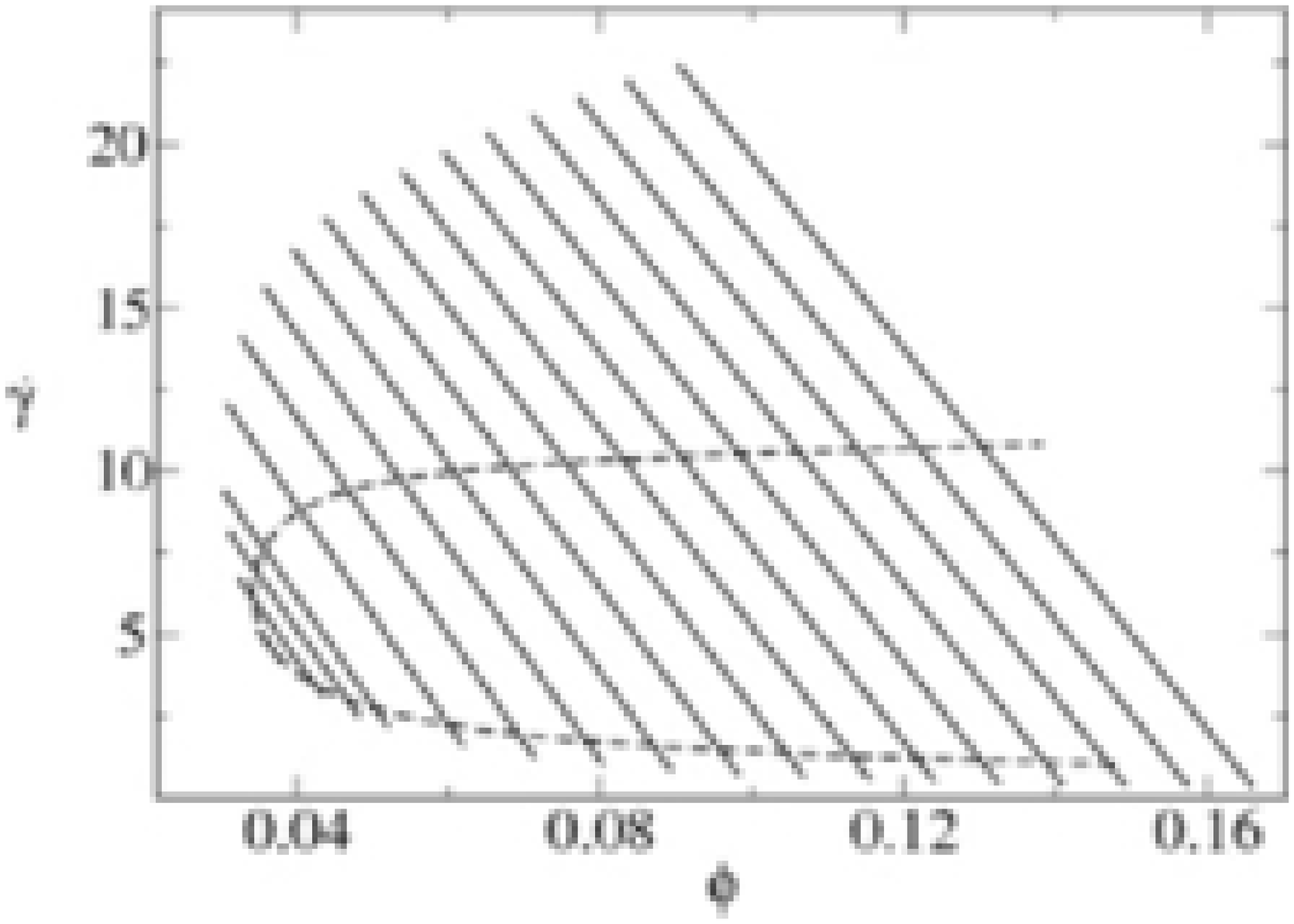}
\caption{Nonequilibrium phase diagram: tie lines show the two-phase (banding) regime. Different tie lines correspond to different values of the stress. Dashed line shows the spinodal limit of linear stability of homogeneous flow. Figure reprinted with permission from Ref.~\cite{olmstedramp}.
\label{fig:gdotVsPhi_alpha1p0e-3_gamma0p0}}
\end{figure}

In recent years, the experimental evidence for shear banding in
wormlike micelles has become
overwhelming~\cite{pine,nmr,nmr2,nmr3,birefringence,berretband}. Reports of
kinks, plateaus and non-monotonicity in the flow curve are now
widespread, while spatially resolved NMR~\cite{nmr,nmr2,nmr3} and
birefringence~\cite{birefringence} data provide direct evidence for
banding in both shear rate and microstructure.
As noted previously, in some cases, the associated stress plateau is
not perfectly flat, but ramps upwards from left to right. In a
cylindrical Couette geometry, there will always be a small slope
caused by a slight stress gradient (absent in the planar case of
Fig.~\ref{fig:bands}) causing the high shear band always to reside
next to the inner cylinder. As this band expands outwards with
increasing applied shear rate into regions of lower stress, the
overall torque must increase to ensure that the interface between the
bands stays at the selected stress $\sigma_{\rm sel}$.
An alternative explanation of the upward slope, independent of cell
geometry, is coupling between flow and concentration, as first
discussed in Ref.~\cite{schmitt95} (see Sec.~\ref{sec:JS} above). If a concentration difference
exists between the bands, the properties of each band must change as
the applied shear rate is tracked through the coexistence regime,
because material is redistributed between them as the high-shear band
expands to fill the gap. This was confirmed by one of us in
Ref.~\cite{olmstedramp} by a numerical study of the dJS$\phi$ model. The
stress now slopes upwards from left to right through the shear banding
regime (Fig.~\ref{fig:fc_alpha1p0e-3_gamma0p0_new}). The concentration
and shear rate in each phase now define a family of tie lines, one for
each banded state, giving the nonequilibrium phase diagram shown in
Fig.~\ref{fig:gdotVsPhi_alpha1p0e-3_gamma0p0}.

\begin{figure}[tbp]
\centering
\includegraphics[width=11cm]{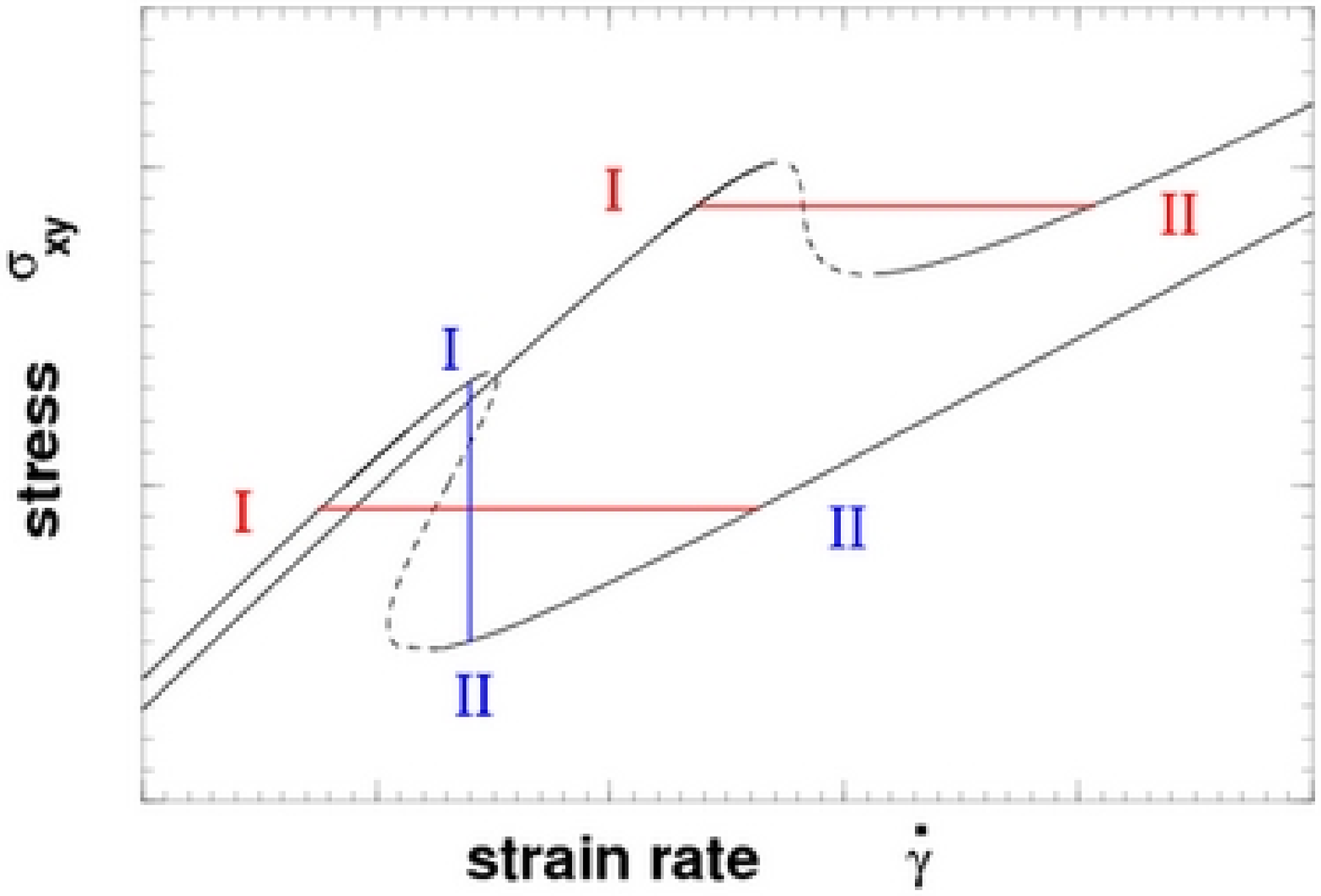}
\caption{Stress--strain-rate curves for the Doi model with
different excluded volume parameters. The dashed line segments are
unstable. The straight lines indicate possible coexistence between
branches I and II under conditions of common stress (horizontal lines)
or strain rate (vertical line). Figure reprinted with permission from Ref.~\cite{OlmsLu99}.
\label{fig:rods}}
\end{figure}
\begin{figure}[bp]
\centering
\includegraphics[width=12cm]{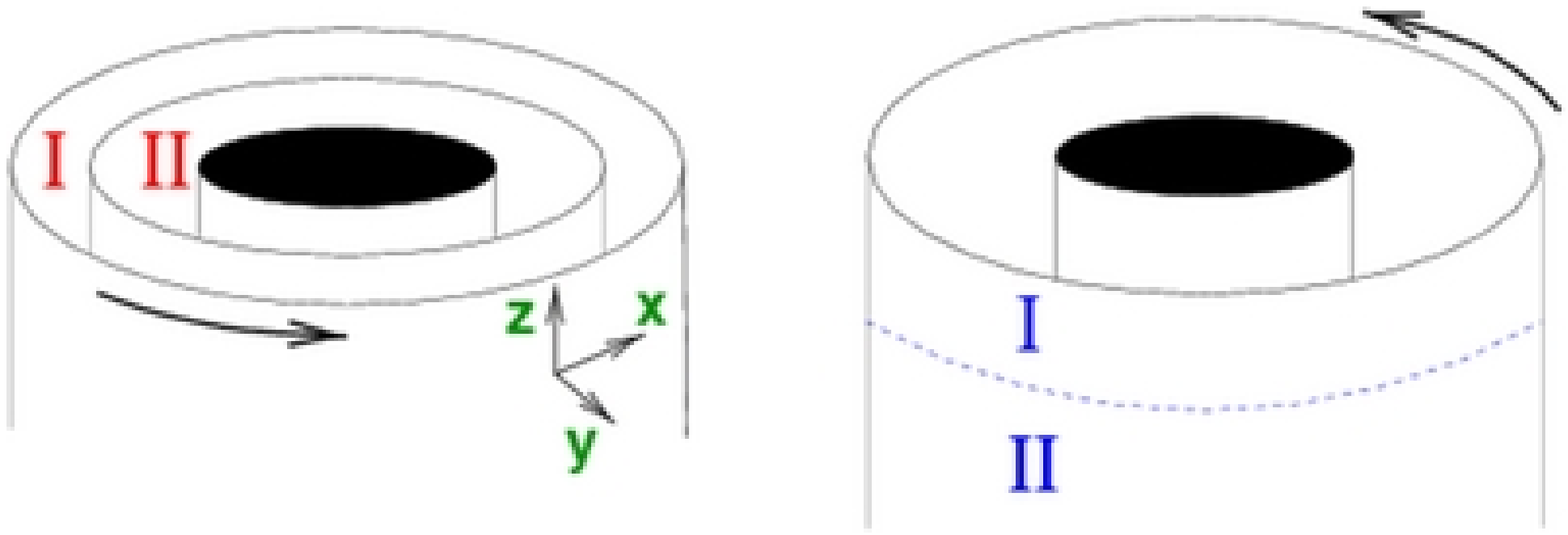}
\caption{Geometries for phase separation at common stress
(left) or strain-rate (right) in a Couette rheometer. At a common
stress (left) phases I and II have different strain rates, while at a
common strain rate (right) they have different stresses. Here $z$ is the
vorticity axis, $x$ is the flow direction, and $y$ is the flow
gradient axis. Figure reprinted with permission from Ref.~\cite{OlmsLu99}.
\label{fig:couette}}
\end{figure}

Beyond the Johnson-Segalman model, shear banding has also been studied
in the Doi model of shear thinning rigid
rods~\cite{doi81,doikuzuu83,OlmsLu99}.
In
this case, the relevant microstructural variables in
(\ref{eqn:basicStress}) are the nematic order parameter $Q_{ij}$ and
the concentration of rods $\phi$.  While this approach obviously
ignores any effects of micellar flexibility, it takes a first step to
incorporating orientational ordering, ignored by dJS$\phi$ and likely
to be important in concentrated micellar solutions close to an
underlying isotropic-nematic transition.
Depending on parameter values, the
constitutive curve for homogeneous flow can now adopt either of the
shapes in Fig.~\ref{fig:rods}. In both cases, the two stable branches
correspond to a low-shear isotropic band (branch I) and a flow-induced
paranematic phase (branch II).  For concentrations inside the
zero-shear biphasic regime, branch II touches down to the origin to
form the zero-shear nematic phase.  Coupling between concentration and
flow arises because more strongly aligned rods in the high shear band
can pack more closely together, giving a higher concentration. In
contrast, the Helfand-Fredrickson coupling in the dJS$\phi$ model
gives a less concentrated high shear band; recall
Fig.~\ref{fig:gdotVsPhi_alpha1p0e-3_gamma0p0}.

The shape of the lower constitutive curve in Fig.~\ref{fig:rods} opens
up a new possibility, shown by the vertical line: that shear bands can
coexist at a common shear rate with a different value of the stress in
each band. In a Couette cell, this corresponds to bands stacked
in the vorticity direction (Fig.~\ref{fig:couette}, right) and is called ``vorticity banding''. In
contrast, the dJS model supports only common-stress banding
(horizontal line in the ($\gdot,\sigma_{xy}$) plane), with the normal
to the banding interface in the flow gradient direction; this is called ``gradient banding''. The latter gives
concentric bands in Couette flow (Fig.~\ref{fig:couette}, left),
reducing to the arrangement of Fig.~\ref{fig:bands} in the limit
of planar shear.

In this section, we have explored steady shear-banded states within models whose underlying constitutive
curve comprises two stable branches separated by an unstable region
of negative slope. In Secs.~\ref{sec:temporal}
and~\ref{sec:spatioTemporal} below, we discuss the effects of (i)
higher dimensionality and (ii) more exotic constitutive curves. As we
will see, either can give rise to unsteady, even chaotic, shear
bands. First, however, we introduce some models of shear thickening.

\subsection{Simple Models for Shear Thickening}
\label{sec:thickening}

\begin{figure}[tbp]
\centering
\includegraphics[width=11cm]{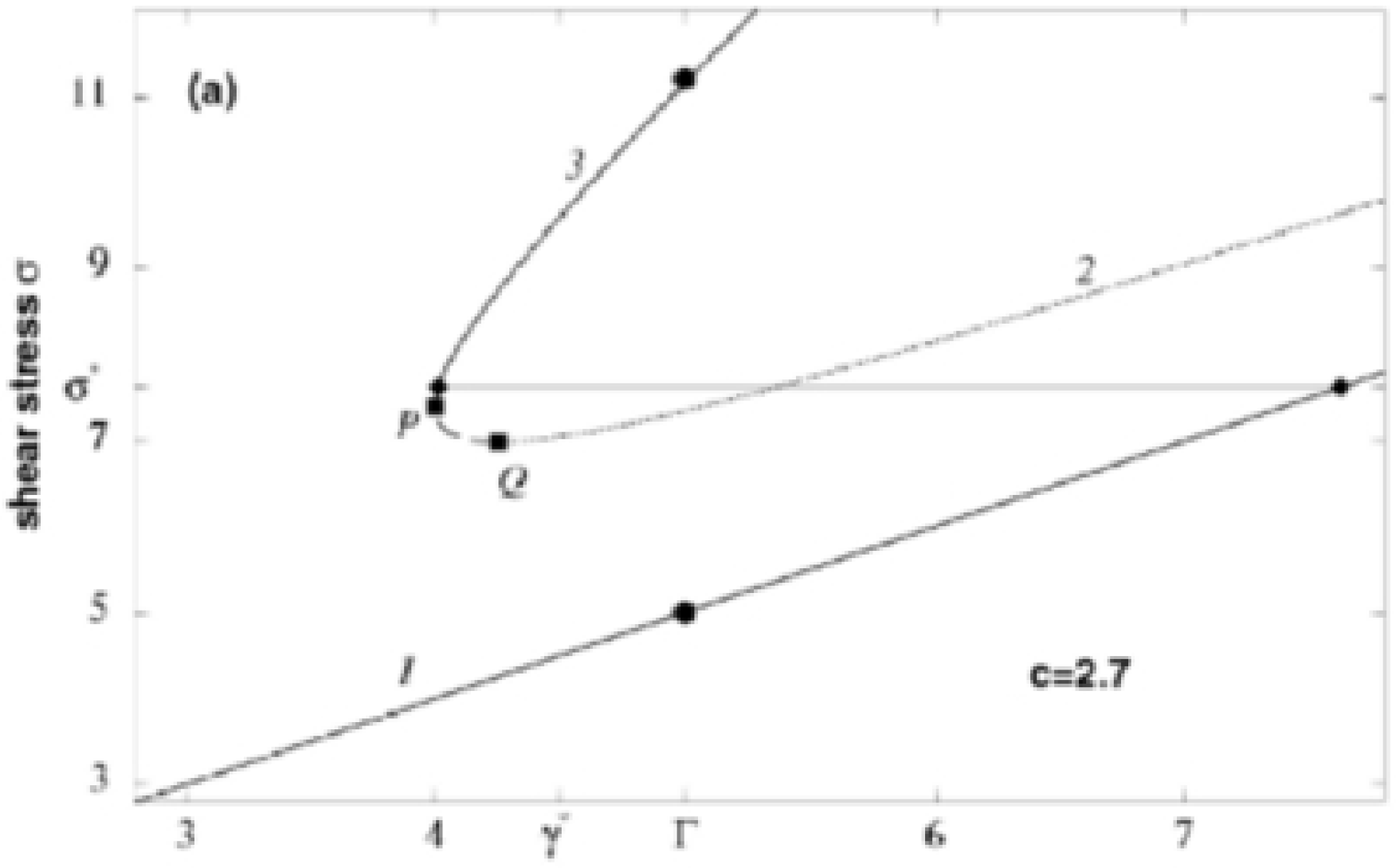}
\caption{Shear thickening constitutive curve. Figure reprinted with permission from Ref.~\cite{goveas1}.
\label{fig:goveas}}
\end{figure}
\begin{figure}[tbp]
\centering
\includegraphics[width=9cm]{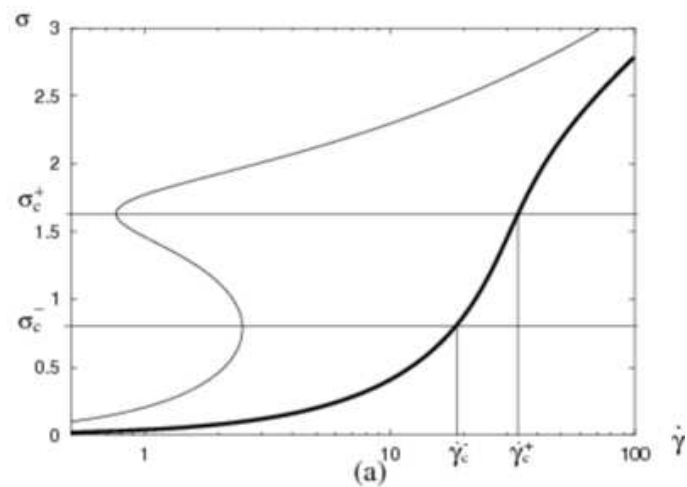}
\caption{Bare (light line) and final (heavy) constitutive curve in the CHA model. The region of instability
$\sigma_c^-<\sigma<\sigma_c^+$ is shown.
Figure adapted from 
Ref.~\cite{cates2002}.
\label{fig:CHAconstit}}
\end{figure}

As discussed in Sec.~\ref{thickening} above, a window of shear
thickening is seen in many micellar solutions, for volume fractions
around the onset of
viscoelasticity~\cite{oelschlagermemory,hoffmannthicken,matthys,oda,pinethick,berretmemory}. After
a period of prolonged shearing, an initially inviscid fluid undergoes
a transition to a notably more viscous state or even a long-lived gel,
with shear banding often implicated in its formation~\cite{pinethick}.

There is no consensus on the microscopic origin of this phenomenon
(see Sec.~\ref{thickening}).  Despite this, several macroscopic
features can be captured within a phenomenological approach~\cite{goveas1,goveas2}
that couples flow to a generalised gelation transition by allowing a
mixture of two species A (sol; concentration $\phi_A=1-\phi$) and B
(gel; concentration $\phi_B=\phi$) to slowly inter-convert under the
influence of shear:
\be
\label{eqn:gelation}
\partial_t\phi=R(\phi,\gdot)+\mathcal{D}\nabla^2\phi
\ee
\be
\label{eqn:R}
R(\phi,\gdot)\equiv|\gdot|(1-\phi)\phi^2-k\phi
\ee
In the absence of shear, $\phi=\phi_B$ relaxes to zero, leaving pure
A. In this way, B is identified as the shear-induced phase (gel).
The model of \cite{goveas1} ignores normal stresses and considers an additive shear
stress $\sigma=\sigma_A+\sigma_B$, with both contributions presumed
Newtonian
\be
\sigma=\left[(1-\phi)\eta_A+\phi\eta_B\right]\gdot
\ee
Hence, all other variables are deemed fast in comparison to the
structural variable $\phi$, an assumption that only strictly becomes
valid in the limit of vanishingly slow inter-conversion. Depending on
the ratio $\eta_B/\eta_A\equiv c$, the model can actually capture
either shear thinning or shear thickening.  Here we focus on
thickening, $c>1$, for which the underlying constitutive curve is
shown in Fig.~\ref{fig:goveas}.

As discussed in Ref.~\cite{goveas1}, for an applied shear rate
$\gdot=\Gamma$ in the non-monotonic regime, the system can in
principle choose between homogeneous states on branches 1 or 3
(circles in Fig.~\ref{fig:goveas}) or it can gradient-band between
these branches at a selected stress $\sigma^*$ (horizontal line). But
when the model equations are evolved numerically at imposed shear
rate, gradient banding is not seen. The system instead always chooses
homogeneous flow: on branch 1 below a critical shear rate
$\gdot=\gdot^*$ (Fig.~\ref{fig:goveas}), and on branch 3 for
$\gdot>\gdot^*$. The sharp vertical jump between branches 1 and 3 at
$\gdot^*$ leaves a range of stresses that is unattainable under
conditions of controlled shear rate. This range is instead accessed by
controlling the stress, and marks a regime of vorticity banding.

A different model of shear thickening was devised by Cates, Head and
Ajdari (CHA) in Ref.~\cite{cates2002}. This has an unspecified,
slowly evolving structural variable, but now with explicit (though
still relatively fast) dynamics for the stress. The distinguishing
feature of this model is that, while the ``instantaneous''
constitutive curve following fast stress equilibration at fixed
structure is non-monotonic (Fig.~\ref{fig:CHAconstit}), the slow
structural evolution restores monotonicity in the true long term
constitutive curve. (In contrast, in Fig.~\ref{fig:goveas} the
non-monotonicity results directly from the structural evolution.) The
short-term tendency to form shear bands is thus frustrated by the
long-term structural evolution, leading to oscillatory or chaotic
dynamics. We return to this model in Sec.~\ref{sec:temporal} below.

It appears that the microscopic mechanisms involved in shear banding
models differ from case to case, and often remain poorly
understood. Nonetheless, the macroscopic phenomenology has many
universal features, including kinks, plateaus or non-monotonicity in
the flow curve. In fact, there are just four fundamental types of
flow curve in shear banding systems: thinning {\em vs} thickening, and
gradient {\em vs} vorticity banding. Ref.~\cite{olmsted99b} collects
these curves, discusses their dependence on concentration, and
explains how a non-equilibrium phase diagram can be reconstructed from
a family of such curves.

\subsection{Other Related Models}
\label{sec:otherModels}

Here we discuss some other approaches to modeling the rheology of
wormlike micelles. In an article this length we have no room to
discuss all of the many nonlinear rheological models that were
developed without micellar systems specifically in mind. Of these,
alongside the JS-type models described extensively above, we mention
only the Giesekus model~\cite{giesekus1982} of shear thinning
polymeric fluids, based on the concept of a deformation dependent
tensorial mobility. This was first applied to micelles in
Ref.~\cite{giesekusadv}; it shows a plateau in the constitutive curve,
but lacks the non-monotonicity required to give a true banding
instability. Any ability to fit the measured flow curve for
shear-banding micelles is therefore somewhat fortuitous.

Among micelle-inspired approaches, Manero and
co-workers~\cite{Manero99,Manero00,GMB} developed a simple model
coupling the evolution of the stress to that of the underlying fluid
structure. For simplicity, they represented the structure by a single
``fluidity'' parameter $\varphi\equiv 1/\eta$, assumed to be the
reciprocal of the fluid viscosity. (Here $\varphi$ is distinct from the
micellar concentration, for which the symbol $\phi$ was used above.) 
According to Ref.~\cite{Manero99}, possible microscopic
interpretations of $1/\varphi$ include the number of bonds, links or
entanglements in a network.
No distinction is made between the viscoelastic and solvent stresses:
instead, the {\em total} stress is directly assumed to obey an
upper convected Maxwell constitutive equation
\be
\label{eqn:man_sigma}
\dot\sigma_{ij} + \frac{1}{G_0\varphi} 
(K_{il}\sigma_{lj} +\sigma_{il}K_{jl})
= \frac{D_{ij}}{\varphi}
\ee
in which
$D_{ij}$ is the
symmetrised shear rate tensor and $G_0$ is the plateau modulus.  The
fluidity is assumed to evolve as
\be
\label{eqn:man_phi}
\frac{d\varphi}{dt} = \frac{\varphi_0-\varphi}{\lambda} + k[\gdot](\varphi_{\infty}-\varphi)\sigma_{ij}D_{ji}
\ee
in which $\varphi_0$ and $\varphi_{\infty}$ are the steady state
fluidities in the limits of zero and infinite shear rates;  $\lambda$
is a structural relaxation time for the build up of structure in zero
shear, while $k[\gdot]$ is a function allowing for the breakdown of
structure under shear.  For $\varphi_{\infty}>\varphi_0$, the model captures
shear thinning. (Predictions for shear thickening,
$\varphi_{\infty}<\varphi_0$, are not discussed here.)

In the simplest case $k$ is assumed constant, independent of shear
rate. Here the constitutive curve is always monotonic, ranging from
Newtonian for $\varphi_\infty=\varphi_0$ through simple shear thinning for
$\varphi_\infty>\varphi_0>0$ to yield stress behaviour for
$\varphi_\infty>\varphi_0=0$. This version of the model therefore lacks a
shear banding instability. Ref.~\cite{Manero99} discusses its
predictions for various rheological tests in homogeneous shear,
including step strain, step stress, shear-rate jumps, and stress
sweeps. In particular, significant hysteresis is predicted in upward
followed by downward stress sweeps.
To allow for shear banding, in Ref.~\cite{Manero00} the rate of
structure breakdown is assigned an additional dependence on shear rate
by taking
\be 
\label{eqn:man_k}
k[\gdot]=k_0(1+\mu_1|\gdot|)
\ee
For large enough values of $\mu_1$ and $\varphi_\infty/\varphi_0$, the
homogeneous constitutive curve is non-monotonic, assuming the basic
form shown for the Johnson-Segalman model in 
Fig.~\ref{fig:flowCurvea0.3}.  The model thus captures a shear banding
instability. 
\if{While appealing in its simplicity, this approach should
be treated with some caution since the function $k[\gdot]$ lacks {\bf what??} the
proper tensorial invariances required in a rheological constitutive
equation.}\fi
The predictions of Eqns.~(\ref{eqn:man_sigma},~\ref{eqn:man_phi},~\ref{eqn:man_k}) for the steady shear banded state are discussed in
Ref.~\cite{Manero02}. In this case, nonlocal terms (recall
Sec.~\ref{sec:steady}) are not invoked to capture a uniquely selected
banding stress. Instead, the authors appeal to extended irreversible
thermodynamics to posit the existence of a Gibbs potential that can
be used to calculate state selection. However, no deeper justification was
offered for this assumption. Shear rate profiles for banded states were
not detailed in Ref.~\cite{Manero02}, but in the absence of nonlocal
terms, these presumably are discontinuous across the
band interface, in contrast to the models discussed in Sec.~\ref{sec:steady}.

Goveas and Pine~\cite{goveas1999a} developed a simple model of
shear-thickening in a wormlike micellar solution. Above a critical
stress, the solution is assumed to undergo a reaction that produces an
insoluble gel phase. Competing against this is the destruction of gel
by, for example, peeling at the interface between gel and
solution. The balance of these determines the relative volume
fractions of gel and sol, and therefore the position of the interface
between the two. The following assumptions were made: (i) a constant
viscosity for the Newtonian sol phase, (ii) no flow in the gel phase,
(iii) a constant rate of gel destruction, and (iv) a rate of gel
production proportional to the product of the stress and the
surfactant concentration in the sol phase. Given these, the model
predicts a re-entrant region in the flow curve that is only accessible
under controlled stress conditions. For controlled shear rate, the
measured flow curve is predicted to be discontinuous. These results
agree qualitatively with the experiments of Pine on shear thickening
micelles~\cite{pinethick}.

In Ref.~\cite{berretband}, Porte, Berret and Harden 
argued for a description of shear-banding
involving a modified thermodynamic transition as
distinct from a mechanical instability.
\if{
claimed the
theoretical description of banding in terms of a ``mechanical
instability'' to be inconsistent with (i) the robustness of the stress
plateau, and (ii) the slow sigmoidal kinetics seen in the regime where
homogeneous flow is metastable with respect to band
formation~\cite{grand}. Instead, they argued these features to be
reminiscent of a thermodynamically driven transition.}
\fi
On this basis,
they developed a thermodynamically inspired model based on a potential
energy $F_s(\gamma_s)$, where $\gamma_s$ is an internal structural
variable related to the local recoverable elastic strain. The local stress is assumed to be directly set by
this strain as
\be
\sigma=\frac{dF_s}{d\gamma_s}
\ee
with the following dynamics assumed
\be
\label{eqn:harden}
\frac{d\gamma_s}{dt}=-\mu(\gamma_s)\frac{dF_s}{d\gamma_s}+\gdot
\ee
in which $\mu$ is a mobility. Here the first term on the right is a
dissipative term; the second is a driving (reactive) term.  Solving
Eqn.~\ref{eqn:harden} in homogeneous steady shear gives 
\be
\sigma=\frac{1}{\mu(\gamma_s)}\gdot
\ee
in which $1/\mu(\gamma_s)$ is the non-Newtonian viscosity. As shown in
Ref.~\cite{berretband}, the homogeneous constitutive curve
$\sigma(\gdot)$ is non-monotonic, leading to banding, whenever the
relation $\sigma(\gamma_s)$ is itself non-monotonic.  Because the
underlying nonlinearity stems from the dissipative term
$-\mu(\gamma_s)dF_s/d\gamma_s$, this approach effectively considers
banding as a flow-induced parameter shift for an underlying thermodynamic transition,
rather than the result of an intrinsically rheological instability.
While this model may indeed be relevant to micellar systems close to
an a true thermodynamic transition (for instance to a state of nematic order)~\cite{berretband}, the suggestion in Ref.~\cite{berretband} that a robust stress plateau and sigmoidal kinetics \cite{grand}
are incompatible with a purely mechanical instability was disproved by subsequent work on 
the dJS model
\if{
 can be written in the form $\sigma_{ij}=GW_{ij}$ with
\be
\label{eqn:JScompare}
\stackrel{\mbox{\scriptsize{$\triangle$}}}W_{ij} = -\frac{1}{\tau}W_{ij} + 2D_{ij},
\ee
in which $W_{ij}$ is a local molecular deformation, analogous to the
recoverable local strain $\gamma_s$. (The diffusive term has been
omitted to highlight the comparison with Eqn.~\ref{eqn:harden}.) 
$\stackrel{\mbox{\scriptsize{$\triangle$}}}W_{ij}$ is a reactive term,
incorporating the LHS and the first two bracketed terms on the RHS of
Eqn.~\ref{eqn:js}. In this case, the dissipative term $-W_{ij}/\tau$ is
purely linear. Instead, the nonlinearity that triggers the banding
instability comes from the nonlinear reactive terms on the LHS
of~\ref{eqn:JScompare}. Despite this, the non local version of the JS
model is clearly capable of producing a robust stress
plateau}\fi
~\cite{olmsted99a,Yuan02}.  

More generally, although a distinction is frequently made
between `thermodynamic' and `mechanical'
instabilities, it is clearly possible to construct
models including ingredients of both. This was done in
Ref.~\cite{olmstedramp}, by coupling the JS model to a nearby demixing
instability. Here, the matrix that determines the linear stability of
homogeneous flow contains a concentration subspace (which can be
subject to a zero-shear demixing transition); and a `mechanical'
subspace (which is subject to a banding instability even without
concentration coupling). Furthermore, the coupling between these
subspaces mixes the two instabilities, and a smooth crossover can be
achieved between the two by varying the model parameters. In the same
way, the distinction between `mechanical' banding instabilities and
shear-induced isotropic-nematic instabilities is likely to be less
clear cut than was once thought.

Dhont introduced a phenomenological model of shear banding
in Ref.~\cite{dhont1999a}, as follows.  At finite Reynolds number in planar
shear, with spatial variations only in the flow-gradient direction
$y$, the Navier-Stokes equations is
\be
\label{eqn:DhontNavier}
\rho\frac{\partial v(y,t)}{\partial t}=\frac{\partial\sigma(y,t)}{\partial y}
\ee
where $v$ is the fluid velocity in the flow direction and $\rho$ is
the mass density. The total shear stress $\sigma$ is assume to respond
adiabatically, such that
\be
\label{eqn:DhontStress}
\sigma(y,t) = \eta(\gdot(y,t))\gdot(y,t) - \kappa(\gdot(y,t))\frac{\partial^2\gdot(y,t)}{\partial y^2}
\ee
The first term on the right describes the nonlinear homogeneous
constitutive relation between stress and shear rate. The second term
encodes the spatial gradients needed to give unique state selection in
the banding regime. Within this term, $\kappa$, called the ``shear curvature
viscosity'' in Ref.~\cite{dhont1999a}, is further assumed to obey
$\lim_{\gdot\to\infty}\kappa(\gdot)=0
$. In steady state, the total shear stress must be a constant,
independent of $y$ (see Eqn.~\ref{eqn:DhontNavier}). Denoting this by
$\sigma_{\rm stat}$, Dhont showed that the banded state obeys
\be
\label{eqn:maxwellAttempt}
\int_{\gdot_-}^{\gdot_+}d\gdot\,{[\sigma_{\rm h}(\gdot)-\sigma_{\rm stat}]}/{\kappa(\gdot)}=0
\ee
where $\sigma_{\rm h}(\gdot)$ is the part of the total stress obtained
by discarding the gradient term in Eqn.~(\ref{eqn:DhontStress}).  If
$\kappa$ were constant,~(\ref{eqn:maxwellAttempt}) would reduce to the
equal area (Maxwell) construction that determines equilibrium phase
coexistence. Note, however, that this is somewhat special property of a
scalar model in which normal stresses are neglected. (In a tensorial
model, even with constant $\kappa$, no equal area construction would be possible in general.)

Dhont then performed a linear stability analysis for fluctuations
about a homogeneous initial state of shear rate $\gdot_0$, to study
the onset of banding, as in Sec.~\ref{sec:onset} above. For a
fluctuation of wavelength $k_n$, he found a growth rate
\be
\omega_n = -\left[\frac{d\sigma_{\rm h}(\gdot_0)}{d\gdot_0}+\kappa(\gdot_0)k_n^2\right]\frac{k_n^2}{\rho}
\ee
As in Eqn.~(\ref{eqn:growthRate}) above, instability is predicted when
the slope of the homogeneous constitutive curve is negative. In
contrast to~(\ref{eqn:growthRate}), however, it predicts a maximum
growth rate at a non-zero wavevector
\be
k_{\rm max}=\sqrt{-\frac{d\sigma_{\rm h}(\gdot_0)/d\gdot_0}{2\kappa(\gdot_0)}}
\ee
due to a competition between diffusion of (conserved) momentum, governed by~(\ref{eqn:DhontNavier}), and the
gradient terms in~(\ref{eqn:DhontStress}). This contrasts with~(\ref{eqn:growthRate}), where the
competition is between the rate at which the  (non-conserved) stress evolves, and the gradient terms. For realistic micellar parameters
and rheometer gap sizes, the latter appears to us more
appropriate.

In a separate study, Dhont and Briels~\cite{dhont2003b} developed an expression for the
stress tensor in inhomogeneous suspensions of rigid rods, in terms of
the flow velocity and the probability density function for the
position and orientation of a rod. By explicitly allowing for large
spatial gradients in the shear rate, concentration, and orientational
order parameter, this approach could potentially be applied to
shear-banding systems.

\subsection{Temporal Instability}
\label{sec:temporal}

The studies discussed so far capture the basic tendency of wormlike
micelles to undergo a transition to shear banded flow. However most of them
fail to address recent reports that the constitutive response to
steady mechanical driving is intrinsically unsteady in some
regimes. In such cases, the stress response to a constant applied
strain rate (or vice versa) does not settle to a constant
value. Instead it shows sustained periodic
oscillations~\cite{hu1998b,wheeler1998,fischer2000,fischer2002,herle2005}
or erratic behaviour suggestive of low-dimensional
chaos~\cite{nmr3,bandyopadhyay2000,bandyopadhyay2001,lopez-Gonzalez2004}. Such
{\em long-time} unsteadiness is distinct from the early-time
instability discussed in Sec.~\ref{sec:onset}; the latter merely
provides the initial trigger for banding, of some sort, to occur.

Hydrodynamic instabilities (Taylor-Couette, Rayleigh-Benard,
turbulence, {\em etc.}) have long been studied in simple
liquids~\cite{faber}, where they stem from the nonlinear inertial term
($\rho v_i\nabla_i v_j$) in the Navier-Stokes equation. For the flows
of interest to us, however, the Reynolds number is virtually zero,
rendering this term negligible. The observed complexity must therefore
instead arise in the constitutive nonlinearity inherent to the
rheology of the micelles themselves, {\em e.g.}, through the coupling
between microstructure and flow. The term ``rheological
chaos''~\cite{cates2002} (or ``rheochaos'') has been coined to
describe this behaviour. Irregular signals have also been reported in
other complex fluids, include so-called `onion' surfactant
phases~\cite{SalColRou02,SalManCol03b} and concentrated
colloids~\cite{lootens2003}. In many cases, evolution of the
microstructure in concert with the rheological signal has been
explicitly observed via birefringence
imaging~\cite{wheeler1998,HilVla02}, light
scattering~\cite{SalColRou02}, or spatially resolved
NMR~\cite{lopez-Gonzalez2004}.

A crucial question is whether these instabilities are spatiotemporal
or purely temporal in character. In wormlike micelles they most often
arise close to the banding regime, suggesting the spatiotemporal
evolution of a heterogeneous ({\em e.g.}, banded) state. Indeed, early
optical experiments on wormlike micelles showed a temporally
oscillating state comprising spatially alternating turbid and clear
bands~\cite{wheeler1998,fischer2000}. More recent advances allowing
spatially and temporally resolved measurements of velocity profiles
have unambiguously revealed fluctuating shear bands in both wormlike
micelles~\cite{lopez-Gonzalez2004} and multilamellar onion
phases~\cite{SalManCol03b,ManSalCol04}.  However, spatial observations
have been made in just a few of these cases, so that in others, the
question remains open. For example, a nearby banding instability could
feasibly play a role in triggering temporal rheochaos, but with
banding itself narrowly averted such that the system stays
homogeneous, as discussed in Ref.~\cite{cates2002}. For systems close
to the nematic transition, another possibility is the purely temporal
director chaos captured theoretically in models of
nemato-dynamics~\cite{GroKeuCreMaf01,RieKroHes02}. In this section, we
therefore discuss purely temporal instability in spatially homogeneous
models, before proceeding to the full spatio-temporal case in
Sec.~\ref{sec:spatioTemporal}. We address thinning and thickening
systems in turn.

\begin{figure}[tbp]
\centering
\includegraphics[width=8cm]{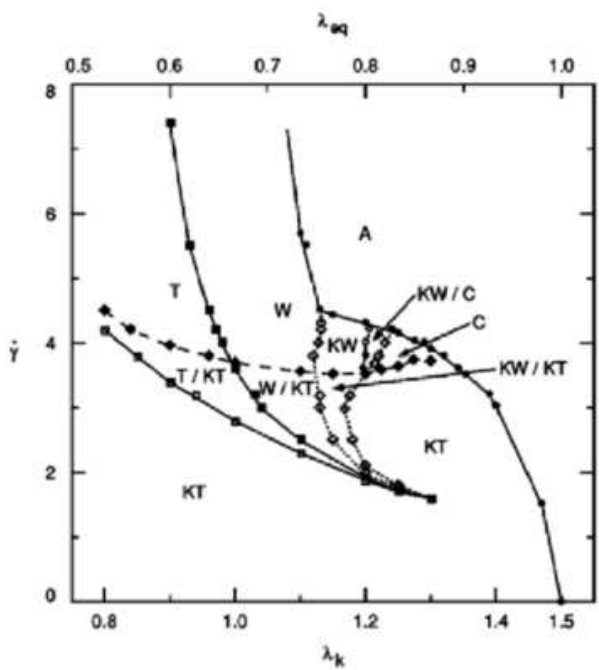}
\caption{Dynamical phases of steady and transient states at the I-N coexistence temperature. The solid line is the border between the in-plane orbits tumbling (T), wagging (W) and aligning (A); the dashed and  dotted lines delimit the regions where the out-of-plane orbits kayaking-tumbling (KT) and kayaking-wagging (KW), respectively, exist. Here $\gdot$, $\lambda_{\rm eq}$ and $\lambda_{\rm k}$ denote dimensionless shear rate and tumbling parameters of the Erickson-Leslie theory of nematic hydrodynamics, with $\lambda_{\rm k}=\lambda_{\rm eq}a_{\rm eq}$. Figure reprinted with permission from Ref.~\cite{RieKroHes02b}.
\label{fig:hess}}
\end{figure}
\begin{figure}[tbp]
\centering
\includegraphics[width=8cm]{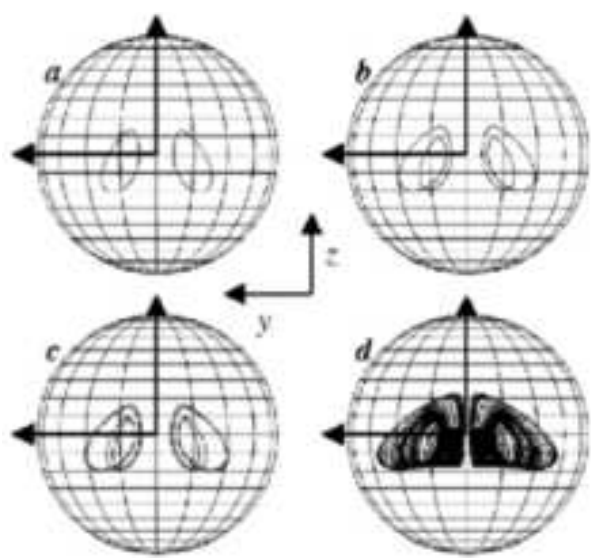}
\caption{Trajectories of the eigenvector corresponding to the largest eigenvalue of $u_iu_j$, after the transient. Orbits are plotted over the unit plane. (a) - (d) correspond to increasing values of the shear rate. Figure reprinted with permission from Ref.~\cite{GroKeuCreMaf01}.
\label{fig:grosso}}
\end{figure}

\subsubsection{Shear Thinning: Rigid Rods}

Models of rigid rods have been widely studied in the context of liquid
crystalline polymers. They capture an isotropic to nematic (I-N)
transition, and make predictions for director dynamics in the nematic
phase under shear. As noted in Sec.~\ref{sec:steady} above, such models
obviously ignore any effects of micellar flexibility or scission; but they do take a first step to incorporating
orientational ordering, relevant for some micellar systems that are close to the nematic
transition. The interplay of shear banding with the I-N
transition was studied in detail (for rigid rods) in
Refs.~\cite{OlmsLu99}; a recent review can be found in Ref.~\cite{DasGel}.

The studies in question, following Hess~\cite{hess76} and
Doi~\cite{doi81}, consider a population of rods, with orientation
vectors $u_i$ chosen from a distribution $\psi(u_i,t)$. This is
assumed spatially homogeneous. Taking account of macroscopic flow,
excluded volume effects and thermal agitation, the evolution of $\psi$
is specified via a Fokker-Plank equation. To solve this numerically,
one must first project it onto a finite number of degrees of
freedom. One method is to expand $\psi$ in a truncated set of
spherical harmonics, giving a set of coupled ordinary differential
equations for the time-dependent expansion
coefficients~\cite{larson1990e}. An alternative is to construct the
second order orientation tensor $Q_{ij}\propto \langle u_i
u_j\rangle_\psi$, and project the dynamics onto it via a closure
approximation~\cite{feng1998} to get an evolution equation for $Q_{ij}(t)$.

The resulting numerical predictions for director dynamics in the
nematic phase under shear can be briefly summarised as follows.
Studies that confine the director to lie in the flow/flow-gradient
plane predict a sequence of transitions from ``tumbling'' through
``wagging'' to ``flow-aligning'' with increasing shear
rate~\cite{larson1990e}, for suitable values of a ``tumbling
parameter'' $\lambda$. (For other values of $\lambda$, flow-alignment
occurs at all shear rates.) In the tumbling and wagging regimes the
director executes periodic motion in the flow/flow-gradient
plane. Studies generalised to allow out-of-plane director
components~\cite{LarsOtti91} predict a richer dynamics, including new
periodic regimes of ``kayaking-tumbling'', ``kayaking-wagging'', as
well as chaos characterised by a positive Lyapunov exponent and a
fractal correlation dimension~\cite{GroKeuCreMaf01,RieKroHes02b}. Both
intermittency and period-doubling routes to chaos are seen.  The
various regimes, calculated via $Q_{ij}$~\cite{RieKroHes02b} are
summarised in the phase diagram of Fig.~\ref{fig:hess}. Director
trajectories, calculated using spherical
harmonics~\cite{GroKeuCreMaf01}, are shown in Fig.~\ref{fig:grosso}.

\subsubsection{Shear Thickening: CHA Model}

As noted in Section \ref{sec:thickening} above,the CHA model of~\cite{cates2002} couples the viscoelastic shear
stress to a slowly evolving structural variable. The dynamics is
defined as follows:
\be
\label{eqn:CHA}
\dot{\sigma}=\gdot -R(\sigma_1)-\lambda\sigma_2
\ee
with the structural evolution modelled by ``retarded stresses''
\be
\sigma_i(t)=\int_{-\infty}^t M_i(t-t')\sigma(t')dt' \;\;\;\mbox{for}\;\;\; i=1,2
\ee
The $M_i(t)$ are memory kernels, each having an integral of unity. In
the absence of relaxation, the first term on the right hand side of
(\ref{eqn:CHA}) causes the stress to increase linearly with straining
(a Hookean solid with a spring constant of unity). The second and
third terms respectively capture nonlinear and linear stress
relaxation.
\begin{figure}[tbp]
\centering
\includegraphics[width=10cm]{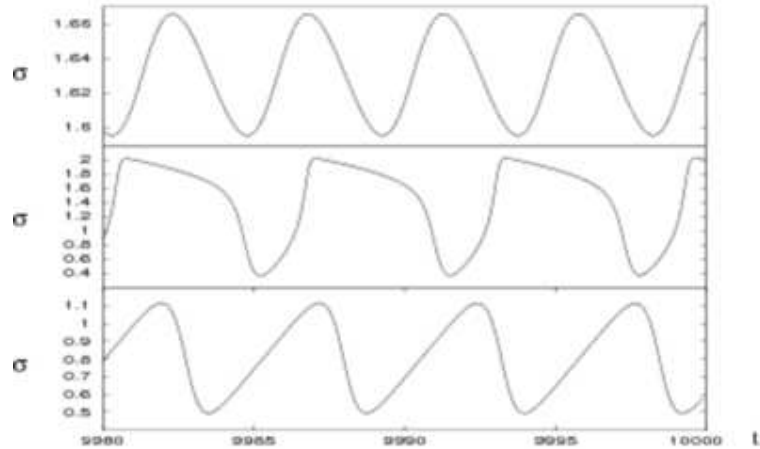}
\caption{Time series of the stress at $\gdot=18.49$, $\gdot=30$, $\gdot=33.38$  (bottom to top) in the CHA model. CHA parameters: $\lambda=20$, $\tau_2=10$ and $R(\sigma)=0.6\sigma^5-3.3\sigma^3+5\sigma$.
Figure adapted from Ref.\cite{cates2002}.\label{fig:CHAosc}}
\end{figure}

In the simplest version of the model, $M_1$ is chosen to be a delta
function such that $\sigma_1(t)=\sigma(t)$ is actually unretarded.
For simplicity, the memory kernel $M_2$ is chosen to be exponential,
$M_2(t)=\tau_2^{-1}\exp\left(-t/\tau_2\right)$. In this case, the system can be rewritten as two coupled differential equations
in the stress $\sigma$ and structural variable $m\equiv\sigma_2$:
\be
\label{eqn:CHAosc1}
\dot{\sigma}=\gdot-R(\sigma)-\lambda m
\;\;\;\;\; ;\;\;\;\;\;
\tau_2\dot{m}=-(m-\sigma)
\ee
In steady state at a given applied shear rate $\gdot$, we find the relation
\be
\label{eqn:CHAconstit}
\gdot=R(\sigma)+\lambda\sigma
\ee
When inverted, this defines the constitutive curve $\sigma(\gdot)$ of
Fig.~\ref{fig:CHAconstit}. Its two components $R(\sigma)$ and
$\lambda\sigma$ stem respectively from rapid nonlinear stress
relaxation on a timescale $t \simeq R^{-1}=O(1)$ and retarded linear
relaxation on a timescale $t=\tau_2\gg 1$. Thus $R(\sigma)$ represents
an ``instantaneous'' constitutive relation, describing the relaxation
of stress at fixed structure. The much slower structural relaxation
eventually recovers the full curve $R(\sigma)+\lambda\sigma$. The
interesting case arises when $R(\sigma)+\lambda\sigma$ is monotonic
but $R(\sigma)$ is not (Fig.~\ref{fig:CHAconstit}). The system then
exhibits a shear banding instability, at short times, in the regime
where $R'(\sigma)<0$. If the linear contribution $\lambda\sigma$ is
sufficient retarded ($\tau_2$ large), it fails to overcome this
instability, despite the monotonic constitutive curve. Accordingly,
the long-term dynamics of the model remain unsteady in a region
$\sigma_c^-<\sigma<\sigma_c^+$ (Fig.~\ref{fig:CHAosc}). The dynamical
system defined by (\ref{eqn:CHAosc1}) undergoes a Hopf bifurcation at
$\sigma_c^+$ and $\sigma_c^-$ (Fig.~\ref{fig:CHAconstit}), signifying
the onset of finite frequency sinusoidal oscillations with an
amplitude varying as $|\gdot-\gdot_c|^{1/2}$.

Chaos requires a phase space of dimensionality at least three; it
cannot occur in the dynamical system (\ref{eqn:CHAosc1}). Without
invoking flow inhomogeneity (which gives infinite dimensionality),
sufficient dimensions can be achieved by assuming $\sigma_1\equiv n$
to be retarded as well as $\sigma_2$, with
$M_1(t)=\tau_1^{-1}\exp\left(-t/\tau_1\right)$. The underlying
constitutive curve is again given by (\ref{eqn:CHAconstit}). In
harmony with the simpler version of the model, one takes $\tau_1\lae
1\ll\tau_2$ and consider the situation where monotonicity of the
constitutive curve is restored only via the more retarded relaxation
term. In the unstable regime, one now finds a period-doubling cascade
leading to temporal chaos.
However, the physical interpretation of the second retarded stress is unclear and the model therefore remains strongly empirical in nature.

The reader may be struck by the naive character of the CHA model when compared with the fully tensorial descriptions outlined in the preceding section for nematics liquid crystals. (Admittedly, the latter field has had an extra half-century to develop its equations!) Note however that the scalar treatment of shear stress is not quite as restrictive as it may appear. Specifically, it does not rule out the existence of normal stresses, but does assume that the shear stress has an autonomous dependence on strain rate in simple shear flows. This holds, for instance, in the upper convected Maxwell model despite the latter's fully tensorial character.
The development of tensorially convincing equations for shear-thickening materials remains a topic for future study, but so far attention has focussed instead on relaxing the assumption of spatial homogeneity of the flow. This we address next.

\subsection{Spatiotemporal Instability; Rheochaos}

\label{sec:spatioTemporal}

Above we have discussed the unsteady rheological response of models
with purely temporal dynamics. Such models assume from the outset that
the sample remains homogeneous, with each point in space following an
identical evolution in time. As noted above, however, reports of
unsteady dynamics in wormlike micelles are most common close to or
inside the shear banding regime. In such cases, a {\em
spatio-}temporal description is essential, to allow for an evolving
state that is heterogeneous ({\em e.g.}, banded) at any
instant. Indeed, recent experiments have unambiguously revealed
fluctuating shear bands in both wormlike
micelles~\cite{lopez-Gonzalez2004} and multilamellar onion
phases~\cite{SalManCol03b,ManSalCol04}. We now turn to models of
spatiotemporal dynamics, considering thinning and thickening systems
in turn.

\begin{figure}[bp]
\centering
\includegraphics[width=10cm]{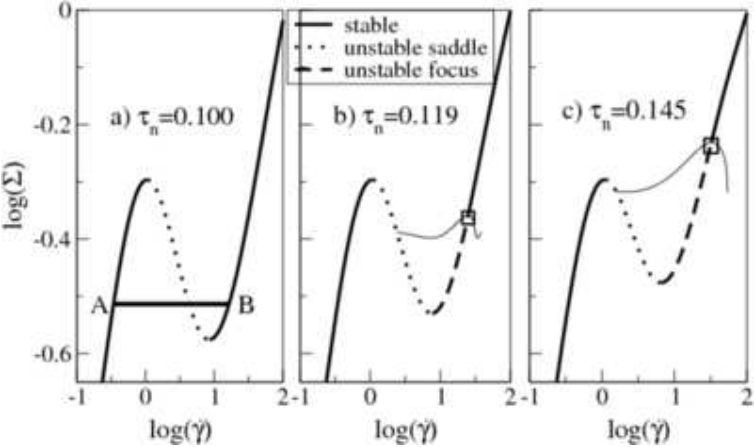}
\caption{Intrinsic constitutive curves for differing degrees of
coupling between flow and micellar length. a) Weak coupling, giving
the standard coexistence of stable low and high shear bands (A and B);
b) moderate coupling; c) strong coupling.  Squares show Hopf
bifurcations. The thin black lines
delimit the periodic orbit of the local model at fixed
$\sigma$. Figure reprinted with permission from Ref.~\cite{fielding2004}.
\label{fig:linear}}
\end{figure}
\begin{figure}[bp]
\centering
\includegraphics[width=10cm]{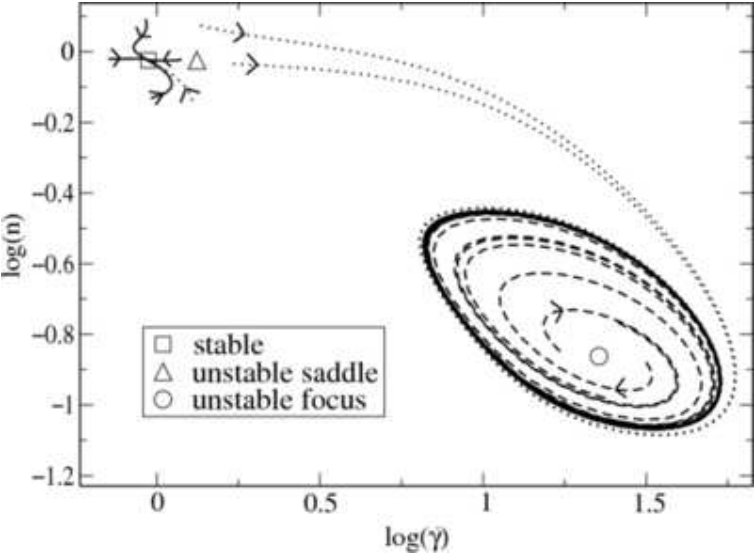}
\caption{Homogeneous dynamics  for fixed stress, $\log(\sigma)=-0.301$, for different initial conditions. Figure reprinted with permission from Ref.~\cite{fielding2004}.
\label{fig:localFlows}}
\end{figure}

\subsubsection{Shear Thinning: A One-Dimensional Model}

As explained previously, homogeneous flow on the negatively sloping
branch of the constitutive curve is unstable with respect to the
formation of shear bands~\cite{Yerushalmi70}. In Sec~\ref{sec:onset},
we further saw that coupling between the flow and an auxiliary
variable such as concentration can enhance this instability, causing
it to extend into regions of positive constitutive slope
(Fig.~\ref{fig:spinodals_with_phi}).  In Ref.~\cite{fielding2004}, one of us
exploited this fact to construct a model in which the high shear band
is itself destabilised, leading to unsteady banding dynamics.

The model is defined as follows. We work in one spatial dimension, the
flow-gradient direction $y$, with a velocity ${v_i}=v(y)\delta_{ix}$ and shear rate $\gdot(y)=\partial_yv(y)$. Normal stresses
are neglected, and the total shear stress is assumed to comprise
additive viscoelastic and Newtonian components. At zero Reynolds
number, $\sigma(t)$ must be uniform across the gap:
\be
\label{eqn:FO1}
\sigmatot(t)=\sigmapol(y,t)+\etasol\gdot(y,t)
\ee
For the dynamics of the viscoelastic component, we use the
scalar model of Sec.~\ref{sec:JS}
\be
\label{eqn:FO2}
\partial_t\sigmapol=-\frac{\sigmapol}{\tau(n)}+\frac{g[\gdot\tau(n)]}{\tau(n)}+\mathcal{D}\partial_y^2\sigma
\ee
with a relaxation time $\tau$ that now
depends on a structural variable $n=n(y,t)$, according to $\tau(n)=\tau_0(n/n_0)^\alpha$.
As before, $g(x)=x/(1+x^2)$ is chosen to ensure a region of negative
constitutive slope.
The auxiliary variable $n$ is taken to represent a nonconserved
quantity. For definiteness it is identified as the mean micellar
length (previously denoted by $\bar{L}$) although there could be other candidates for its interpretation. Coupling of $n$ to the flow is completed by
assuming that it evolves with its own relaxation time
$\tau_n$, distinct from $\tau(n)$:
\be
\label{eqn:FO3}
\partial_t n = -\frac{n}{\tau_n}+\frac{N(\gdot\tau_n)}{\tau_n}
\ee
Here the coupling term $N(x) = n_0/(1+x^\beta)$ represents (say) shear-induced scission.
\begin{figure}[tbp]
  \centering 
\subfigure{
  \includegraphics[width=4.8cm]{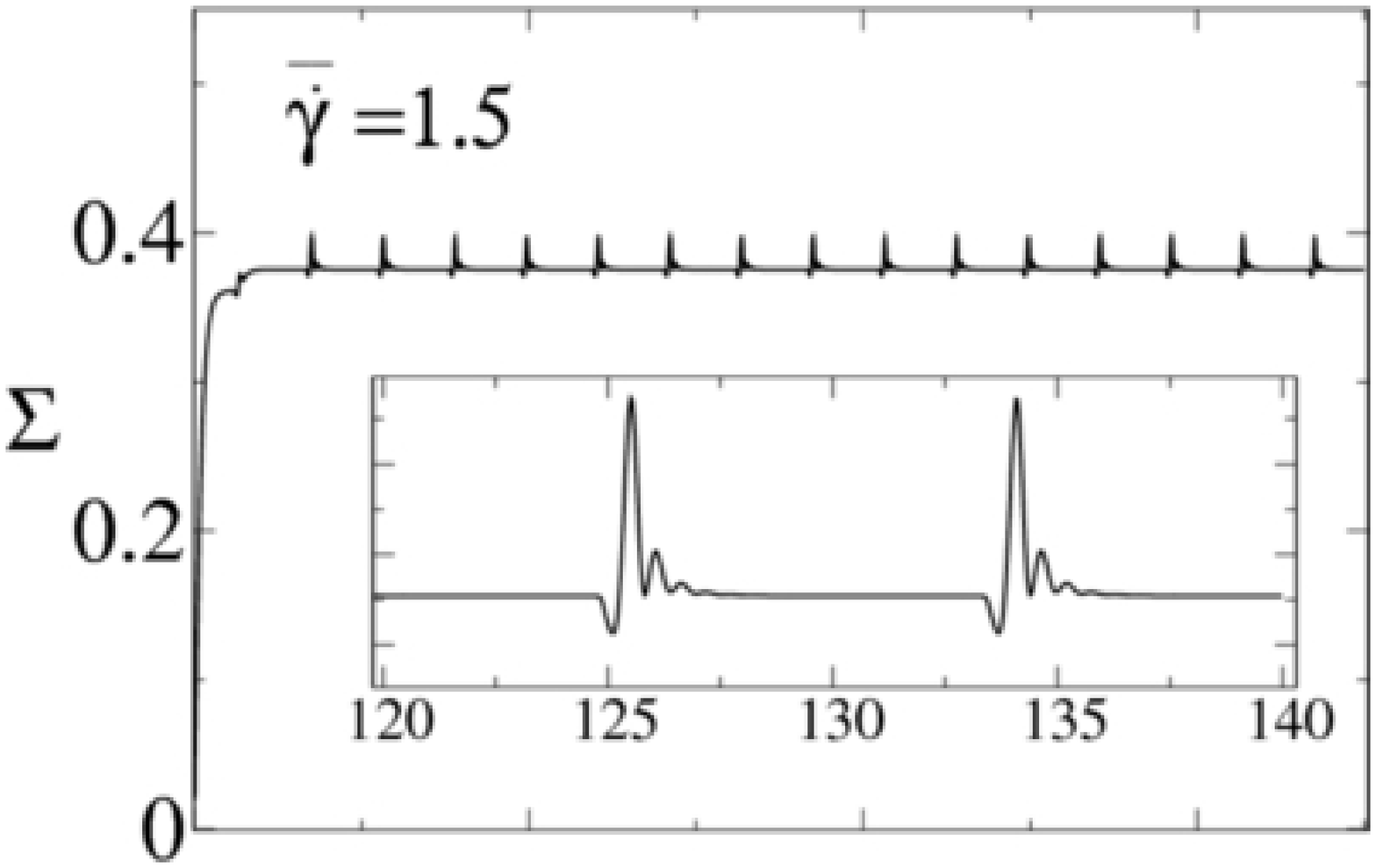}
\label{fig:test:a}
}
\hspace{0.2cm}
\subfigure{
\includegraphics[width=4.0cm]{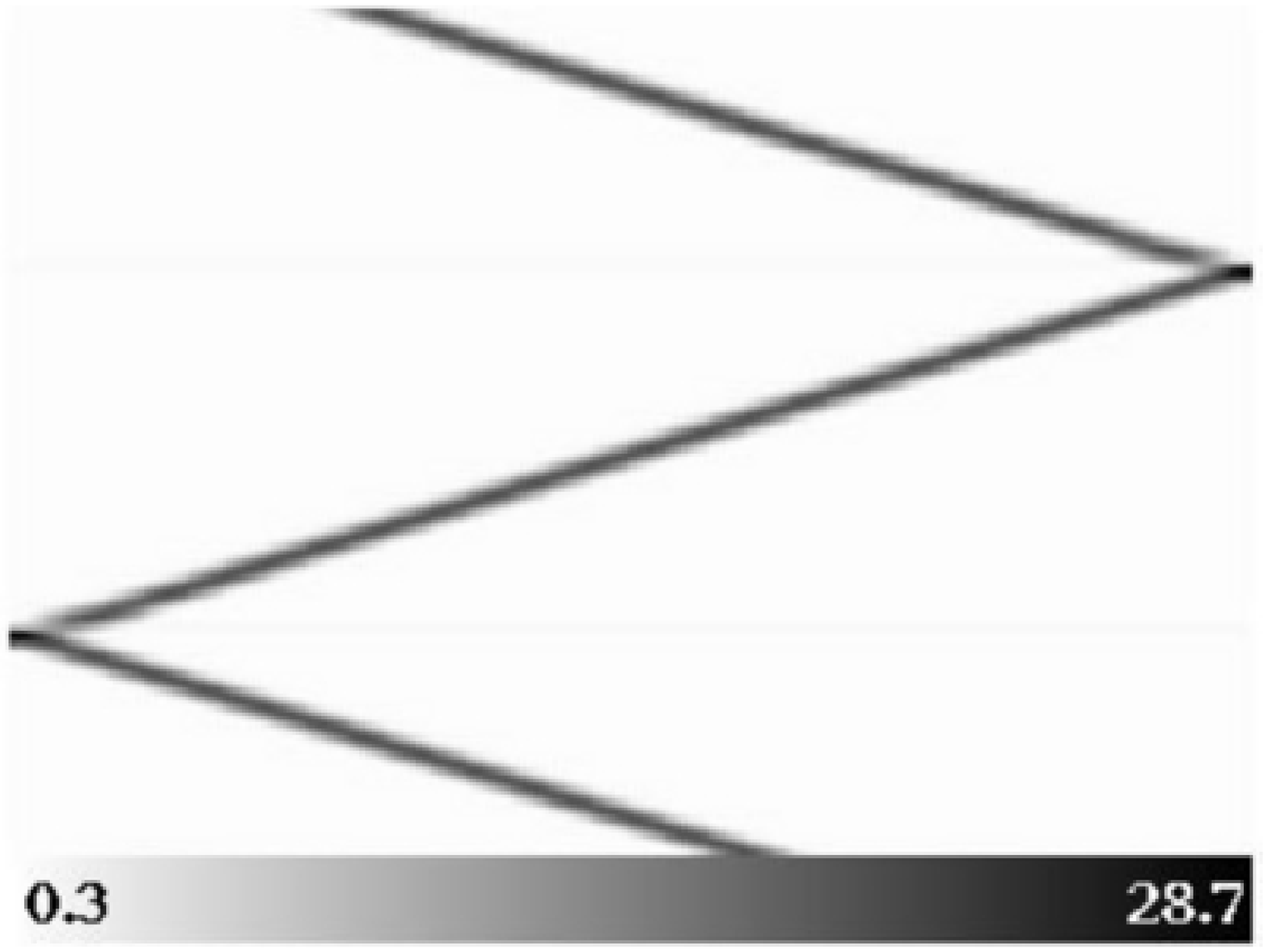}
\label{fig:test:b}
}\\
\subfigure{
  \includegraphics[width=4.8cm]{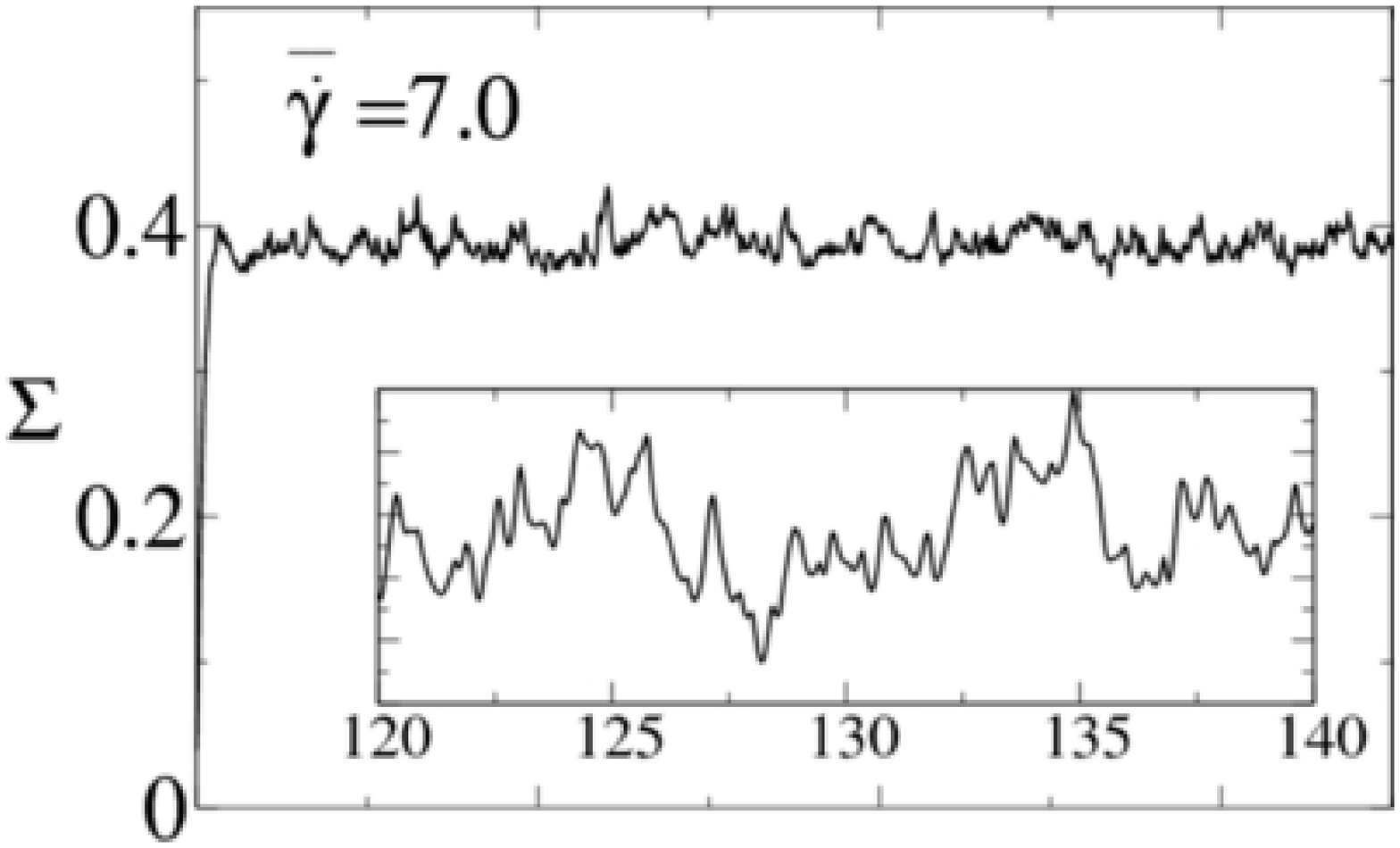}
\label{fig:test:c}
}
\hspace{0.2cm}
\subfigure{
  \includegraphics[width=4.0cm]{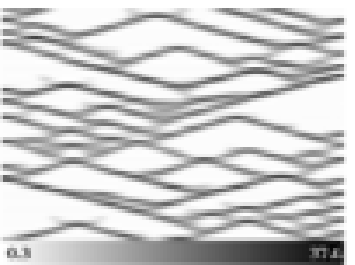}
\label{fig:test:d}
}\\
\subfigure{
\includegraphics[width=4.8cm]{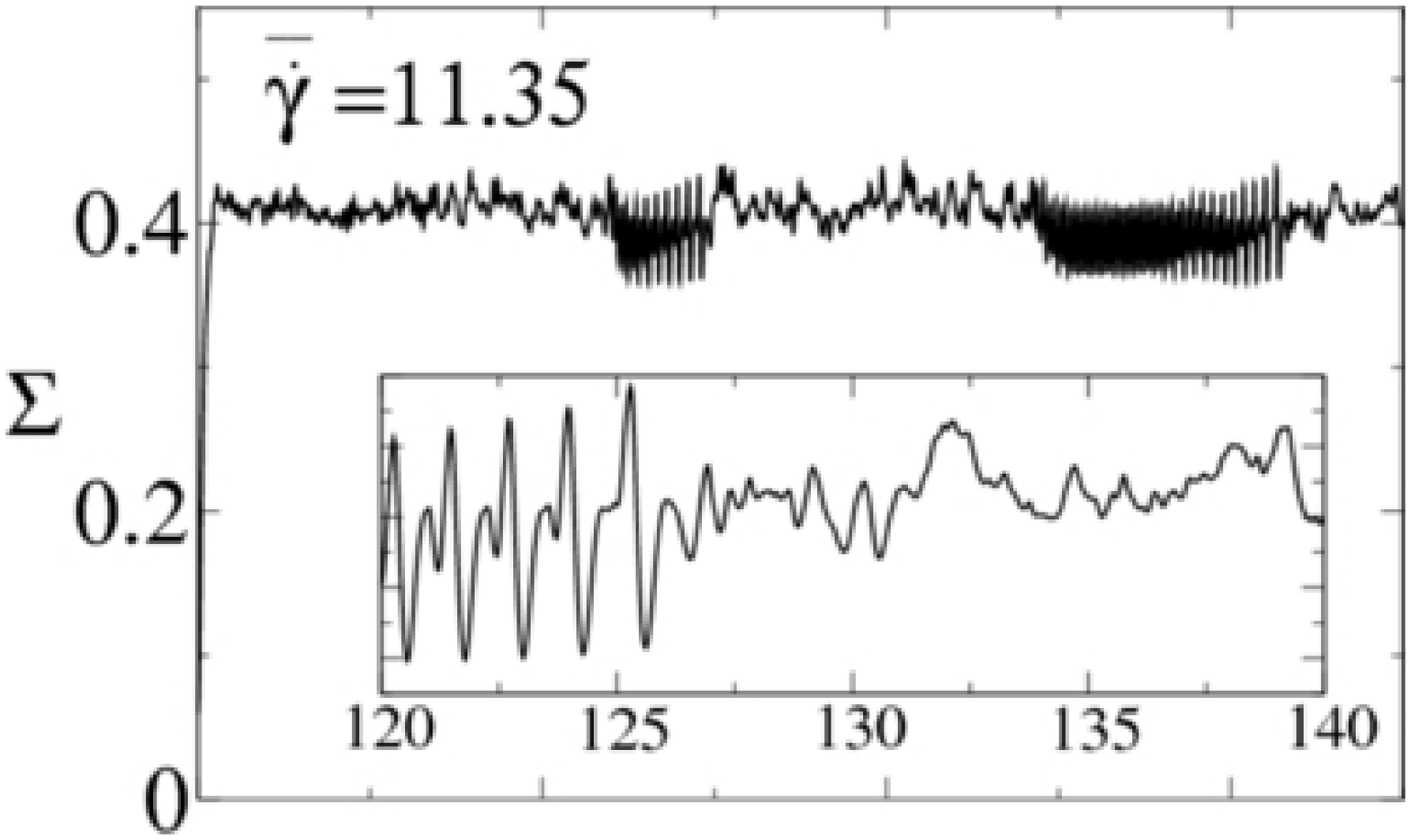}
\label{fig:test:e}
}
\hspace{0.2cm}
\subfigure{
  \includegraphics[width=4.0cm]{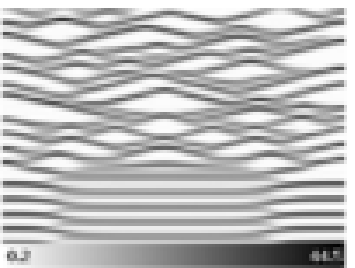}
\label{fig:test:f}
}\\
\subfigure{
  \includegraphics[width=4.8cm]{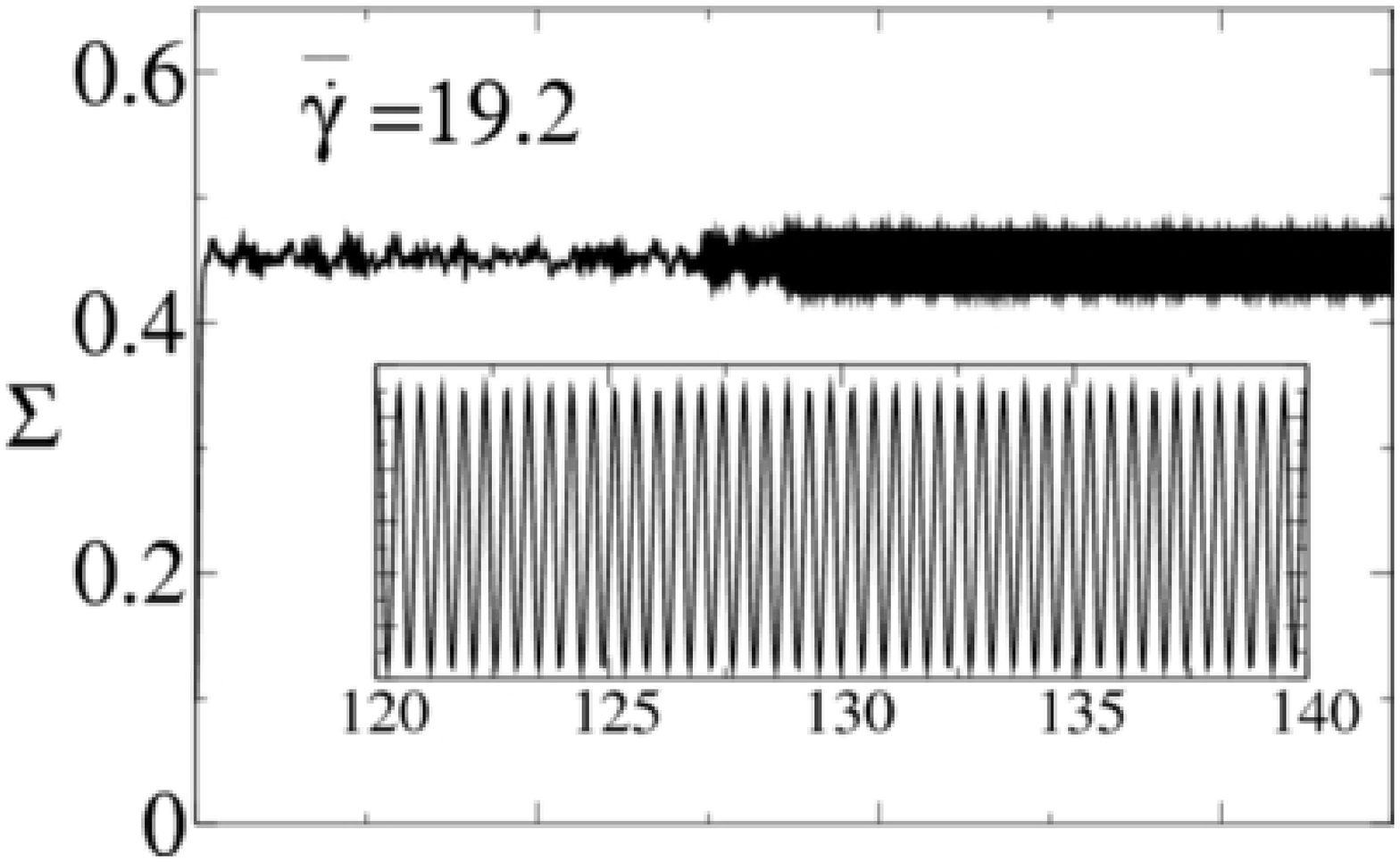}
\label{fig:test:g}
}
\hspace{0.2cm}
\subfigure{
\includegraphics[width=4.0cm]{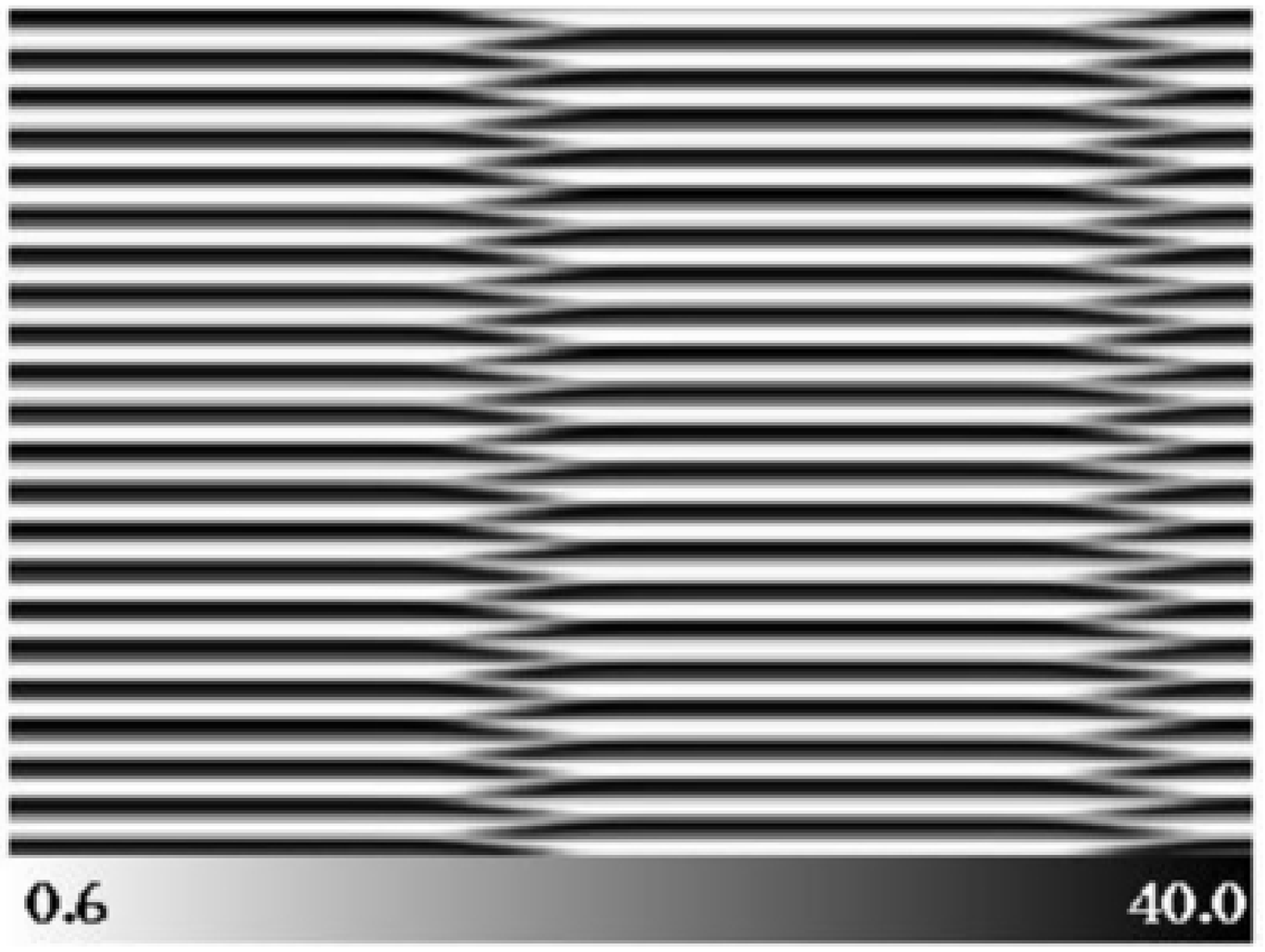}
\label{fig:test:h}
}
\caption{Right panels: spacetime plots showing the time evolution of the shear rate profile for different average shear rates (as identified in the left panels). The data is for the 1D model with parameters set at $\tau_n=0.145$, $\mathcal{D}=0.0016$. The space coordinate $0<y<1$ runs left to right and time  $120<t<140$ runs bottom to top; the shading denotes local shear rate with regions of high shear rate showing as dark zones (as per the grey-scale bar at the bottom of each panel). Taking a horizontal slice through the spacetime plot at a given time shows as dark the regions of high shear rate; a slice higher up the plot shows these regions at a later time. Hence a localized band of high shear rate, moving across the system with velocity $v$, shows up as a grey stripe of slope $v$ on the spacetime plots. Left panels: time series showing the corresponding stress {\em vs.} time. The average shear rate $\gdotbar$ is denoted $\gdot$ in the main text. Figures adapted from Ref.~\cite{fielding2004} and reprinted with permission.}
\label{fig:taun0.145}
\vspace{-0.5cm}
\end{figure}

As discussed in Sec.~\ref{sec:onset} above, a state of initially
homogeneous flow on the underlying constitutive curve is unstable
(growth rate $\omega_n>0$) if $b<0$ or $c<0$ in
the dispersion relation $\omega_n^2+b\omega_n+c=0$, where $c=\tilde{c}S'$ and $S'$ is the
slope of the constitutive curve. The coefficients $b$ and $\tilde{c}$
depend on the model parameters and the applied shear rate. As in the
dJS model, it can be shown that $\tilde{c}>0$; this model therefore
shows the familiar banding instability when $S'<0$, at intermediate shear rates. It also shows
a new instability, $b<0$, not seen in dJS. This destabilises the high
shear branch to a degree that depends on $\tau_n$, eventually
terminating in a Hopf bifurcation shown by squares in
Fig.~\ref{fig:linear}.

\begin{figure}[t]
  \centering 
\subfigure{
    \includegraphics[width=5.5cm]{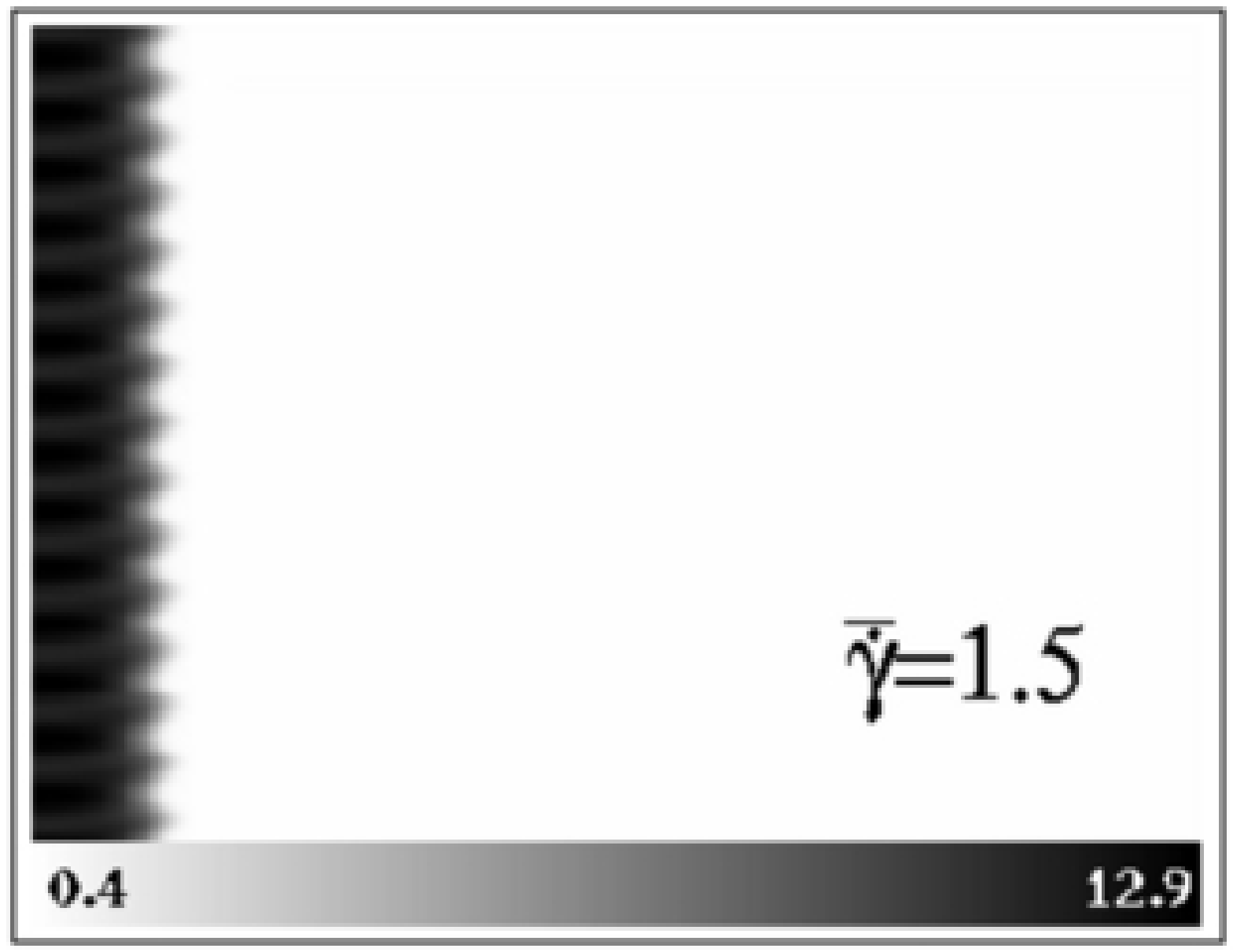}
\label{fig:testing:a}
}
\hspace{0.1cm}
\subfigure{
\includegraphics[width=5.5cm]{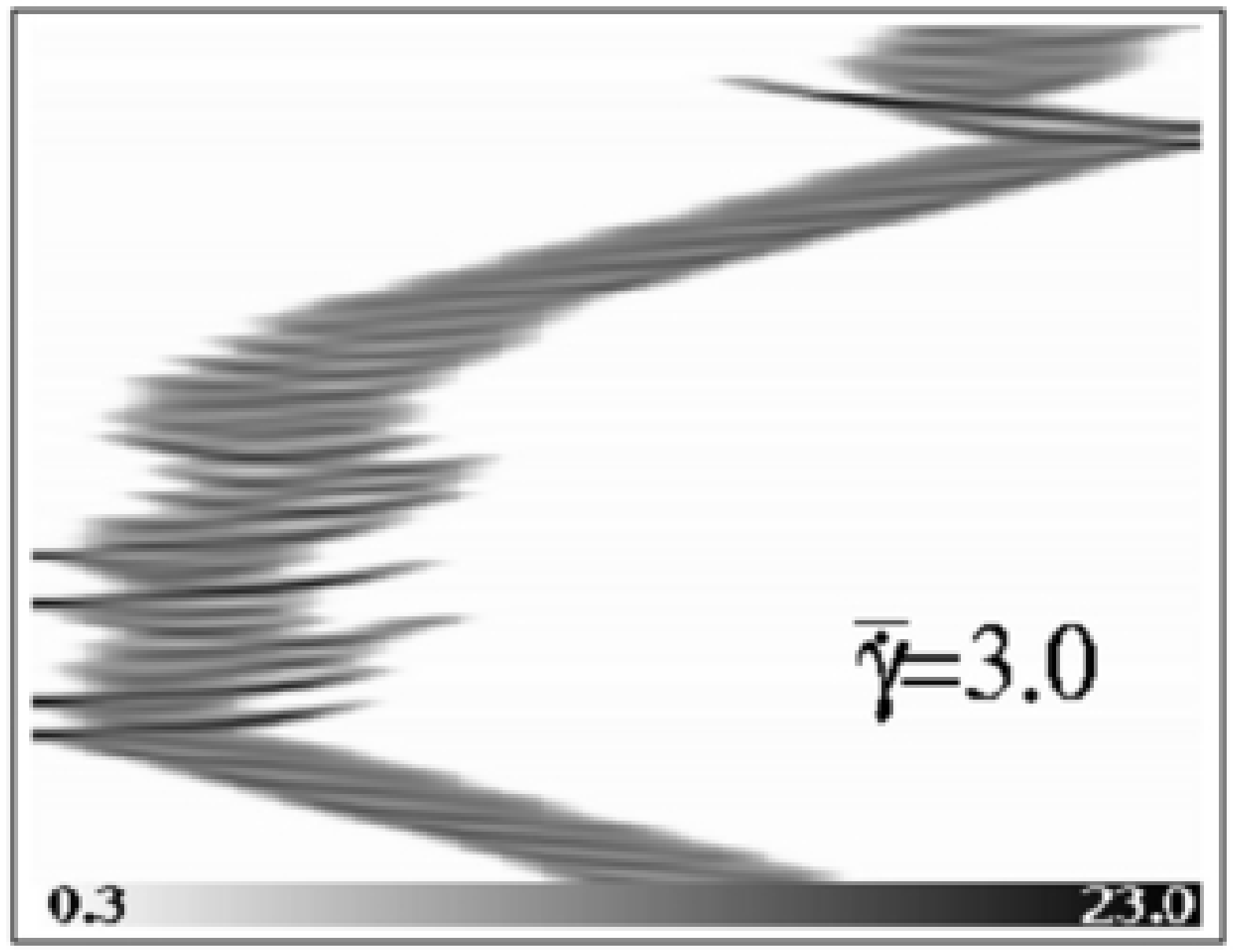}
\label{fig:testing:b}
}\\
\caption{Shear rate intensity plots for $\tau_n=0.13$, $D=0.0016$. Figure reprinted with permission from Ref.~\cite{fielding2004}}.
\label{fig:taun0.13}
\end{figure}

The resulting model's {\em
homogeneous} dynamics are explored in Fig.~\ref{fig:localFlows}, which
shows parametric phase portraits $\gdot(t)$, $n(t)$ at a fixed value
of the stress $\sigma$ for various initial conditions. Here, any
possibility of spatial structuring has been artificially suppressed so
that $\gdot$ and $n$ can depend only on time. The new instability at high shear rates corresponds to an
unstable focus in the $\gdot, n$ plane, associated with a pair of
complex conjugate eigenvalues. Trajectories originating near this
point spiral outwards to attain a stable limit
cycle. 
\if{Periodic orbits such as this represent the most complex
behaviour possible for the homogeneous dynamics, since it only has two
degrees of freedom, $\gdot(t)$ and $n(t)$. (Chaos requires at least three
degrees of freedom, as noted above.)}\fi

Turning to the full spatio-temporal dynamics at imposed
shear rate $\gdot=\int dy\,\gdot(y,t)$, for small $\tau_n$, we find stable
shear bands, connecting points A and B in the constitutive curve of
Fig.~\ref{fig:linear}a. For higher $\tau_n$, the high-shear
band is unstable, leading to unsteady dynamics of the banded state; see
Fig.~\ref{fig:taun0.145}, where
several regimes are evident. At low applied shear rate a thin pulse or
band of locally high shear ricochets back and forth across the
cell. (A thin fluctuating high shear band, away from the rheometer
wall, was seen experimentally in Ref.~\cite{nmr3}.)  At larger
shear rates, we find two or more such pulses. For two pulses (not
shown), we typically find a periodically repeating state with the
pulses alternately bouncing off each other (mid-cell) and the cell
walls.  Once three pulses are present ({\em e.g.}, $\gdot=7.0$),
periodicity gives way to chaotic behaviour.
At still higher mean shear rate, $\gdot=19.2$, we find regular
oscillations of spatially extended bands pinned at a stationary
defect. The local shear rates span both the low and high shear
constitutive branches.  (Oscillating vorticity bands were seen
experimentally in Ref.~\cite{HilVla02,FisWheFul02}.) For the
intermediate value $\gdot=11.35$ we find intermittency between
patterns resembling those for $\gdot=7.0$ and
$\gdot=19.2$. Finally for $\gdot=23.0,31.0$ we find oscillating
bands separated by {\em moving} defects, with the flow now governed only by the high shear
constitutive branch: in each band, the shear rate cycles round the
periodic orbit of the local model (Fig.~\ref{fig:localFlows}).

For different $\tau_n$ we find other interesting
phenomena~\cite{fielding2004}. For example, for weaker instability
($\tau_n=0.13$) at low applied shear rates we see a high shear band
that pulsates in width while adhering to the rheometer wall, or
meanders about the cell (Fig.~\ref{fig:taun0.13}). The former
behaviour resembles that sometimes seen in
micelles~\cite{nmr2,HBP98} and onion
phases~\cite{WunColLenArnRou01}.

\subsubsection{Shear Thinning: Higher Dimensional Model}

In Sec.~\ref{sec:JS} we introduced the diffusive Johnson-Segalman
(dJS) model and in Sec.~\ref{sec:steady} we discussed its predictions
for steady shear banded states. Those calculations were restricted to
1D variations only (in the flow gradient direction $y$) and implicitly
assume that the interface remains perfectly flat at all times. Recent
experiments suggest more complex interfacial
dynamics~\cite{lopez-Gonzalez2004}, and a natural question is
whether these 1D states remain stable in higher dimensions, or whether
they give way to more complex behaviour.  This question was recently
addressed by one of us in a study comprising two separate stages:
first, a linear stability analysis of an initially 1D banded state~\cite{fielding2005}, and then
a numerical study of nonlinear interfacial dynamics~\cite{upcoming_2d}. We summarise each of these in
turn.

\begin{figure}[t]
\centering
\includegraphics[width=10cm]{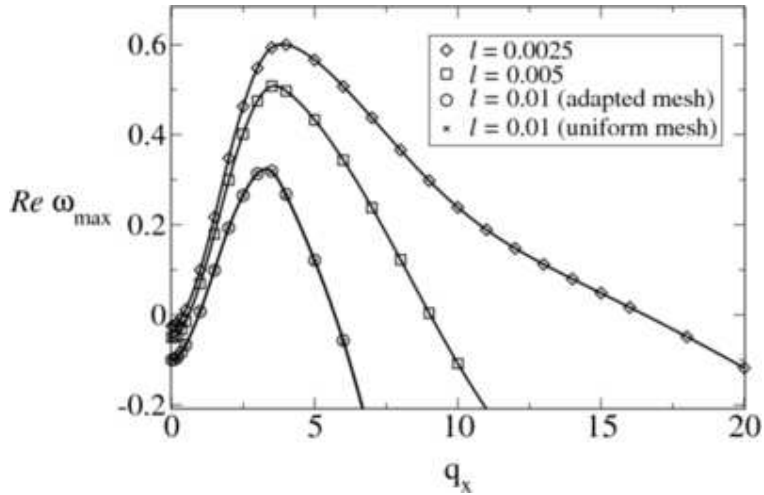}
\caption{Real part of the eigenvalue of the most unstable mode; $a=0.3$,  $\eta=0.05$,  $\gdot=2.0$, Reynolds number $\rho/\eta=0$. The data for $l=0.01$ correspond to the base profile in Fig.~\ref{fig:profile}. Symbols: data. Solid lines: cubic splines. Figure reprinted with permission from Ref.~\cite{fielding2005}.
\label{fig:maxEigen} } 
\end{figure}

{\it Linear stability analysis:} As discussed in
Sec.~\ref{sec:JS}, a 1D flow-gradient calculation of planar shear
within the dJS model predicts the steady state flow curve of
Fig.~\ref{fig:flowCurvea0.3}, with a corresponding banded profile
shown in Fig.~\ref{fig:profile}.
To study the linear stability of this 1D ``base'' profile with respect
to small fluctuations with wavevector in the ($xz$) plane of the interface, we linearise the model for small perturbations
(lower case) about the (upper case) base profile:
$\vecv{\tilde{\Phi}}(x, y, z, t)=\vecv{\Phi} (y)+
\vecv{\phi}_{\vecv{q}}(y)\exp(\omega_{\vecv{q}}t+ iq_x x+iq_z z)$.
The vector $\vecv{\Phi}$ comprises all components
$\vecv{\Phi}=(W_{\alpha\beta}, V_\alpha)$, the pressure being
eliminated by incompressibility. This gives a linear eigenvalue equation.
\if{
 with an operator $\mathcal{L}$ and eigenfunction $\vecv{\phi}_{\vecv{q}}(y)$:
\be
\label{eqn:eigen}
\omega_{\vecv{q}}\vecv{\phi}_{\vecv{q}}(y)=\mathcal{L}({\vecv{\Phi}(y), {\vecv{q}}}, \partial_y, \partial_y^2...)\vecv{\phi}_{\vecv{q}}(y).
\ee
We}\fi
We
are interested in the eigenvalue $\ommax(\vecv{q})$ with the
largest real part, $\Re\,\ommax(\vecv{q})$: in particular, whether it
is stable, $\Re\,\ommax<0$, or unstable, $\Re\,\ommax>0$.
We find (Fig.~\ref{fig:maxEigen}) that at
any $q_x$, $\Re\ommax$ increases with decreasing values of $l\equiv \sqrt{\mathcal{D}\tau}$, which sets the width of the interface between the bands.  For small enough $l$ the
dispersion relation is positive over a range of wavevectors, rendering
the 1D profile unstable. This applies to shear rates across most of the
stress plateau of the flow curve, and is furthermore robust to variations in the JS parameter $a$.  Because the $l$
values accessed here ($l\sim 1-10\mu {\rm m}$ for a 1mm rheometer gap)
are even larger than those expected physically ($l \sim 100{\rm nm}$) this study suggests that, experimentally, the entire stress plateau region
will be unstable to perturbations away from a flat interface between shear bands.

%
%
%

%
\begin{figure}[thb]
\centering
\includegraphics[width=8cm]{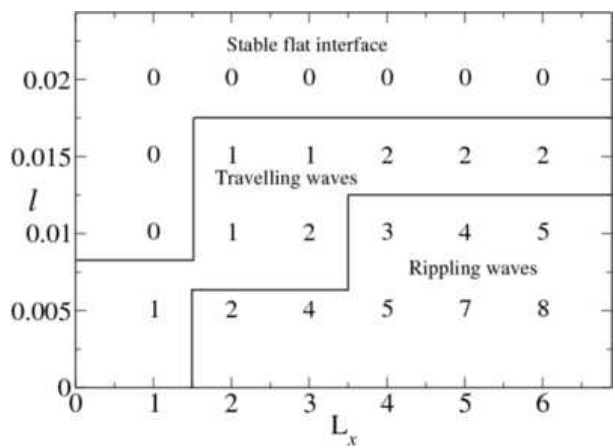}
\caption{Nonequilibrium phase diagram showing regimes of nonlinear interfacial dynamics in the dJS model together with the numbers of
   linearly unstable modes. ($L_x=\Lambda_x$ in the present notation.)
   The slip parameter $a=0.3$ and applied shear rate
   $\gdot=2.0$. Figure reprinted with permission from Ref.~\cite{upcoming_2d}.
\label{fig:phase} } 
\end{figure}
\begin{figure}[t]
  \centering
\includegraphics[width=8.5cm]{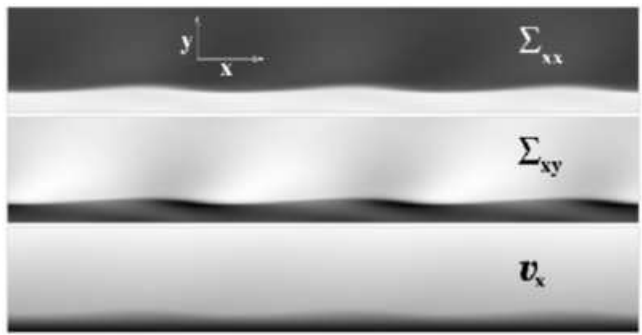}
 \caption{Greyscale of order parameters for travelling wave in the
 $(x,y)$ plane for $l=0.015$, $\Lambda_x=6$, and upper wall velocity
 $V\equiv \gdot \Lambda_y=2$ to the right. (Notation differs from the text:
 $\Sigma_{ij}=W_{ij}$.) Figure reprinted with permission from Ref.~\cite{upcoming_2d}.
\label{fig:travel}}
\end{figure}
\begin{figure}[th]
\centering
\includegraphics[width=7cm]{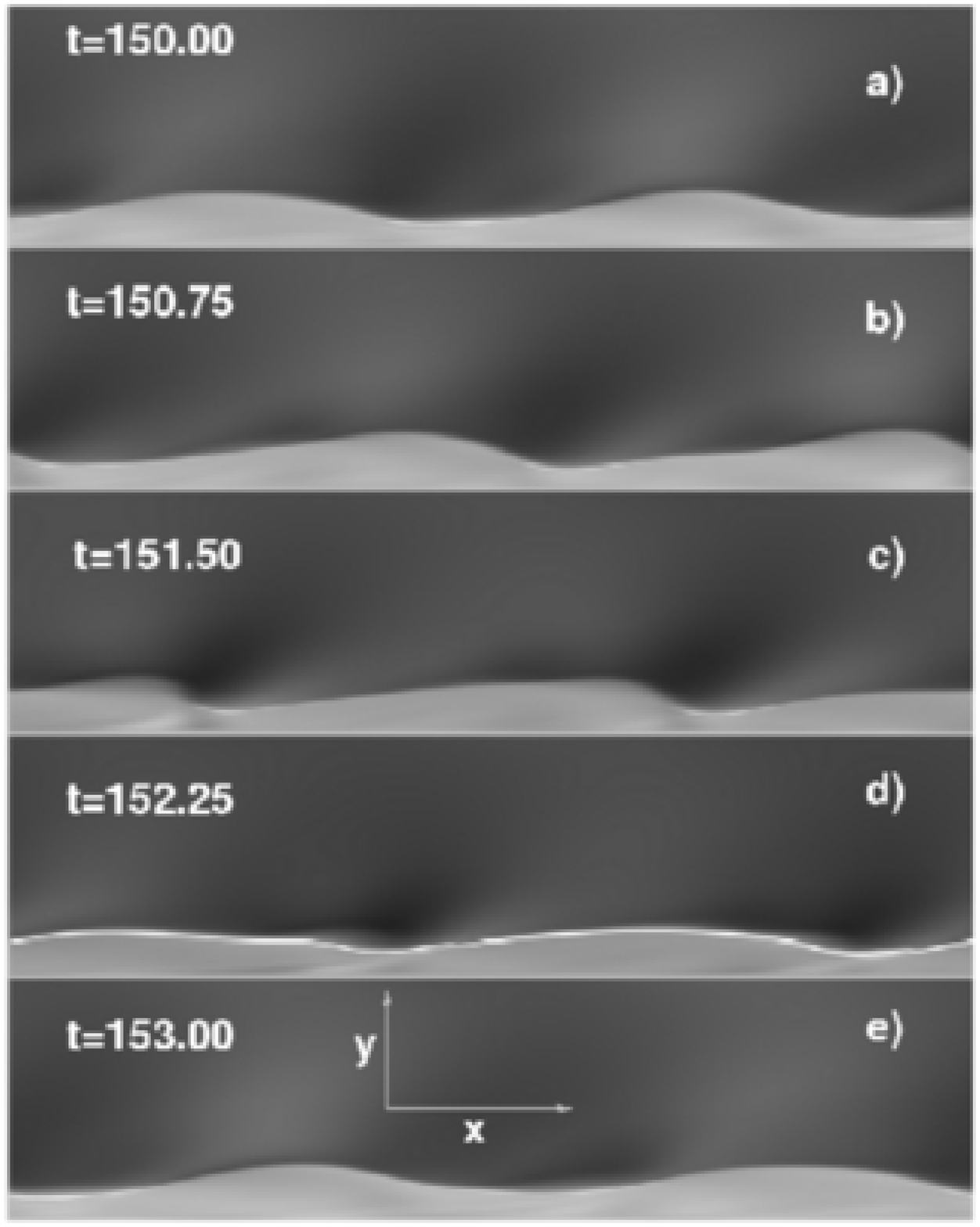}
\caption{Rippling wave  at $l=0.005, \Lambda_x=4, \gdot=2$.  
  Greyscale of $\Sigma_{xx}(x,y)$.  Upper wall moves to the
  right. White line in d): interface height. Figure reprinted with permission from
  Ref.~\cite{upcoming_2d}.
\label{fig:break}}
\end{figure}

{\it Nonlinear interfacial dynamics:} Once the undulations
attain a finite amplitude, nonlinear effects become important, and the
linear calculation then breaks down: one
must perform a full nonlinear study of the dJS model in the
flow/flow-gradient ($xy$) plane~\cite{upcoming_2d}.
The results are summarised in Fig.~\ref{fig:phase}. At large values of
$l$, for which the dispersion relation of the linear
analysis is negative at all wavevectors, the 1D base profile remains
stable as expected: the interface stays flat at all times. For smaller
values of $l$, the dispersion relation is positive over a window of
$q_x$ (Fig.~\ref{fig:maxEigen}); at fixed $l$,
the number $N$ of linearly unstable modes  increases with the system size
$\Lambda_x$ (Fig.~\ref{fig:phase}).
For small $l/\Lambda_x$, just inside the unstable
regime, the
ultimate attractor comprises a travelling wave
(Fig.~\ref{fig:travel}). The wall-averaged shear stress is
constant in time, with a value $\overline{W}_{xy,ss}$
that depends on $l$ and $\Lambda_x$ and is slightly higher than the
selected value $W_{xy}^{\rm{sel}}$ of the 1D calculation. 
For $l/\Lambda_x$ values deeper inside the unstable regime in
Fig.~\ref{fig:phase}, we see a new regime in which the travelling wave
now periodically ``ripples'' (Fig.~\ref{fig:break}).  The corresponding
wall-averaged stress $\overline{W}_{xy}$ is periodic in time,
with variations of the order of one
percent, and an average value larger than the 1D selected stress
$W_{xy}^{\rm{sel}}$. The interface height $h(x,t)$ is shown as a white
line in Fig.~\ref{fig:break}d.

The results just discussed were obtained in numerical runs starting
from an initial state comprising two adjacent shear bands separated by
a flat interface. Shear startup from rest can give more complicated
results, allowing multiple interfaces to form, with complex dynamical
interactions suggesting low dimensional chaos. As discussed in
Ref.~\cite{upcoming_2d}, cell curvature (almost always present
experimentally) is likely to eliminate this degeneracy, restoring the
single-interface scenario described above.  

Note that in earlier work, Yuan {\em et al}~\cite{Yuan99} used a mixed
Lagrangian-Eulerian algorithm to evolve a related non local JS model
in 2D planar shear. Contrary to Ref.~\cite{upcoming_2d}, they
apparently found a stable interface. However, the banded profiles
reported in Ref.~\cite{Yuan99} were averaged along the
flow-direction. In any case, rather short cell lengths $\Lambda_x$
were used, likely to be in the stable regime of Fig.~\ref{fig:phase}.

\subsubsection{Shear Thinning: Other Relevant Approaches}

Here we touch on two other theoretical approaches to spatio-temporal
rheochaos in generic viscoelastic fluids, which may be
relevant to micelles. The first generalises models of nematodynamics
(Sec.~\ref{sec:temporal}) to allow spatial structuring, and finds
spatio-temporal rheochaos setting in via
intermittency~\cite{das2005,chakrabarti2004}. Although the existing
study neglects backflow effects, a hydrodynamic equivalent could be
relevant in concentrated wormlike micelles close to the nematic
transition, where the highly sheared band may have a high degree of
nematic ordering. However, any direct link with wormlike micelles
remains to be established; for more details see a recent
review~\cite{DasGel}.

The second approach is based on the concept of ``elastic
turbulence''. For simple Newtonian liquids, it has long been known
that the nonlinear inertial term of the Navier Stokes equation can
cause smooth, laminar flow to destablise at high Reynolds number,
giving way to more complicated flow profiles~\cite{Drazin,Mullin}.
For complex fluids, including wormlike micelles, inertia is usually
negligible. However, analagous instabilities (which are somewhat distinct from those involving shear-banding as discussed above) can arise directly from constitutive nonlinearity. As reviewed in Ref.~\cite{LarsonReview}, this
opens up the possibility of purely elastic instabilities that occur
even in inertialess flows at zero Reynolds number. 

For example, the
Oldroyd B model discussed in Sec.~\ref{sec:ucmm} predicts purely
elastic instabilities triggered by hoop stresses in both curved
Couette~\cite{larson1990c} and torsional~\cite{PhanThien} flows. Such phenomena
have been observed experimentally in model ``Boger'' fluids, which
comprise high molecular weight polymers in viscous solvents, in both
curved Couette~\cite{MulLarSha} and plate-plate~\cite{MagLar88}
flow. The basic observation is that, above a critical shear rate, the
laminar base flow destablises, accompanied by significant increase in mean shear stress. In a
remarkable development, a transition series linking the base flow at
low shear rates to fully developed elastic
turbulence~\cite{larson2000c,groisman2000,mckinley1991} at high shear
rates was uncovered. This involves multiple dynamic states including
stationary rings, competing spirals, multi-spiral chaos, spiral bursts
and eventually fully developed elastic turbulence.  Planar shear lacks
hoop stresses, so the linear viscoelastic instabilities discussed
above are absent. Nonetheless, a {\em nonlinear} equivalent is still
predicted to arise, since fluctuations cause the originally parallel
streamlines to become curved, and so subject to instability at 
nonlinear order~\cite{morozov2005}. The relevance to micelles of these generic findings on elastic instabilities remains to be explored.

\subsubsection{Shear Thickening}

\begin{figure}[t]
  \centering
\includegraphics[width=8.5cm]{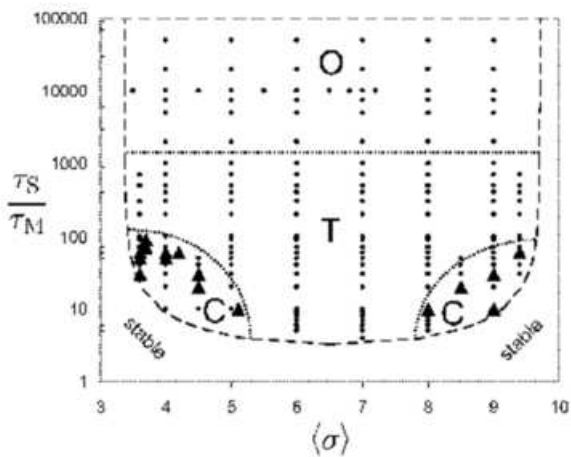}
 \caption{Nonequilibrium phase diagram of the shear-thickening model of Ref.~\cite{upcoming_CHA} when $\tau_{\rm s}$ and
 $\avstress$ are varied. Black triangle: chaotic points, black circle:
 periodic point. Three main regimes are observed: (O)~oscillating
 shear bands, (T)~travelling shear bands, (C)~chaotic regions. The
 outer dashed line is the linear stability limit. 
Figure reprinted with permission from Ref.~\cite{upcoming_CHA}. \label{fig:CHAphase}}
\end{figure} 

In Sec.~\ref{sec:temporal} above, we discussed the temporal dynamics
of the CHA model of shear thickening, in which the {\em instantaneous}
constitutive relation at fixed structure is nonmonotonic
(Fig.~\ref{fig:CHAconstit}). This creates a short-term tendency to
form bands of differing shear stress, coexisting in the vorticity
direction at a common value of the shear rate. However this short term
tendency is opposed by the long term structural evolution. In a full
spatio-temporal scenario, the interplay of these two effects can give
rise to shear banding with complex dynamics.

The CHA model of Sec.~\ref{sec:temporal} was recently extended to
allow for spatial structuring~\cite{aradian2005,upcoming_CHA},
allowing the stress to vary
along the vorticity direction: $\sigma(t)\to\sigma(z,t)$. A spatial gradient term
$\mathcal{D}\nabla^2\sigma$ was added, in line with the discussion of
Sec.~\ref{sec:steady} above. For simplicity, only $\sigma_2$ was taken as retarded, by extending
(\ref{eqn:CHAosc1}) as follows:
\be
\label{eqn:CHAspatial1}
\dot{\sigma}(z,t)=\gdot(t)-R(\sigma)-\lambda m + \mathcal{D}\nabla^2\sigma
\ee
\be
\label{eqn:CHAspatial2}
\dot{m}(z,t)=-\frac{m-\sigma}{\tau_{\rm s}}
\ee
As before the instantaneous nonlinear relaxation term was chosen as
\be
R(\sigma)=a\sigma-b\sigma^2+c\sigma^2
\ee
giving in linear response the
Maxwell time $\tau_{\rm M}=1/a$.
Note that the shear rate $\gdot(t)$ in (\ref{eqn:CHAspatial1}) is now {\em
uniform}: the moving wall of the rotor imposes the same velocity for
all heights $z$, so that $\gdot(z,t)=\gdot(t)$ only. The model is
studied under conditions of imposed torque, {\em i.e.}, under an imposed
value of the spatial mean of the stress $\avstress$. The two main
control parameters are taken to be $\avstress$ and the ratio
$\tau_{\rm s}/\tau_{\rm M}$. Depending on the values of these
parameters, the model shows a rich variety of oscillatory and chaotic
banding dynamics, as summarised in the phase diagram of
Fig.~\ref{fig:CHAphase}.

\begin{figure*}
\includegraphics[width=\textwidth]{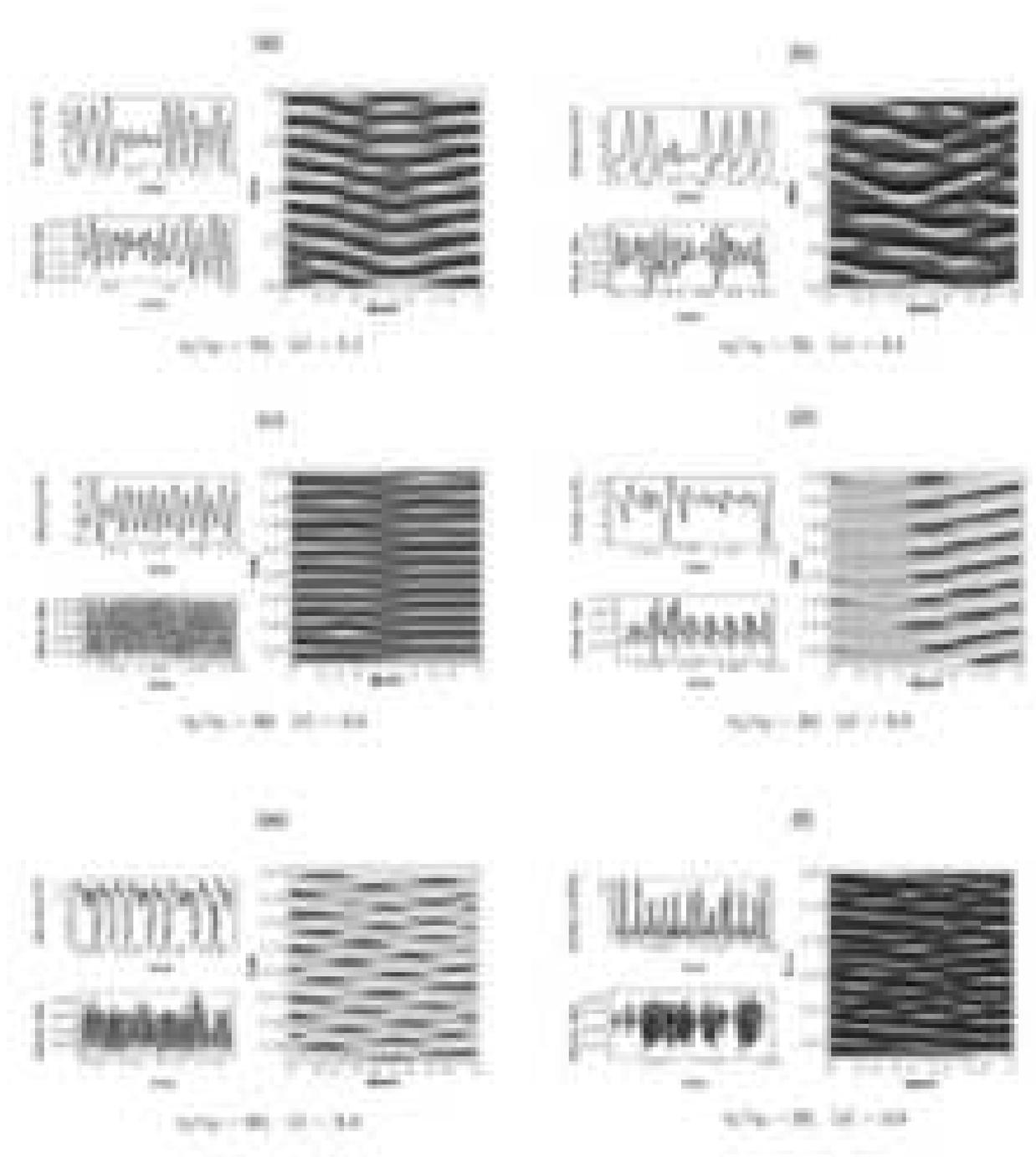}
\caption{\label{RheochaosExamples}
Various types of spatiotemporal rheochaos observed in the
shear thickening model of Ref.~\cite{aradian2005,upcoming_CHA}. Left: time series for a local shear rate and for stress. Right: spacetime plots (as explained in Fig.~\ref{fig:taun0.145}). Figure adapted from Ref.~\cite{upcoming_CHA} and reprinted with permission.\label{fig:CHAchaos}}
\end{figure*}

In the regime where the structural evolution is much slower than the
stress relaxation (marked `O' in Fig.~\ref{fig:CHAphase}), the model predicts oscillating shear bands. Varying the imposed stress
$\avstress$ along any horizontal line of fixed $\tau_{\rm s}/\tau_{\rm
M}$ in this regime, the associated waveforms range from simple to very
complex. Near the middle of the line one sees simple `flip-flopping'
bands, with the cell divided equally between a high-stress and a
low-stress band, the identities of which repeatedly switch with a
period of order the structural time $\tau_{\rm s}$.  Moving slightly
off-centre along the line, the interface between the bands now adopts
a zig-zagging motion, superposed on the flip-flopping motion just
described. Near the edges of the line, very complex oscillations are
seen. The regime $10\le\tau_{\rm s}/\tau_{\rm M}\le 10^3$ marked `T'
in Fig.~\ref{fig:CHAphase} is characterised by a periodic nucleation
of shear bands, which then cross the cell with roughly constant
velocity. In the two disconnected pockets marked `C', at moderately
long values of $\tau_{\rm s}$ relative to $\tau_{\rm M}$, and strongly
off-centered values of $\avstress$, the complex oscillations of
regimes `O' and `T' give way to true rheochaos, characterised by a
positive Lyapunov exponent. Various examples of rheochaotic flows are
shown in Fig.~\ref{fig:CHAchaos}.
This model also admits an interesting
low-mode truncation in which only the lowest two non-homogeneous
Fourier modes are retained for each of $\gdot$ and $n$, giving 4 modes
in all~\cite{upcoming_CHA}. The basic structure of the phase diagram of
Fig.~\ref{fig:CHAphase} was found to be preserved by this truncation,
showing that rheochaos is robust within the model, and is not
dependent on the presence of sharp interfaces between the
bands. Within
this truncation, the chaos in the pockets `C' was
found to set in via a classical period doubling scenario.

\subsection{Rheochaos: Relation of Macroscopic Theories to Experiment}

From the preceding sections it is clear that a wide range of oscillatory, irregular and chaotic behaviour can be found within relatively simple models of both shear-thinning and shear-thickening fluids, once structural memory is allowed for. Some of these models were directly intended to address giant micelles; others were not. 

Over the past several years, there have been a range of experimental studies, many of them referred to above, in which broadly similar flow behaviour was reported in micellar systems. 
There is an emerging consensus that in micelles, spatiotemporal rather than purely temporal instability is the norm.
However, it is in the very nature of unstable and chaotic systems that detailed prediction of particular trajectories is a near-futile task; these trajectories depend on both model parameters and initial conditions in an erratic fashion. The more important task, in confronting theory and experiment, is to develop robust tests of whether the physical content of the models is supported by or explains the experimental data. Such tests are likely to concern {\em scenarios} (such as whether the route to chaos is by period doubling, or some type of intermittency) rather than trajectories. At the time of writing, such detailed comparative work has barely begun. 

In many related fields, including work on inertial hydrodynamic instabilities in Newtonian fluids, the task is hampered by the difficulty of acquiring clean experimental data that can expose the route to chaos and/or the low-dimensional stucture of the underlying attractors on first entering the chaotic regime. Interestingly, careful recent work shows such measurements to be possible with very high precision in certain micellar systems; see for instance \cite{man04,lopez-Gonzalez2004}. The recent data of Ganapathy and Sood \cite{ganpath} deserves special mention. It shows a transition to chaos via a type II intermittency route, and gives direct evidence for coupling between concentrations and shear stress. 
This raises the hope that an increasing body of high quality experimental data will soon allow discrimination between the various theoretical models outlined above, and lead to better ones being formulated.

\section{Summary and Outlook}

In this article, we have reviewed theoretical modelling efforts that address the rheology of giant micelles. In the well-entangled regime, the linear viscoelastic spectra, often close to pure-Maxwell in character, are robustly explained by the reptation-reaction model which couples micellar kinetics to the tube dynamics of entangled objects.
The occurrence of a single relaxation time, despite the presence of exponential polydispersity in micellar lengths, arises from the rapidness of micellar reactions on the time scale of stress relaxation and the consequent averaging over micellar reaction dynamics of the local rates for stress decay.

In addressing the concentration dependence and ionic strength dependence of the Maxwell time, the reptation-reaction model has had relatively mixed success; however this is in part attributable to difficulties connecting the input parameters of the model (micellar end-cap energy, reaction rates, etc.) to quantities more directly measurable in experiment. 
Setting aside these difficulties, once one allows for micellar branching ---which, perhaps surprisingly, assists rather than impedes stress relaxation--- the model does seem to account for the main observational trends in the regime of linear viscoelasticity. 

For well-entangled micellar materials, shear-banding is routinely observed in the nonlinear flow regime. This was an early, robust prediction of the reptation-reaction model, which gives a nonmonotonic dependence of shear stress on strain rate in steady shear flows. (The model can also rationalize the ever-increasing normal stress observed as the mean shear.) The original reptation-reaction model was however based on a `first generation' description of the tube which also predicts nonmonotonic flow curves in monodisperse, unbreakable polymers.
In the past decade, `second generation' tube models, incorporating the concept of convective constraint release have been developed. So long as the phenomenological parameter describing the strength of this effect is neither too large nor too small, these `second generation' models can restore the observed monotonicity of the flow curve for unbreakable polymers, while retaining nonmonotonicity for micelles. The primary rheological predictions of the reptation-reaction model thereby remain intact.

Beyond explaining the basic tendency to form shear bands, it has not yet proved practical to relate the rather complex nature and dynamics of such bands to micellar constitutive models directly.
One must instead turn to macroscopic models which couple features of the reptation-reaction approach to spatially varying order parameters. These models currently offer several competing descriptions of a wide range of phenomena involving unsteady or chaotic banded flow, many of which were observed in recent experiments. Further detailed experimental work is highly desirable so as to enable better discrimination between the various models on offer. In particular one would like to clarify the interdependence between the following effects: local stress relaxation; concentration coupling (which causes the micellar volume fraction to vary between bands); stress diffusion (which can control the physics of band interfaces); normal stress effects at such interfaces;
curvature of streamlines (causing `purely elastic' instabilities); and structural memory. The last term refers to the evolution of internal degrees of freedom, such as the mean micellar length, on a time scale that can be at least as large as the characteristic time for stress relaxation. More than one of this list of physical ingredients could be implicated in the destabilization of steady shear bands, but we are not sure yet what is the balance of power among them. 
Additionally, the flows seen in entangled micelles can sometimes include bands of extremely high shear rate, where there could be very strong molecular alignment at the micellar scale. This could allow molecular physics of a nonuniversal kind (beyond that captured by coupling to liquid crystalline order parameters and/or shear-rate dependent micellar reaction rates) to influence the macroscopic rheology. This could occur in regimes where the {\em average} shear rate would not lead one to suspect any such effect. 

The position is even more complicated for weakly entangled micellar systems in the concentration window close to the onset of viscoelasticity. Here, drastic shear-thickening can be seen, such that at high flow rates the system becomes much more strongly entangled (or at least, much more viscoelastic) than at rest. The resulting shear-induced gel-like phase can be long lived, suggesting a role for structural memory in this case also. Further evidence comes from the fact that the thickening behaviour is often accompanied by shear banding and/or unsteady flows, with complex sample-history dependences in some cases. Microscopic modelling of this esoteric but fascinating regime  remains speculative, and offers limited guidance to formulating macroscopic models. 
However, one candidate, which can plausibly explain various of these experimental features, is a scenario in which the presence of micellar rings comes to dominate the rheology. Whether or not this eventually proves to be the correct picture, models in which the local Maxwell time is coupled to slowly evolving structural variables (such as the mean micellar chain length, or a gel fraction parameter) may offer a promising way forward in addressing the bafflingly complex experimental physics of giant micellar systems close to the onset of viscoelasticity.

On the experimental side also, the onset regime has tended to be neglected, perhaps because the fully entangled case is more relevant for applications. (The latter, not reviewed here, include personal care products, oilbore fluids, and creation of new biocompatible materials \cite{GMB}.) Another long-neglected area has been the study of entangled micelles in strongly elongational flows. Study of such flows has, in recent years, been a major driver in the development of improved constitutive models for unbreakable polymers, where these flows are crucial in process areas such as fibre-spinning. Clarifying the similarities and differences between micelles and conventional polymers in such flows could help to guide future modelling efforts for both classes materials.

\bibliographystyle{plain}


\begin{thebibliography}{99}

\bibitem{israelachvili}
Israelachvili, J. {\em Intermolecular and Surface Forces}, 2nd Ed., Academic Press, London, 1992.

\bibitem{herb} {\em Structure and Flow of Surfactant Solutions}, Herb, C. A., Prud'homme, R. K., Eds., ACS, Washington 1994 (Symp. Ser. 578).

\bibitem{rehagehoffmann} Rehage, H., Hoffmann, H., {\em Mol. Phys.} 1991, 74, 933.

\bibitem{GMB} Zana, R., and Kaler E., Eds., {\em Giant Micelles}, Taylor and Francis, London, 2007 in press.
 
\bibitem{degennes} de Gennes, P.-G. {\em Scaling Concepts in Polymer Physics}, Cornell University Press, Ithaca, 1979.

\bibitem{doiedwards} Doi, M., Edwards, S. F. {\em The Theory of Polymer Dynamics}, Clarendon Press, Oxford, 1986.

\bibitem{degennesnobel} de Gennes, P.-G. {\em Soft Matter (Nobel Lecture)}, {\em Rev. Mod. Phys.}
1992, 64, 645.

\bibitem{rrm} Cates, M. E. {\em Macromolecules} 1987, 20, 2289.


\bibitem{scott} Scott, R. L. {\em J. Phys. Chem.} 1965, 69, 261.



\bibitem{mukerjee} Mukerjee, P. {\em J. Phys. Chem.} 1972, 76, 565.


\bibitem{catescandau} Cates, M. E., Candau, S. J.
{\em J. Phys. Cond. Matt.} 1990, 2, 6869.

\bibitem{oelschlagermemory} Oelschlaeger, C.,  Waton, G., Buhler, E., Candau, S. J., Cates, M. E. {\em Langmuir} 2002, 18, 30276;
Oelschlaeger, C., Waton, G., Candau, S. J., Cates, M. E. {\em Langmuir} 2002, 18, 7265.

\bibitem{mackintosh} MacKintosh, F. C., Safran, S. A., Pincus, P. A. {\em Europhys. Lett.} 1990, 12, 697; Porte, G., Marignan, J., Bassereau, P., May, R. {\em J. Physique Paris} 49, 511; Odijk, T. {\em J. Phys. Chem.} 1989, 93, 3888.

\bibitem{drye} Drye, T. J., Cates, M. E. {\em J. Chem. Phys.} 1992, 96, 1367.

\bibitem{lequeux} Lequeux, F. {\em Europhys. Lett.} 1992, 19, 675.

\bibitem{networkrefs} Khatory, A., Kern, F., Lequeux, F., Appell, J., Porte, G., Morie, N., Ott, A., Urbach, W., {\em Langmuir} 1993, 9, 933.

\bibitem{portesatnet} Appell, J.; Porte, G. {\em Europhys. Lett.} 1990, 12, 185; Appell, J.; Porte, G.; Khatory, A.; Kern, F.; Candau, S. J. {\em J. Physique Paris II}, 1992, 2, 1045.

\bibitem{ringpolymers} Klein, J. {\em Macromolecules} 1986, 19, 105; Cates, M. E. and Deutsch, J., {\em J. de Physique} 1986, 47, 2121; Obukhov, S. P., Rubinstein, M. and Duke, T. {\em Phys. Rev. Lett.} 1994, 73, 1263; Hur, K., Winkler, R. G. and Yoon., D. Y. {\em Macromolecules} 2006, 39, 3775. 
\bibitem{ppw} Petscheck, R. G.;  Pfeuty, P.; Wheeler, J. C. {\em Phys. Rev. A} 1986, 34, 2391; Cordery, R., {\em Phys. Rev. Lett} 1981, 47, 457.

\bibitem{bose} Kubo, R. {\em Statistical Mechanics}, North Holland, Amsterdam, 1965.

\bibitem{ringstatpaps} Jacobson, H.; Stockmayer, W. H. {\em J. Chem. Phys.} 1950, 18, 1600; Porte, G. {\em J. Phys. Chem.} 1983, 87, 3541.

\bibitem{in} In, M., Aguerre-Chariol, O., Zana, R., {\em J. Phys. Chem. B} 1999, 103, 7747; Bernheim-Groswasser, A., Zana, R., Talmon, Y.
{\em J. Phys. Chem. B} 2000, 104, 4005.

\bibitem{rings} Cates, M. E., Candau, S. J., {\em Europhys. Lett.} 2001, 55, 887.

\bibitem{catesjphysique} Cates, M. E. {\em J. Physique Paris} 1988, 49, 1593. 

\bibitem{catesjpl} Cates, M. E., {\em J. Physique Paris Lett.} 1985, 46, 1059.

\bibitem{beno} O'Shaughnessy, B., Yu, J. {\em Phys. Rev. Lett.} 1995, 74, 4329; Friedman, B., O'Shaughnessy, B. {\em Macromolecules} 1993, 26, 5726; O'Shaughnessy, B. {\em Phys. Rev. Lett.} 1993, 71, 3331.

\bibitem{marquesstep} Marques, C. M., Turner, M. S., Cates, M. E. {\em J. Chem. Phys.} 1993, 99, 7260.

\bibitem{db} van Kampen, N. G. {\em Stochastic Processes in Physics and Chemistry}, North Holland, Amsterdam, 1981.

\bibitem{turnerjump} Turner, M. S., Cates, M. E., {\em J. Physique Paris} 1990, 51, 307.

\bibitem{marquesjump} Marques, C. M., Cates, M. E.,  {\em J. Physique Paris II} 1991, 1, 489.

\bibitem{kern} Kern, F., Lemarechal, P., Candau, S. J., Cates, M. E. {\em Langmuir} 1992, 8, 437.

\bibitem{bint} Turner, M. S., Marques, C. M.,  Cates, M. E., {\em Langmuir} 1993, 9, 695.

\bibitem{rehagebi} Shikata, T., Hirata, H., Kotaka, T., {\em Langmuir} 1987, 3, 1081; {\em Langmuir} 1988, 4, 354; {\em Langmuir} 1989, 5, 398; Shikata, T., Hirata, H., Takatori, E., Osaki, K. {\em J. Non-Newtonian Fluid Mech.} 1988, 28, 171.

\bibitem{pine} Pine, D. J. In {\em
Soft and Fragile Matter: Nonequilibrium Dynamics, Metastability and Flow},  Cates, M. E., Evans, M. R., Eds., IOP Publishing, Bristol 2000.

\bibitem{callaghan} Callaghan, P. T.,  {\em Repts. Prog. Phys.} 1999, 62, 599. 



\bibitem{allen} Allen, M. P., Tildesley, D. J., {\em Computer Simulation of Liquids}, 2nd Ed, Oxford Univ. Press 1989.

\bibitem{kremer} Kremer, K., In {\em
Soft and Fragile Matter: Nonequilibrium Dynamics, Metastability and Flow},  Cates, M. E., Evans, M. R., Eds., IOP Publishing, Bristol 2000.

\bibitem{rehagefigure} Rehage, H., Hoffmann, H. {\em J. Phys. Chem.} 1988, 92, 4712.

\bibitem{mcleish02} McLeish, T.C.B., {\em Adv. in Phys.} 2002, 51, 1379.



\bibitem{tirrell} Merrill, W. W., Tirrell, M., Tassin, J. F. and Monnerie, L., {\em Macromolecules} 22, 896 (1989).

\bibitem{milner} Likhtman, A. E. and McLeish, T. C. B., {\em Macromolecules} 2002, 35, 6332;  Milner, S. T., McLeish, T. C. B., {\em Phys. Rev. Lett.} 1998, 81, 725. For earlier work see also Klein, J. {\em, Macromolecules} 1978, 11, 852 and Graessley, W. W. {\em Adv. Polym. Sci.} 1982, 47, 67.




\bibitem{larson} Larson, R. G. {\em The Structure and Dynamics of Complex Fluids}, Clarendon Press, Oxford 1999.


\bibitem{graessley} Graessley, W. W., McLeish, T. C. B. In {\em Stealing the Gold: A Celebration of the Pioneering Physics of Sam Edwards}, Goldbart, P. M., Goldenfeld, N., Sherrington, D., Eds., Clarendon Press, Oxford, 2004.


\bibitem{giesekusadv} Holz, T., Fischer, P., Rehage, H. {\em J. Non-Newtonian Fluid Mech.} 1999, 88, 133; Fischer, P., Rehage, H. {\em Rheol. Acta} 1997, 36, 13.

\bibitem{recentpoly} Leygue, A., Bailly, C. and Keunings, R., {\em J. Non-Newtonian Fluid Mech.}, 2006, 133, 28; Maier, D., Eckstein, A., Friedrich, C. and Honerkamp, J. {\em J. Rheol.} 1998, 42, 1153.

\bibitem{jphyschem} Cates, M. E. {\em J. Phys. Chem.} 1990, 94, 371.

\bibitem{majid} Magid, L. J. {\em J. Phys. Chem. B} 1998, 102, 4064.

\bibitem{hoffmannrev} Hoffmann, H., Loebl, M., Rehage, H., Wunderlich, I. {\em Tenside Detergents} 1986, 22, 290;   Hoffmann, H., Ebert, G., {\em Ange. Chemie} 1988, 27, 902;
Hoffmann, H., Platz, G., Rehage, H.,  
Schorr, W., Ulbricht, W., {\em Ber. Bunsen-Ges. Phys. Chem. Chem. Phys.}, 1981, 85, 255.

 
\bibitem{turnerlang91} Turner, M. S., Cates, M. E., {\em Langmuir} 1991, 7, 1590.

\bibitem{granek} Granek, R., Cates, M. E. {\em J. Chem. Phys.} 1992, 96, 4758; Granek, R., {\em Langmuir} 1994, 10, 1627.

\bibitem{berretcole} Berret, J. F., Appell, J., Porte, G. {\em Langmuir} 1993, 9, 2851.

\bibitem{zanacole} Kern, F., Zana, R., Candau, S. J. {\em Langmuir} 1991, 7, 1344.

\bibitem{oelschlagertjump} Oelschlaeger, C., Waton, G., Candau, S. J. {\em Langmuir} 2003, 19, 10495.

\bibitem{oelschlagertjump2} C. Oelschlaeger, Ph.D. Thesis, University of Strasbourg, 2003.

\bibitem{nmr} Mair, R. W., Callaghan, P. T. {\em Europhys. Lett.} 1996, 36, 241; {\em J. Rheol.} 1997, 41, 901;  Fischer, E., Callaghan, P. T. {\em Phys. Rev. E} 2001, 64, 011501. 

\bibitem{nmr2} Britton, M. M., Callaghan, P. T. {\em Eur. Phys. J. B} 1999,7 237.

\bibitem{nmr3} Holmes, W. M., Lopez-Gonzales, M. R., Callaghan, P. T. {\em Europhys. Lett.} 2003, 64, 274.
\bibitem{birefringence} Decruppe, J.P., Cressely, R., Makhloufi, R., Cappalaere, E. {\em Colloid and Polym. Sci.} 1995, 273, 346; Makhloufi, R., Decruupe, J. P., Aitali, A., Cressely, R. {\em Europhys. Lett.} 1995, 32, 253; Lerouge, S., Decruppe, J. P., Humbert, C. {\em Phys. Rev. Lett.} 1998, 81, 5457.

\bibitem{catesherb} Cates, M. E. In {\em Structure and Flow of Surfactant Solutions}, Herb, C. A., Prud'homme, R. K., Eds., ACS, Washington 1994 (Symp. Ser. 578).

\bibitem{spenley} Spenley, N. A., Cates, M. E., McLeish, T. C. B. {\em Phys. Rev. Lett.} 1993, 71, 939.

\bibitem{spenleythesis} Spenley, N. A., Ph.D Thesis, University of Cambridge (1994).

\bibitem{banding} Olmsted, P. D., {\em Curr. Opinion in Colloid Interface Sci.} 1999), 4, 95.

\bibitem{olmstedluball} 
Lu, C. Y. D, Olmsted, P. D., Ball R. C. {\em Phys. Rev. Lett.} 2000, 84, 642.


\bibitem{grand} Grand, C., Arrault, J., Cates M. E., {\em J. Physique Paris II} 1997, 7, 1071.

\bibitem{berret} 
Berret, J. F., Porte, G., Decruppe, J. P. {\em Phys. Rev. E.} 1997, 55 1668.

\bibitem{cmm} Cates, M. E., McLeish, T. C. B., Marrucci, G. {\em Europhys. Lett.} 1993, 21, 451.


\bibitem{berretband} 
Berret, J. F., Roux, D. C., Porte, G., Lindner, P. {\em Europhys. Lett.} 1994, 25, 521; Porte, G., Berret, J. F., Harden, J. L. {\em J. Physique Paris II}, 1997, 7, 459.

\bibitem{olmstedramp} Fielding, S. M., Olmsted, P. D. {\em Eur. Phys. J E} 2003, 11, 65.

\bibitem{polymerband} Tapiada, P., Wang, S. Q. {\em Phys. Rev. Lett.} 2003, 91, 198301; {\em Macromolecules} 2004, 37, 9083.

\bibitem{CCR} Ianniruberto, G., and Marrucci, G. {\em J. Non-Newtonian Fluid Mech.} 1996, 65, 241; ibid, 2000, 95, 363; Mead, D. W., Larson, R. G., and Doi, M. {\em Macromolecules} 1998, 31, 7895.

\bibitem{milner2} Milner, S. T., McLeish, T. C. B., Likhtman, A. E., {\em J. Rheol.} 2001, 45, 539.



\bibitem{canscale} Candau, S. J., Hirsch, E., Zanan R., Delsanti, M. {\em Langmuir} 1989, 5, 1525.



\bibitem{leng} Leng, J., Egelhaaf, S. U., Cates, M. E. {\em Biophys. J.} 2003, 85, 1624.

\bibitem{hoffmannthicken} Rehage, H., Wunderlich, I., Hoffmann, H. {\em Prog. Colloid Polym. Sci.} 1986, 72, 11; Hoffmann, H., Rehage, H., Wunderlich, I., {\em Rheol. Acta} 1987, 26, 532; Rehage, H., Hoffmann, H. {\em Rheol. Acta}, 1982, 21, 561; Hoffmann, H., Rasucher, A., Hoffmann, H. {Ber. Bunsen-Ges. Phys. Chem. Chem. Phys.} 1991, 95, 153.

\bibitem{matthys} Hu, Y. T., Matthys, E. F. {\em J. Rheol.} 1997, 41, 151.

\bibitem{oda} Oda, R., Panizza, P., Schmutz, M., Lequeux, F. {\em Langmuir}, 1997, 13, 6407.

\bibitem{pinethick} Boltenhagen, P., Hu, Y. T., Matthys, E. F., Pine, D. J. {\em Phys. Rev. Lett.} 1997, 79, 2369; {\em Europhys. Lett.} 1997, 38, 389.

\bibitem{berretmemory} Berret, J. F., Gamez-Corrales, R., Lerouge, S., Decruppe, J. P. {\em Eur. Phys. J. E} 2000, 2, 343.


\bibitem{aggregation} Cates, M. E., Turner, M. S. {\em Europhys. Lett.} 1990, 7, 681; Wang, S. Q., Gelbart, W., Ben-Shaul, A. {\em J. Phys. Chem.} 1990, 94, 2219.

\bibitem{goveas1} Goveas, J. L., Olmsted, P. D. {\em Eur. Phys. J. E} 2001, 6, 79.

\bibitem{goveas2} Picard, G., Ajdari, A., Bocquet, L., Lequeux, F. {\em Phys. Rev. E} 2002, 66, 051501; Ajdari, A., {\em Phys. Rev. E} 1998, 58, 6294.


\bibitem{liu} Barentin, C., Liu, A. J. {\em Europhys. Lett.} 2001, 55, 432.































\bibitem{milner93}
Milner, S. T.,
\newblock {\em Phys.~Rev. E} 1993, 48, 3674.

\bibitem{dhont2003b}
Dhont, J. K. and Briels, W. J,
\newblock {\em J. Chem. Phys.} 2003, 118, 1466.

\bibitem{johnson77}
Johnson, M. and Segalman, D.,
\newblock {\em J. Non-Newtonian Fluid Mech.} 1977, 2, 255.

\bibitem{olmsted99a}
Olmsted, P. D., Radulescu, O. and Lu, C.-Y. D.,
\newblock {\em J. Rheology} 
2000, 44, 257.

\bibitem{fielding2003c}
Fielding, S. M., and Olmsted, P. D.,
\newblock {\em Phys. Rev. Lett.} 2003, 90, 224501.

\bibitem{rossi}
Cook, L. P., and Rossi, L. F., {\em J. Non-Newtonian Fluid Mech.} 
2004, 116, 347.

\bibitem{rossi1}
Rossi, L. F., McKinley, G., and Cook, L. P. {\em J. Non-Newtonian Fluid Mech.} 
2006, 136, 79.



\bibitem{brochdgen77}
Brochard, F.  and de~Gennes, P.-G.,
\newblock {\em Macromolecules} 1977, 10, 1157.

\bibitem{milner91}
Milner, S. T.,
\newblock {\em Phys. Rev. Lett.} 1991, 66, 1477.

\bibitem{deGen76}
de~Gennes, P.-G.,
\newblock {\em Macromolecules} 1976, 9, 587.

\bibitem{Brochard83}
Brochard, F.,
\newblock {\em J.Physique. (Paris)} 1983, 44, 39.

\bibitem{WPD91}
Wu, X. L., Pine, D. J. and Dixon, P. K.,
\newblock {\em Phys. Rev. Lett.} 1991, 66, 2408.

\bibitem{GerHigCla99}
Gerard, H., Higgins, J. S., and Clarke, N.,
\newblock {\em Macromolecules} 1999, 32, 5411.

\bibitem{WheIzuFul96}
Wheeler, E., Izu, P., and Fuller, G. G.,
\newblock {\em Rheol. Acta} 1996, 35, 139.

\bibitem{KadEgm97}
Kadoma, I. A., and van Egmond, J. W.,
\newblock {\em Langmuir} 1997, 13, 4551.

\bibitem{HelfFred89}
Helfand W., and Fredrickson, G. H.,
\newblock {\em Phys. Rev. Lett.} 1989, 62, 2468.


\bibitem{schmitt95}
Schmitt, V., Marques, C. M., and F.~Lequeux, F.,
\newblock {\em Phys.~Rev. E} 1995, 52, 4009.

\bibitem{DecLerBer01}
Decruppe, J. P., Lerouge, S., and Berret, J.-F.,
\newblock {\em Phys. Rev. E} 1001, 63, 022501.

\bibitem{yuan2002}
Yuan, X.-F., and Jupp, L., {\em Europhys Lett.} 2002, 60, 691.

\bibitem{spenley96}
Spenley, N. A., Yuan, X.-F., and Cates, M. E.,
\newblock {\em J.~Physique (Paris)~II} 1996, 6, 551.

\bibitem{Yerushalmi70}
Yerushalmi, J., Katz, S., and Shinnar, R.,
\newblock {\em Chem. Eng. Sci.} 1970, 25, 1891.


\bibitem{DE4}
Doi, M. and Edwards, S. F.,
\newblock{\em J. Chem. Soc. Faraday Trans. II} 1979, 75, 38.

\bibitem{fielding2003a}
Fielding, S. M., and Olmsted, P. D.,
\newblock {\em Phy. Rev. E} 2003, 68, 036313.

\bibitem{fielding2005}
Fielding, S. M.,
\newblock {\em Phys. Rev. Lett.} 2005, 95, 134501.

\bibitem{doi81}
Doi, M.,
\newblock {\em J.~Polym. Sci: Polym. Phys.} 1981, 19, 229.



\bibitem{doikuzuu83}
Kuzuu, N.,  and Doi, M.,
\newblock {\em J.~Phys. Soc. Jap.} 1983, 52, 3486.


\bibitem{OlmsLu99}
Olmsted, P. D., and Lu, C.-Y. D.,
\newblock {\em Phys. Rev. E} 1999, 60, 4397.

\bibitem{cates2002}
Cates, M. E., Head, D. A., and Ajdari, A.,
\newblock {\em Phys. Rev. E} 2002, 66, 025202.

\bibitem{olmsted99b}
Olmsted, P. D.,
\newblock {\em Europhys. Lett.} 1999, 48, 339.

\bibitem{giesekus1982}
Giesekus, H.,
\newblock {\em J. Non-Newtonian Fluid Mech.} 1982, 11, 69.

\bibitem{Manero00}
Bautista, F.,  Soltero, J. F. A., Perez-Lopez, J. H., Puig, J. E., and Manero, O., {\em J. Non-Newtonian Fluid Mech.} 2000, 94, 57.

\bibitem{Manero99}
Bautista, F., de Santos, J. M., Puig, J. E., Manero, O., {\em J. Non-Newtonian Fluid Mech.} 1999, 80, 93.

\bibitem{Manero02}
Bautista, F., Soltero, J. F. A., Macias, E. R., Puig, J. E., Manero, O., {\em J. Phys. Chem. B} 2002, 106, 13018.

\bibitem{goveas1999a}
Goveas, J. L.,  and Pine, D. J.,
\newblock {\em Europhys. Lett.} 1999, 48, 706.

\bibitem{Yuan02}
Yuan, X.-F. {\em Europhys. Lett} 2002, 60, 691.




\bibitem{dhont1999a}
Dhont, J. K.,
\newblock {\em Phys. Rev. E} 1999, 60, 4534.


\bibitem{hu1998b}
Hu,Y. T., Boltenhagen, P., Matthys, E., and Pine, D. J.,
\newblock {\em J. Rheology} 1998, 42, 1209.


\bibitem{wheeler1998}
Wheeler, E. K., Fischer, P., and Fuller G. G.,
\newblock {\em J. Non-Newtonian Fluid Mech.} 2998, 75, 193.

\bibitem{fischer2000}
Fischer, E., and Callaghan, P. T.,
\newblock {\em Europhys. Lett.} 1000, 50, 803.

\bibitem{fischer2002}
Fischer, E., Callaghan, P. T., Heatley, F., and Scott, J. E.,
\newblock {\em J. Mol. Struct.} 2002, 602, 303.

\bibitem{herle2005}
Herle, V., Fischer, P., and Windhab, E. J.,
\newblock {\em Langmuir} 2005, 21, 9051.

\bibitem{bandyopadhyay2000}
Bandyopadhyay, R., Basappa, G., and Sood, A. K.,
\newblock {\em Phys. Rev. Lett.} 2000, 84, 2022.

\bibitem{bandyopadhyay2001}
Bandyopadhyay, R., and Sood, A. K.,
\newblock {\em Europhys. Lett.} 2001, 56, 447.


\bibitem{lopez-Gonzalez2004}
Lopez-Gonzalez, M. R., Holmes, W. M., Callaghan, P. T., and Photinos, P. J.,
\newblock {\em Phys. Rev. Lett.} 2004, 93, 268302.

\bibitem{faber} Faber, T. E., {\em Fluid Mechanics for Physicists}, Cambridge Univ. Press, Cambridge (1995).

\bibitem{SalColRou02}
Salmon, J. B., Colin, A., and Roux, D.,
\newblock {\em Phys. Rev. E} 2002, 66, 031505.

\bibitem{SalManCol03b}
Salmon, J. B., Manneville, S., and Colin, A.,
\newblock {\em Phys. Rev. E} 2003, 68, 051504.

\bibitem{lootens2003}
Lootens, D., Van~Damme, H., andHebraud, P.,
\newblock {\em Phys. Rev. Lett.} 2003, 90, 178301.

\bibitem{HilVla02}
Hilliou, L., and Vlassopoulos, D.,
\newblock {\em Ind. Eng. Chem. Res.} 2002, 41, 6246.

\bibitem{ManSalCol04}
Manneville, S., Salmon, J. B., and A~Colin, A.,
\newblock {\em Eur. Phys. J. E} 2004, 13, 197.

\bibitem{GroKeuCreMaf01}
Grosso, M., Keunings, R., Crescitelli, S., and Maffettone, P. L..
\newblock {\em Phys. Rev. Lett.} 2001, 86, 3184.



\bibitem{RieKroHes02}
Rienacker, G., Kroger, A., and S~Hess, S.,
\newblock {\em Physica A} 2002, 315, 537.

\bibitem{DasGel}
Das, M., Bandyopadhyay, R., Chakrabarti, B., Ramaswamy, S., Dasgupta, C. and Sood, A. K., in {\em Molecular Gels}, Eds., Terech, P. and Weiss, R. G., Kluwer, 2006.


\bibitem{hess76}
Hess, S.,
\newblock {\em Z. Naturforsch.} 1976, 31a, 1507.


\bibitem{larson1990e}
Larson, R. G.,
\newblock {\em Macromolecules} 1990, 23, 3983.

\bibitem{feng1998}
Feng, J., Chaubal, C. V., and Leal, L. G.,
\newblock {\em J. Rheology} 1998, 42, 1095.



\bibitem{LarsOtti91}
Larson, R. G.,  and Ottinger, H. C.,
\newblock {\em Macromolecules} 1991, 24, 6270.

\bibitem{RieKroHes02b}
Rienacker, G., Kroger, M., and Hess, S.,
\newblock {\em Phys. Rev.  E} 2002, 66, 040702.

\bibitem{fielding2004}
Fielding, S. M.,  and Olmsted, P. D.,
\newblock {\em Phys. Rev. Lett.} 2004, 92, 084502.

\bibitem{FisWheFul02}
Fischer, P., Wheeler, E. K., and G~G Fuller, G. G., 
\newblock {\em Rheol. Acta} 2002, 41, 35.







\bibitem{HBP98}
Hu, Y. T., Boltenhagen, P., and Pine, D. J.,
\newblock {\em J. Rheology} 1998, 42, 1185.

\bibitem{WunColLenArnRou01}
Wunenburger, A. S., Colin, A., Leng, J., Arneodo, A., and Roux, D.,
\newblock {\em Phys. Rev.  Lett.} 2001, 86, 1374.

\bibitem{upcoming_2d}
Fielding, S. M.,  and Olmsted, P. D. {\em Phys. Rev. Lett.} 2006, 96, 104502.

\bibitem{Yuan99}
Yuan. X.-F., {\rm Europhys. Lett} 1999, 46, 542.



\bibitem{das2005}
Das, M., Chakrabarti, B., Dasgupta, C., Ramaswamy, S., and Sood, A. K., 
\newblock {\em Phys. Rev. E} 2005, 71, 021707.

\bibitem{chakrabarti2004}
Chakrabarti, B., Das, M., Dasgupta, C., Ramaswamy, S., and Sood, A. K.,
\newblock {\em Phys. Rev. Lett.} 2004, 92, 055501.

\bibitem{Drazin}
Drazin, P. and Reid, W. H., {\em Hydrodynamic stability} (Cambridge
University Press, Cambridge, 2004).

\bibitem{Mullin}
Hof, B., Juel, A., and Mullin, T., {\em Phys. Rev. Lett.} 2003, 91, 244502.

\bibitem{LarsonReview}
Larson, R. G.,
\newblock {\em Rheological Acta} 1992, 31, 213.



\bibitem{larson1990c}
Larson, R. G., Shaqfeh, E. S., and Muller, S. J.,
\newblock {\em J. Fluid Mech.} 1990, 218, 573.




\bibitem{PhanThien}
Phan-Thien N., {\em J. Non-Newtonian Fluid Mech.} 1983, 13, 325.

\bibitem{MulLarSha}
Muller, S. J., Larson, R. G., and Shaqfeh, E. S. G., {\em Rheol. Acta}
1989, 28, 499.

\bibitem{MagLar88}
Magda, J. J. and Larson, R. G., {\em J. Non-Newtonian Fluid Mech.} 1988, 30, 1.



\bibitem{larson2000c}
Larson, R. G.,
\newblock {\em Nature} 2000, 405, 27.




\bibitem{groisman2000}
Groisman, A., and Steinberg, V.,
\newblock {\em Nature} 2000, 405, 53.


\bibitem{mckinley1991}
Mckinley, G. H., Byars, J. A., Brown, R. A., and Armstrong, R. C.,
\newblock {\em J. Non-Newtonian Fluid Mech.} 1991, 40, 201.

\bibitem{morozov2005}
Morozov A. N., and van Saarloos, W.,
\newblock {\em Phys. Rev. Lett.} 2005, 95, 024501.

\bibitem{aradian2005}
Aradian, A., and Cates, M. E.,
\newblock {\em Europhys. Lett.} 2005, 70, 397.

\bibitem{upcoming_CHA}
Aradian, A.,  and Cates, M. E., {\em Phys. Rev. E}, 2006, 73, 041508.



\bibitem{man04} Becu, L., Manneville, S., and Colin, A.,
\newblock {\em Phys. Rev. Lett.} 2004, 93, 018301.

\bibitem{ganpath} Ganapathy, R., and Sood, A. K., 
\newblock {\em Phys. Rev. Lett.} 2006, 96, 108301.



\end{thebibliography}

\end{document}